\documentclass[12pt,a4paper,oneside,english,twoside]{book}
\usepackage[T1]{fontenc}
\usepackage[latin1]{inputenc}
\usepackage{amssymb}

\makeatletter

\def\pp#1{\partial_#1}
\usepackage{exscale}
\usepackage{amssymb,amsthm}
\def\obar{\overline}
\def\one{\mbox{1 \kern-.59em {\rm l}}}
\usepackage{graphicx}

\def\a{\alpha}         

\def\b{\beta}           

\def\c{\gamma} 
\def\C{\Gamma}  

\def\d{\delta}

\def\e{\epsilon}              
\def\vare{\varepsilon}

\def\g{\gamma}  
\def\G{\Gamma}

\def\la{\lambda}
\def\m{\mu}

\def\cA{\mathcal{A}}
\def\C{{\mathbb C}}
\def\cN{\mathcal{N}}

\def\R{{\mathbb R}}
\def\C{{\mathbb C}}

\def\Z{{\mathbb Z}}
\def\one{\mbox{1 \kern-.59em {\rm l}}}
\newcommand{\torus}{{\mathbb T}}

\def\nn{\nonumber}
\def\bit{\begin{itemize}}
\def\eit{\end{itemize}}

\def\tens{\otimes}

\def\reps{representations }

\def\pp#1{\partial_#1}
\def\mc#1{\mathcal{#1}}
\def\sl#1{\: \raisebox{1pt}{$\slash$} \hspace{- 7 pt} #1}
\def\dd{\partial}

\usepackage{epsfig}

\usepackage{fancyhdr}
\fancypagestyle{plain}{\fancyhf{} \fancyhead[LE,RO]{\thepage}}

\input{epsf}
\usepackage{ngerman}

\usepackage{babel}
\makeatother
\begin{document}
\begin{titlepage}

$\mbox{}$

\vspace{70pt}

\begin{center} \huge \bfseries

Noncommutative Gauge Theory \\ {\Large beyond the} \\Canonical Case

\end{center}

\vspace{20pt}

\begin{center} \small

Dissertation der Fakultät für Physik \\ der  \\ LudwigMaximiliansUniversität München

\end{center}

\vspace{10pt}

\begin{center} \small

vorgelegt von

\end{center}

\vspace{10pt}

\begin{center} 

Wolfgang Behr

\end{center}

\begin{center} \small

\vspace{25pt}

Juli 2005

\end{center}

\vspace{40pt}


\end{titlepage}

\thispagestyle{empty}

\newpage

$\mbox{}$

 \vspace{18cm}

{\small

Wolfgang Behr 

MPI f\"ur Physik und ASC for Theoretical Physics, M\"unchen 

behr@theorie.physik.uni-muenchen.de }

\newpage

$\mbox{}$

\vspace{18cm}

\thispagestyle{empty}

\pagenumbering{roman}

Tag der m\"undlichen Pr\"ufung: 3. November 2005

\vspace{0,5cm}

$\begin{array}{rl}
\mbox{erster Gutachter:} & \mbox{Prof. Dr. Julius Wess}\\
\mbox{zweiter Gutachter:} & \mbox{Prof. Dr. Ivo Sachs}\end{array}$\newpage

\thispagestyle{empty}

$\mbox{}$

\newpage

$\mbox{}$

\vspace{20pt}

{

{\Huge \bf Abstract}

\vspace{25pt}

Canonically deformed spacetime, where the commutator of two coordinates is a constant, is the most commonly studied noncommutative space.  Noncommutative gauge theories that have ordinary gauge theory as their commutative limit have been constructed there. But these theories have their drawbacks: First of all, constant noncommutativity can only be an approximation of a realistic theory, and therefore it is necessary to study more complicated space-dependent structures as well. Secondly, in the canonical case, the noncommutativity didn't fulfill the initial hope of curing the divergencies of quantum field theory. Therefore it is very desirable to understand noncommutative spaces that really admit finite QFTs.

These two aspects of going beyond the canonical case will be the main focus of this thesis. They will be addressed within two different formalisms, each of which is especially suited for the purpose.

In the first part noncommutative spaces created by $\star$-products are studied. In the case of nonconstant noncommutativity, the ordinary derivatives possess a deformed Leibniz rule, i.e. $\dd_i (f\star g)\neq \dd_i f \star g + f \star \dd_i g $. Therefore we construct new objects that still have an undeformed Leibniz rule. These derivations of the $\star$-product algebra can be gauged much in the same way as in the canonical case and lead to function-valued gauge fields. By linking the derivations to frames (vielbeins) of a curved manifold, it is possible to formulate noncommutative gauge theories that admit nonconstant noncommutativity and go to gauge theory on curved spacetime in the commutative limit. We are also able to express the dependence of the noncommutative quantities on their corresponding commutative counterparts by using Seiberg-Witten maps.

In the second part we will study noncommutative gauge theory in the matrix theory approach. There, the noncommutative space is the ground state of a matrix action, the fluctuations around this ground state creating the gauge theory. In the canonical case the matrices used are infinite-dimensional (they are the Fock-space representation of the Heisenberg algebra), leading to a number of problems, especially with divergencies. Therefore we construct gauge theory using finite dimensional matrices (fuzzy spaces). This gauge theory is finite, goes to gauge theory on a 4-dimensional manifold in the commutative limit and can also be used to regularize the noncommutative gauge theory of the canonical case. In particular, we are able to match parts of the known instanton sector of the canonical case with the instantons of the finite theory.

}

\newpage 

\thispagestyle{empty}  $\mbox{}$\newpage

\chapter*{Acknowledgements}

I want to express my gratitude to all those who made this thesis possible. 

\vspace{30pt}

\noindent  I especially want to thank

\vspace{20pt}

\noindent Prof. Dr. Julius Wess for his guidance and steady support,\vspace{10pt}

\noindent Andreas Sykora, Frank Meyer and Harold Steinacker for the many hours
spent together in front of the blackboard, the fruitful discussions
and their collaboration on parts of the material included in this
thesis, \vspace{10pt}

\noindent all and everyone in the group for the congenial, relaxed and productive
atmosphere\vspace{10pt}

\noindent and last but not least the MPI for funding me during these last three
years.

\fancyhead[LE,RO]{\thepage} 

\fancyhead[LO,RE]{\nouppercase{\leftmark}} 

\newpage \thispagestyle{empty}

\tableofcontents{}

\cleardoublepage

\fancyhead[LE,RO]{\thepage} 

\fancyhead[LO,RE]{\leftmark}

\pagenumbering{arabic}

\setcounter{page}{1}

\chapter{Introduction}

There is a simple Gedankenexperiment showing that any quantum theory
including gravity will make it impossible to measure distances smaller
than the Planck length: Trying to measure smaller and smaller distances,
we are forced to use test particles with more and more energy. But
this energy will affect the geometry of space itself, creating black
holes which finally become bigger than the distances we wanted to
measure (see e.g. \cite{Doplicher:1994tu}). Below the Planck length,
distance looses its meaning.

In the absence of a consistent formulation of quantum gravity, we
do not know the exact nature of quantized spacetime, but it is clear
that the usual notion of a differentiable manifold should be replaced
by something reflecting the quantum nature of spacetime at very small
distances. Following the well known ideas of quantum mechanics, the
uncertainty in the measurement of the coordinates leads directly to
the notion of noncommutative spaces.

There is another motivation for the introduction of noncommutative
spacetime, this time coming from quantum field theory. There, the
divergencies appearing in the quantization are UV-effects, and therefore
related to small distances. The introduction of noncommutativity could
work as a ultraviolet cut-off, making QFT finite. Even though the
UV-divergencies are now well under control through the renormalization
programme, they nevertheless suggest that spacetime should change
its nature at very small distances. 

To make spacetime noncommutative, the commutative algebra of functions
is usually replaced by a noncommutative algebra generated by coordinates
$\hat{x}^{i}$ with commutation relations \begin{equation}
[\hat{x}^{i},\hat{x}^{j}]=i\theta^{ij}.\label{erste}\end{equation}
In the canonical case, this commutator is a constant, i.e. $\theta^{ij}\in\mathbb{R}$.
Gauge theory on this space was studied in great detail in the last
few years, mainly due to its appearance in string theory. But if we
think that noncommutativity is an effect of quantum gravity, the canonical
case can only be the simplest example. Other, more complicated structures
should be studied, especially structures that are related to curved
backgrounds. But also in view of our second motivation, the canonical
case proved to be disappointing: it doesn't cure the infinities of
QFT, it rather adds new ones.

The aim of this thesis will be to extend noncommutative gauge theory
beyond the canonical case, in the two directions mentioned above:
towards noncommutative gauge theory on curved backgrounds in part
I, and towards gauge theory models which are actually finite in part
II. For these two goals, we will use two different approaches, two
different ways in which noncommutative gauge theory can be formulated
already in the canonical case: one using $\star$-products, the other
one matrices. \vspace{10pt}

\noindent

The notion of a $\star$-product came first up when Groenewold \cite{Groenewold:1946kp}
and Moyal \cite{Moyal:1949sk} used Weyl's quantization prescription
\cite{Weyl:1927vd} to pull back the noncommutativity of the quantum
mechanical position and momentum operators onto the classical phase
space. Later on, it was generalized in the framework of deformation
quantization \cite{Bayen:1977ha,Bayen:1977hb} to arbitrary symplectic
and Poisson manifolds. A $\star$-product is an associative noncommutative
product acting on functions on a manifold, the noncommutativity being
controlled by a deformation parameter. Expanded in this parameter,
one can write\begin{equation}
f\star g=f\cdot g+\theta^{ij}\dd_{i}f\dd_{j}g+\mc O(\theta^{2})\end{equation}
To zeroth order, the $\star$-product reproduces ordinary pointwise
multiplication, higher orders are bidifferential operators acting
on the functions. The first order term corresponds to a Poisson structure.
While every $\star$-product corresponds to a Poisson structure, the
opposite is also true: For on every Poisson manifold there is a $\star$-product
quantizing the Poisson structure \cite{Kontsevich:1997vb}. 

Interest in noncommutative gauge theory formulated with the help of
$\star$-products triggered when it became clear that it appears in
string theory as the low energy limit of open strings with a background
$B$-field \cite{Chu:1998qz}. In this picture, the endpoints of the
open strings on the D-brane cease to commute, and depending on the
regularization used, their behavior can be described either by noncommutative
Yang-Mills theory or by commutative Yang-Mills with background B-field.
These two descriptions can be linked by a map from the noncommutative
quantities to the commutative ones, the Seiberg-Witten map \cite{Seiberg:1999vs}.

The approach to noncommutative gauge theory most important to this
thesis was developed in Munich in a series of papers \cite{Madore:2000en,Jurco:2000ja,Jurco:2001rq},
noticing that multiplication with a noncommutative coordinate is no
longer a covariant operation. Then, coordinates have to be gauged
much in the same way that derivatives have to be gauged in commutative
gauge theory, leading to covariant coordinates. As the $\star$-product
in the canonical case behaves very much like an ordinary product with
respect to differentiation and integration, the covariant coordinates
can be used to formulate noncommutative gauge theory in close analogy
to the commutative case. The noncommutative theory is finally linked
to the commutative one by using Seiberg-Witten maps, allowing to deal
with nonabelian gauge theory as well. This way, it was possible to
construct a noncommutative version of the standard model \cite{Calmet:2001na,Melic:2005fm}
and study its phenomenological implications \cite{Behr:2002wx,Ohl:2004tn,Melic:2005hb,Melic:2005fm,Melic:2005am}.
The extension to supersymmetry is somewhat more complicated as the
SW-maps in general become nonlocal \cite{Mikulovic:2003sq}, but for
the case of a reduced $N=\frac{1}{2}$ supersymmetry it is still possible
\cite{Mikulovic:2004qj}. Lately, it was even possible to formulate
noncommutative gravity \cite{Aschieri:2005yw} for the canonical case.

As the noncommutative gauge transformations contain translations in
space, there can't exist local observables in noncommutative gauge
theory. But it was realized that in momentum space, certain Wilson
loops with fixed momentum are actually gauge invariant \cite{Ishibashi:1999hs,Gross:2000ba,Das:2000md}.
These Wilson loops do not quite close, but contain a gap corresponding
to the noncommutativity and the momentum, which is why they are also
referred to as open Wilson lines. These open Wilson lines were used
to construct the inverse SW-map for the field strength to all orders
\cite{Okawa:2001mv}. Other approaches to calculating SW-maps include
a solution for abelian gauge theory to all orders using the Kontsevich
formality map \cite{Kontsevich:1997vb,Jurco:2000fs,Jurco:2001my},
a cohomological procedure within the BRST formalism \cite{Brace:2001fj}
and a refined analysis of its internal structure \cite{Cerchiai:2002ss}.

In the quantization of noncommutative gauge theory, the legs of diagrams
can no longer be exchanged, leading to a distinction between planar
and non-planar diagrams \cite{Filk:1996dm}. The planar diagrams have
the same high energy behavior as their commutative counterparts, but
the nonplanar diagrams lead to what is called IR/UV-mixing \cite{Minwalla:1999px}.
The diagrams are made finite in the UV by oscillatory factors, but
only for finite momentum. For vanishing momenta, the divergencies
reappear, therefore mixing the UV and the IR behavior of the theory.
There are many studies on the renormalization properties of such theories
(see \cite{Douglas:2001ba,Szabo:2001kg} for references), but so far
the only consistently renormalizable theory is $\phi^{4}$-theory
with a special potential term added \cite{Grosse:2004ik}. 

There are several lines of research going beyond the canonical case
\cite{Calmet:2003jv}. Covariant coordinates and SW-maps can be constructed
for arbitrary Poisson manifolds \cite{Jurco:2000fs,Jurco:2001kp,Jurco:2001my},
but the limit to commutative gauge theory no longer is clear. On $\kappa$-deformed
spacetime, it was possible to establish noncommutative gauge theory,
the nonconstant commutator of the coordinates leading to derivative
valued gauge fields \cite{Dimitrijevic:2003wv,Dimitrijevic:2003pn,Dimitrijevic:2004vv}.
Somewhat closer to our approach, gauge theory on the $E_{q}(2)$-covariant
plane was studied using frames \cite{Meyer:2003wj}. More recently,
there have been attempts using coordinate transformations from the
canonical case to more complicated algebras \cite{Correa:2004cm,Fosco:2004yz,Pinzul:2005nh}.

\vspace{10pt}

The first part of this thesis will be devoted to expanding noncommutative
gauge theory to more general $\star$-products and relating it to
gauge theory on curved spacetime. 

In chapter 2, we will first discuss the canonical case. We introduce
the $\star$-product usually used in this case, the Moyal-Weyl $\star$-product.
With this, noncommutative gauge theory is formulated in the standard
way. As this approach can only deal with $U(N)$ gauge groups, we
introduce Seiberg-Witten maps to accommodate for general gauge groups.
We end this chapter with discussing noncommutative observables.

In chapter 3, we start with the general definition of $\star$-products,
and show how they arise out of ordering prescriptions of algebras.
For a special ordering, the Weyl- (or symmetric) ordering, we then
calculate the corresponding $\star$-product to second order for general
algebras, a result already published in \cite{Behr:2003qc} together
with Andreas Sykora. Two other $\star$-products are presented as
well, the Jambor-Sykora $\star$-product \cite{Jambor:2004kc} and
Kontsevich's formality $\star$-product \cite{Kontsevich:1997vb}.
After discussing integration on such $\star$-product algebras, we
end with concrete examples.

In chapter 4, we discuss derivatives and derivations on $\star$-product
algebras. For the canonical case, the usual derivatives still had
the undeformed Leibniz rule. For general $\star$-products, this is
no longer the case. The derivatives acquire a nontrivial coproduct,
which means that their Leibniz rule is deformed. But for our construction
of gauge theory we will need objects that still have the usual Leibniz
rule, i.e. derivations of the $\star$-product algebra. We are able
to identify such objects by linking them to vector fields commuting
with the Poisson structure corresponding to the $\star$-products.
We explicitly construct these derivations for the three $\star$-products
introduced in chapter 4, and end with the continuation of the example
from chapter 3.

In chapter 5, we use the derivations to construct gauge theory. As
the derivations have the usual Leibniz rule, they can be gauged in
full analogy to the canonical case, leading to function-valued gauge
fields and field strength. As we want the noncommutative gauge theory
to have a meaningful commutative limit, we link it to gauge theory
on curved spacetime by introducing frames. On the commutative side,
frames can be introduced to diagonalize the metric. If they fulfill
a compatibility condition with the Poisson structure of the noncommutative
space, we can lift them to derivations of the $\star$-product algebra.
Then we use these derivations to build a noncommutative gauge theory
that in the commutative limit reduces to gauge theory on curved spacetime.
We give an example where the spacetime of the commutative limit is
a manifold with constant curvature. To deal with general gauge groups,
we again introduce SW-maps from the noncommutative to the commutative
quantities. For the Weyl-ordered $\star$-product, we calculate the
SW-maps for all relevant quantities up to second order. For the formality
$\star$-product we are able to construct the SW-maps to all orders
for abelian gauge theory. The results of this chapter (and parts of
the preceding chapter) have already been published in \cite{Behr:2003qc}
together with Andreas Sykora.

In chapter 6, the last chapter on the $\star$-product approach, we
start with noticing that covariant coordinates can be defined for
any $\star$-product, and use them to construct noncommutative analogs
of Wilson lines. These can then be used to build noncommutative observables
and to extend the construction of the inverse SW-map of \cite{Okawa:2001mv}
to general $\star$-products with nondegenerate Poisson structure.
This has been published in \cite{Behr:2003hg}, again together with
Andreas Sykora.

\vspace{10pt}

But $\star$-products aren't the only way to express the noncommutativity
(\ref{erste}). In the canonical case, the algebra of the coordinates
is nothing but the well known Heisenberg algebra, and we can use the
creator and annihilator formalism to represent it. The coordinates
then become infinite-dimensional matrices acting on a Fock space,
the derivatives commutators with the coordinates and integration the
trace over the Fock space. Gauge transformations are now unitary transformations,
and we again have to gauge the coordinates $x^{i}$ to get covariant
coordinates $X^{i}=x^{i}+A^{i}$. The gauge theory action\begin{equation}
S=c\:{\bf tr}\:([X^{i},X^{j}]-i\theta^{ij})^{2}\label{zwei}\end{equation}
 can be expressed entirely in terms of the dynamical matrix variables
$X^{i}$, reproducing the noncommutative space as the ground state,
with the fluctuations forming the gauge theory. In the canonical case,
this description is equivalent to the $\star$-product approach, but
it is the better framework to address nonperturbative questions such
as topological solutions.

The instanton sector of noncommutative gauge theory is very rich,
and many classical constructions can be reformulated on the noncommutative
side. In two dimensions, all instantons have been classified \cite{Gross:2000ss},
but in four dimensions the picture is far more complicated. There
are the generalizations of the two-dimensional instantons (which will
become important in this thesis), but there are many other instantons
as well, which can be found by using a noncommutative ADHM-construction
or Nahm's equations (see \cite{Douglas:2001ba} for references). 

The quantization of the model of course is troubled by the same divergencies
as the one constructed via $\star$-products, but the exact definition
is quite nontrivial for another reason as well: the theory contains
sectors with any rank of the gauge group $U(n)$ \cite{Gross:2000ss}.
To have a well-defined theory and quantization prescription, a regularization
of gauge theory on $\R_{\theta}^{d}$ is therefore very desirable.

Luckily, there is a number of cases (in particular certain quantized
compact spaces such as fuzzy spheres and tori), which have finite
dimensional matrix representations of size $N$. In the limit $N\rightarrow\infty$,
they nevertheless approach a commutative space. Gauge theory on these
spaces can be introduced much in the same way as in (\ref{zwei}),
but now the covariant coordinates $X_{i}$ are finite-dimensional
Hermitian matrices of size $N$. The conventional gauge theory is
then correctly reproduced in the limit $N\rightarrow\infty$. This
leads to a natural quantization prescription by simply integrating
over these matrices, making everything finite and well defined.

In the 2-dimensional case, this matrix-model approach to gauge theory
has been studied in considerable detail for the fuzzy sphere $S_{N}^{2}$
\cite{Madore:1992,Carow-Watamura:1998jn,Steinacker:2003sd,Imai:2003vr,Castro-Villarreal:2004vh}
and the noncommutative torus $\torus_{\theta}^{2}$ \cite{Ambjorn:1999ts,Paniak:2002fi,Paniak:2003xm,Griguolo:2003kq},
both on the classical and quantized level. It is well-known that $\R_{\theta}^{2}$
can be obtained as the scaling limit of these spaces $S_{N}^{2}$
and $\torus_{N}^{2}$ at least locally, which suggests a correspondence
also for the gauge theories. This correspondence of gauge theories
has been studied in great detail for the case of $\torus_{\theta}^{2}\rightarrow\R_{\theta}^{2}$
\cite{Paniak:2002fi,Griguolo:2001ce,Griguolo:2004jp} on the quantized
level, exhibiting the role of certain instanton contributions. 

In 4 dimensions, the quantization of gauge theory is more difficult,
and a regularization using finite-dimensional matrix models is particularly
important. The most obvious 4-dimensional spaces suitable for this
purpose are $\torus^{4},S^{2}\times S^{2}$ and $\C P^{2}$. On fuzzy
$\C P_{N}^{2}$ \cite{Grosse:1999ci,Alexanian:2001qj,Carow-Watamura:2004ct},
such a formulation of gauge theory was given in \cite{Grosse:2004wm}.
This can indeed be used to obtain $\R_{\theta}^{4}$ for the case
of $U(2)$ -invariant $\theta^{ij}$. The case of $\R^{2}\times S_{N}^{2}$
as regularization of $\R_{\theta}^{4}$ with degenerate $\theta^{ij}$
was considered in \cite{Ydri:2004im,Ydri:2004vq}, exhibiting a relation
with a conventional non-linear sigma model. A formulation of lattice
gauge theory for even-dimensional tori has been discussed in \cite{Ambjorn:2000nb,Ambjorn:2000cs,Griguolo:2003kq}.
Related {}``fuzzy'' solutions of the string-theoretical matrix models
\cite{Ishibashi:1996xs} were studied e.g. in \cite{Iso:2001mg,Kitazawa:2002xj},
see also \cite{Kimura:2002nq}.

\vspace{10pt}

The second part of this thesis will be devoted to the construction
of gauge theory on such a 4-dimensional fuzzy space, the product of
two fuzzy spheres $S_{N}^{2}\times S_{N}^{2}$. Besides introducing
fermions as well, we will use this model to regularize gauge theory
in the canonical case, i.e. on $\mathbb{R}_{\theta}^{4}$, with a
special interest in the behavior of the instanton sector.

For this, we will again study the canonical case in chapter 7, this
time using the matrix-model approach. The coordinates become annihilation
and creation operators on a Fock space, and gauge theory can be formulated
as an infinite-dimensional matrix model having the space as its ground
state. We explain why this theory contains sectors for every rank
$n$ of the gauge group $U(n)$, and construct the 4-dimensional generalization
of the instantons found in \cite{Gross:2000ss}.

In chapter 8, we first present the fuzzy sphere $S_{N}^{2}$ introduced
by John Madore in \cite{Madore:1991bw}. To go to 4 dimensions, we
use the product of two such spheres to get to $S_{N}^{2}\times S_{N}^{2}$,
and show how to get to the canonical case of $\mathbb{R}_{\theta}^{4}$
in a double scaling limit.

In chapter 9, we give a definition of $U(n)$ gauge theory on fuzzy
$S_{N}^{2}\times S_{N}^{2}$. The action is a generalization of the
approach of \cite{Steinacker:2003sd} for fuzzy $S_{N}^{2}$. It differs
from similar string-theoretical matrix models \cite{Ishibashi:1996xs}
by adding a constraint-term, which ensures that the vacuum solution
is stable and describes the product of 2 spheres. The fluctuations
of the covariant coordinates then correspond as usual to the gauge
fields, and the action reduces to ordinary Yang-Mills theory on $S^{2}\times S^{2}$
in the limit $N\rightarrow\infty$. 

We then discuss some features of the model, in particular a hidden
$SO(6)$ invariance of the action which is broken explicitly by the
constraint. This suggests some alternative formulations in terms of
collective matrices, which are assembled from the individual covariant
coordinates. This turns out to be very useful to construct a Dirac
operator, and may help to eventually study the quantization of the
model explicitly. The stability of the model without constraint is
also discussed, and we show that the only flat directions of the $SO(6)$
-invariant action are fluctuations of the constant radial modes of
the 2 spheres. The quantization of the model is defined by a finite
integral over the matrix degrees of freedom, which is shown to be
convergent due to the constraint term. We also give a gauge-fixed
action with BRST symmetry.

We also include charged fermions in the fundamental representation
of the gauge group, by giving a Dirac operator $\widehat{D}$ which
in the large $N$ limit reduces to the ordinary gauged Dirac operator
on $S^{2}\times S^{2}$. This Dirac operator inherits the $SO(6)$
symmetry of the embedding space $S^{2}\times S^{2}\subset\R^{6}$,
and exactly anti-commutes with a chirality operator. The 4-dimensional
physical Dirac spinors are obtained by suitable projections from 8-dimensional
$SO(6)$ spinors. This projection however commutes with $\widehat{D}$
only in the large $N$ limit, and is achieved by giving one of the
2 spinors a large mass. Weyl spinors can then be defined using the
exact chirality operator. An alternative version of chirality is given
by defining a Ginsparg-Wilson system.

As a further test of the proposed gauge theory, we study topologically
non-trivial solutions (instantons) on $S_{N}^{2}\times S_{N}^{2}$.
We find in particular a simple class of solutions which can be interpreted
as $U(1)$ instantons with quantized flux, combined with a singular,
localized flux tube. They are related to the {}``fluxon'' solutions
of $U(1)$ gauge theory on $\R_{\theta}^{4}$ \cite{Gross:2000ss}
discussed in chapter \ref{fuzzy canonical case}. Solutions which
can be interpreted as 2-dimensional spherical branes wrapping one
of the two spheres are also found. 

In chapter 10, we then study the relation of the model on $S_{N}^{2}\times S_{N}^{2}$
with Yang-Mills theory on $\R_{\theta}^{4}$, and demonstrate that
the usual Yang-Mills action on $\R_{\theta}^{4}$ is recovered in
the appropriate scaling limit. We show in detail how the $U(1)$ instantons
(fluxons) on $\R_{\theta}^{4}$ of chapter 7 arise as limits of the
above non-trivial solutions on $S_{N}^{2}\times S_{N}^{2}$. In particular,
we are able to match the moduli space of $n$ fluxons, corresponding
to their location on $\R_{\theta}^{4}$ resp. $S_{N}^{2}\times S_{N}^{2}$.
We find in particular that even though the field strength in the bulk
vanishes in the limit of $\R_{\theta}^{4}$, it does contribute to
the action on $S_{N}^{2}\times S_{N}^{2}$ with equal weight as the
localized flux tube. This can be interpreted on $\R_{\theta}^{4}$
as a topological or surface term at infinity. Another unexpected feature
on $S_{N}^{2}\times S_{N}^{2}$ is the appearance of certain superselection
rules, restricting the possible instanton numbers. In other words,
not all instanton numbers on $\R_{\theta}^{4}$ are reproduced for
a given matrix size $\cN$, however they can be found by considering
matrices of different size. This depends on the precise form of the
constraint term in the action, which is hence seen to imply also certain
topological constraints. 

Most of the results of the second part of this thesis have already
been published in \cite{Behr:2005wp}, together with Frank Meyer and
Harold Steinacker.

\part{The $\star$-product approach}

$\mbox{}$

\vspace{80pt}

The use of $\star$-products made noncommutativity more accessible
to physicists, as they can be applied very intuitively without reference
to any strong (and complicated) mathematical background. We can still
work with ordinary functions on ordinary commutative space-time, introducing
the noncommutativity through the $\star$-product. The $\star$-product
reproduces the ordinary pointwise product to zeroth order in some
deformation parameter, the higher orders are differential operators
acting on the functions and produce the noncommutativity. Therefore,
$\star$-products are a very convenient tool for deforming commutative
theories. The naive prescription for constructing noncommutative theories
would then be to take the commutative theory and replace ordinary
multiplication by $\star$-multiplication. As the deformation depends
on a parameter, we can get back the commutative theory by letting
it go to zero. Corrections to the commutative theory can be calculated
order by order.

As we will see, this simple prescription works surprisingly well in
the canonical case where the commutator of two coordinates is a constant.
This is mainly due to the fact that in this case the $\star$-product
still behaves very much like the commutative product with respect
to differentiation and integration. But if we go to more complicated
structures, this is no longer the case. Derivatives acquire a deformed
Leibniz rule and ordinary integration no longer has the trace property.
Therefore, the recipe of just replacing ordinary multiplication with
$\star$-multiplication no longer works. In order to nevertheless
construct noncommutative gauge theory on these more complicated spaces,
it will be necessary to first have a closer look especially at the
behavior of the derivatives. We will be able to identify objects that
still have an undeformed Leibniz rule (we will call them derivations
of the $\star$-product algebra), using them as building blocks for
gauge theory. By linking them to frames on a curved spacetime, we
can also make sense of the measure function we have to introduce in
order to make integration cyclic again. \newpage

\thispagestyle{empty}

\chapter{The canonical case}

Noncommutative gauge theory in the canonical case, where the commutator
of two coordinates is a constant, has been studied extensively in
the last few years (see e.g. \cite{Douglas:2001ba,Szabo:2001kg} for
reviews), mainly due to its appearance in string theory \cite{Seiberg:1999vs}.
It would be beyond the scope of this thesis to review all the aspects
of this fascinating field, so we will have to concentrate on what
will be important for going beyond the canonical case in the chapters
to follow. We will start with the most commonly used $\star$-product
for the canonical case, the Moyal-Weyl $\star$-product. Only the
most important features of this $\star$-product will be presented
here, but we will come back to it at the beginning of chapter \ref{sec: general starproducts}
with a more detailed analysis. After a quick look at commutative gauge
theory, an introduction into how noncommutative gauge theory can be
formulated with the help of this $\star$-product is given. This introduction
will mainly follow the approach developed here in Munich \cite{Madore:2000en,Jurco:2000ja,Jurco:2001rq,Calmet:2001na}
using Seiberg-Witten maps. Finally we will present the noncommutative
observables found in \cite{Das:2000md,Gross:2000ba,Ishibashi:1999hs},
as we will be able to generalize them later on in chapter \ref{sec: covariant coordinates general}.

\section{The Moyal-Weyl $\star$-product}

In the canonical case, the noncommutative coordinates fulfill commutation
relations \begin{equation}
[\hat{x}^{i},\hat{x}^{j}]=i\theta^{ij}\label{canonical comm rel}\end{equation}
with the constant noncommutativity parameter $\theta\in\mathbb{R}$.
The noncommutative algebra generated by the noncommutative coordinates
can be represented on the space of functions on $\mathbb{R}^{n}$
by introducing a noncommutative product, the Moyal-Weyl%
\footnote{actually, Groenewold-Moyal $\star$-product would be the more appropriate
name, as Groenewold was the first to introduce the $\star$-product
in \cite{Groenewold:1946kp}, but to avoid misunderstandings, we will
nevertheless stick to the term usually used in the literature.%
} $\star$-product \cite{Groenewold:1946kp,Moyal:1949sk} \begin{equation}
f\star g=m\cdot e^{\frac{i}{2}\theta^{ij}\dd_{i}\otimes\dd_{j}}f\otimes g=fg+\frac{i}{2}\theta^{ij}\dd_{i}f\dd_{j}g+O(2),\label{moyal-weyl star prod}\end{equation}
 with $m\cdot(f\otimes g)=fg$ and $\dd_{i}=\frac{\dd}{\dd x_{i}}$.
The product is associative, as 

\begin{eqnarray}
(f\star g)\star h & = & m\cdot e^{\frac{i}{2}\theta^{kl}\dd_{k}\otimes\dd_{l}}(m\cdot e^{\frac{i}{2}\theta^{ij}\dd_{i}\otimes\dd_{j}}f\otimes g)\otimes h\\
 & = & m\cdot m\cdot e^{\frac{i}{2}\theta^{kl}(\dd_{k}\otimes1\otimes\dd_{l}+1\otimes\dd_{k}\otimes\dd_{l})}e^{\frac{i}{2}\theta^{ij}\dd_{i}\otimes\dd_{j}\otimes1}f\otimes g\otimes h\nonumber \\
 & = & m\cdot m\cdot e^{\frac{i}{2}\theta^{ij}(\dd_{i}\otimes1\otimes\dd_{j}+\dd_{i}\otimes\dd_{j}\otimes1)}e^{\frac{i}{2}\theta^{kl}1\otimes\dd_{k}\otimes\dd_{l}}f\otimes g\otimes h\nonumber \\
 & = & m\cdot e^{\frac{i}{2}\theta^{ij}\dd_{i}\otimes\dd_{j}}(f\otimes(m\cdot e^{\frac{i}{2}\theta^{kl}\dd_{k}\otimes\dd_{l}}g\otimes h))\nonumber \\
 & = & f\star(g\star h)\nonumber \end{eqnarray}
and obviously reproduces (\ref{canonical comm rel}). Furthermore,
as $\theta$ is antisymmetric, usual complex conjugation is still
an involution\begin{equation}
\overline{f\star g}=m\cdot e^{-\frac{i}{2}\theta^{ij}\dd_{i}\otimes\dd_{j}}\overline{f}\otimes\overline{g}=\overline{g}\star\overline{f}\end{equation}
and integration has the trace property\begin{equation}
\int d^{n}x\; f\star g=\int d^{n}x\; g\star f,\end{equation}
 if the functions $f$ and $g$ vanish sufficiently fast at infinity
(of course $f\star g$ has to be integrable in the first place).

Differentiation on this space is an inner operation, i.e. we have
\begin{equation}
i\theta^{\mu\nu}\dd_{\nu}\;\;=\;\;[x^{\mu},\,\cdot\,],\end{equation}
 which can easily be calculated from (\ref{moyal-weyl star prod}).
This also means that the derivatives still have the usual Leibniz
rule, i.e. we have\begin{equation}
\dd_{i}(f\star g)=\dd_{i}f\star g+f\star\dd_{i}g.\end{equation}

\section{Commutative gauge theory}

Let us now recall some properties of a general commutative gauge theory.
A non-abelian gauge theory is based on a Lie group with Lie algebra\begin{equation}
[T^{a},T^{b}]=i\, f^{ab}{}_{c}T^{c}.\end{equation}
 Matter fields transform under a Lie algebra valued infinitesimal
parameter\begin{equation}
\lambda=\lambda_{a}T^{a}\label{com_g_field}\end{equation}
in the fundamental representation as \[
\delta_{\lambda}\psi=i\lambda\psi.\]
It follows that\begin{equation}
(\delta_{\lambda}\delta_{\xi}-\delta_{\xi}\delta_{\lambda})\psi=\delta_{i[\xi,\lambda]}\psi.\label{com_consitency_condition}\end{equation}
 The commutator of two consecutive infinitesimal gauge transformation
closes into an infinitesimal gauge transformation. As differentiation
isn't a covariant operation, a Lie algebra valued gauge potential
$a_{i}=a_{ia}T^{a}$ is introduced with the transformation property\begin{equation}
\delta_{\lambda}a_{i}=\partial_{i}\lambda+i[\lambda,a_{i}].\label{com_gt_a}\end{equation}
 With this the covariant derivative of a field is\begin{equation}
D_{i}\psi=\partial_{i}\psi-ia_{i}\psi.\end{equation}
 The field strength of the gauge potential is defined to be the commutator
of two covariant derivatives\begin{equation}
f_{ij}=i[D_{i},D_{j}]=\partial_{i}a_{j}-\partial_{j}a_{i}-i[a_{i},a_{j}].\end{equation}
 For nonabelian gauge theory, the field strength is not invariant
under gauge transformations, but rather transforms covariantly, i.e.\begin{equation}
\delta_{\lambda}f=i[\lambda,f].\end{equation}
 The same is true for the Lagrangian density $f_{ij}f^{ij}$. In order
to get a gauge invariant action, we have to use the trace over the
representation of the gauge fields. As the trace is cyclic, the commutator
with the gauge parameter vanishes and the action \begin{equation}
S=\int dx^{n}{\bf tr}f_{ij}f^{ij}\end{equation}
becomes invariant.

\section{\label{sec: canonical NC gauge theory}Noncommutative gauge theory}

To do noncommutative gauge theory in the $\star$-product approach,
we can simply mimic the commutative construction, replacing the ordinary
pointwise product with the $\star$-product.

Fields should now transform as\begin{equation}
\delta_{\Lambda}\Psi=i\Lambda\star\Psi.\end{equation}
 The commutator of two such gauge transformations should again be
a gauge transformation, i.e we want \begin{equation}
(\delta_{\Lambda}\delta_{\Xi}-\delta_{\Xi}\delta_{\Lambda})\Psi=\delta_{i[\Xi\stackrel{\star}{,}\Lambda]}\Psi,\label{NC cons cond}\end{equation}
which is only possible for gauge groups $U(N)$, as for $\Lambda=\Lambda_{a}T^{a}$
and $\Xi=\Xi_{a}T^{a}$ the commutator\begin{equation}
[\Xi\stackrel{\star}{,}\Lambda]=\frac{1}{2}[\Xi_{a}\stackrel{\star}{,}\Lambda_{b}]\{ T^{a},T^{b}\}+\frac{1}{2}\{\Xi_{a}\stackrel{\star}{,}\Lambda_{b}\}[T^{a},T^{b}]\label{star commutator doesnt close}\end{equation}
 will only close into the Lie algebra for $u(N)$ in the fundamental
representation. But general gauge groups can be implemented by using
Seiberg-Witten maps (see chapter \ref{canonical case SW map}).

As coordinates do not transform under gauge transformations, multiplication
from the left with coordinates no longer is a covariant operation,
i.e.\begin{equation}
\delta_{\Lambda}(x_{i}\star\Psi)=x_{i}\star\Lambda\star\Psi\neq\Lambda\star x_{i}\star\Psi.\end{equation}
This is very much like the situation in commutative gauge theory,
where acting with a derivative from the left isn't a covariant operation.
Following the procedure there, we introduce covariant coordinates
$X^{i}$ by adding a gauge field $A_{i}$ as\begin{equation}
X^{i}=x^{i}+\theta^{ij}A_{j}.\label{cov coord 1}\end{equation}
 To make the $X^{i}$ covariant, i.e. $\delta_{\Lambda}X^{i}=i[\Lambda\stackrel{\star}{,}X^{i}]$,
the gauge field has to transform as\begin{equation}
\delta_{\Lambda}(\theta^{ij}A_{j})=-i[x^{i}\stackrel{\star}{,}\Lambda]+i[\Lambda\stackrel{\star}{,}\theta^{ij}A_{j}]\end{equation}
and therefore\begin{equation}
\delta_{\Lambda}A_{i}=\dd_{i}\Lambda+i[\Lambda\stackrel{\star}{,}A_{i}],\end{equation}
 in exact analogy to the commutative case. The commutator with the
coordinate produces the derivative on the gauge parameter, as $[x^{i}\stackrel{\star}{,}f]=i\theta^{ij}\dd_{j}f$.
More generally we can introduce a covariantizer $D$ that applied
to a function $f$ renders it covariant \cite{Jurco:2000fs}\begin{equation}
\delta_{\Lambda}(D(f))=i[\Lambda\stackrel{\star}{,}D(f)].\end{equation}

We can now go on to formulate noncommutative gauge theory much in
the same way as we formulated commutative gauge theory.

The covariant derivative $D_{i}$ can be introduced as\begin{equation}
D_{i}\Psi=\dd_{i}\Psi-iA_{i}\star\Psi,\end{equation}
 the field strength $F_{ij}$ as\begin{equation}
F_{ij}=i[D_{i}\stackrel{\star}{,}D_{j}]=\dd_{i}A_{j}-\dd_{j}A_{i}-i[A_{i}\stackrel{\star}{,}A_{j}].\end{equation}
 The relation to the covariant coordinates subsists at this level
with\begin{equation}
-i([X^{i}\stackrel{\star}{,}X^{j}]-i\theta^{ij})=\theta^{ik}\theta^{jl}F_{kl}.\end{equation}
 For nondegenerate $\theta$, the two descriptions - either at the
level of covariant coordinates or covariant derivatives - are clearly
equivalent. 

In noncommutative gauge theory, the field strength $F$ is not gauge
invariant, even for gauge group $U(1)$. It rather transforms covariantly
under gauge transformations, i.e.\begin{equation}
\delta_{\Lambda}(F_{\mu\nu}\star F^{\mu\nu})=i[\Lambda\stackrel{\star}{,}F_{\mu\nu}\star F^{\mu\nu}].\end{equation}
Therefore even Abelian noncommutative gauge theory looks more like
nonabelian gauge theory. But just inserting a trace over the representation
of the gauge group no longer guarantees gauge invariance. To get gauge
invariant expressions, we have to use the trace property of the integral.
If we set the action for noncommutative gauge theory as \begin{equation}
S=\int d^{n}x\:{\bf tr}\, F_{\mu\nu}\star F^{\mu\nu},\end{equation}
 this expression will transform as\begin{equation}
\delta_{\Lambda}S=i\int d^{n}x\:{\bf tr}\,[\Lambda\stackrel{\star}{,}F_{\mu\nu}\star F^{\mu\nu}]=0,\end{equation}
 because the cyclicity of the integral annihilates the $\star$-part
of the commutator, and the cyclicity of the trace annihilates the
nonabelian part. This means that we cannot separate the trace over
the representation of the gauge group and the integration as in the
commutative case, we need both to get a gauge invariant action.

\section{\label{canonical case SW map}The Seiberg-Witten map}

Up to now, we could only do noncommutative gauge theory for gauge
groups $U(n)$ because of (\ref{star commutator doesnt close}). We
will now show how to implement general gauge groups by using Seiberg-Witten
maps \cite{Seiberg:1999vs,Jurco:2000ja}.

As we have seen, the commutator of two noncommutative gauge transformations
no longer closes into the Lie algebra for general gauge groups. The
noncommutative gauge parameter and the noncommutative gauge potential
will therefore have to be enveloping algebra valued. In principle,
this should mean that we are left with infinitely many degrees of
freedom. But the enveloping algebra valued parameters will only depend
on their commutative counterparts, therefore preserving the right
number of degrees of freedom. These Seiberg-Witten maps $\Lambda$,
$\Psi$ and $A$ are now functionals of their classical counterparts
and additionally of the gauge potential $a$. 

They will transform as\begin{equation}
\delta_{\lambda}\Psi_{\psi}[a]=i\Lambda_{\lambda}[a]\star\Psi_{\psi}[a]\label{NC gauge transform Psi}\end{equation}
and\begin{equation}
\delta_{\lambda}A_{i}[a]=\dd_{i}\Lambda_{\lambda}[a]+i[\Lambda_{\lambda}[a]\stackrel{\star}{,}A_{i}[a]].\end{equation}
The covariantizer $D[a]$ will now transform as \begin{equation}
\delta_{\lambda}(D[a](f))=i[\Lambda_{\lambda}[a]\stackrel{\star}{,}D[a](f)].\end{equation}
Their dependence on the commutative fields is given by the requirement
that their noncommutative transformation properties should be induced
by the commutative ones (\ref{com_g_field}) and (\ref{com_gt_a})
like\begin{eqnarray}
\Psi_{\psi}[a]+\delta_{\lambda}\Psi_{\psi}[a] & = & \Psi_{\psi+\delta_{\lambda}\psi}[a+\delta_{\lambda}a],\nonumber \\
A_{i}[a]+\delta_{\lambda}A_{i}[a] & = & A_{i}[a+\delta_{\lambda}a],\label{SW condition 1}\\
\Lambda_{\lambda}[a]+\delta_{\xi}\Lambda_{\lambda}[a] & = & \Lambda_{\lambda}[a+\delta_{\xi}a].\nonumber \end{eqnarray}
This means that it doesn't matter if we transform the noncommutative
fields under the noncommutative gauge transformations or if we transform
the commutative fields they depend on under commutative gauge transformations.
This is why we do not differentiate in our notation between commutative
and noncommutative gauge transformations, using $\delta_{\lambda}=\delta_{\Lambda_{\lambda}[a]}$.
Additionally, to zeroth order in the deformation parameter, the noncommutative
fields should be equal to their commutative counterparts, i.e.\begin{eqnarray}
\Psi_{\psi}[a] & = & \psi+\mc O(\theta),\nonumber \\
A_{i}[a] & = & a_{i}+\mc O(\theta),\label{SW condition 2}\\
\Lambda_{\lambda}[a] & = & \lambda+\mc O(\theta).\nonumber \end{eqnarray}

The SW-maps (Seiberg-Witten maps) can be found order by order in the
deformation parameter. Alternatively they can be calculated via a
consistency condition. Although the gauge transformations do not have
to close into the Lie algebra, there is still the requirement that
the commutator of two Seiberg-Witten gauge transformations (\ref{NC gauge transform Psi})
should again be a Seiberg-Witten gauge transformation (\ref{NC gauge transform Psi}),
i.e. \begin{equation}
(\delta_{\lambda}\delta_{\xi}-\delta_{\xi}\delta_{\lambda})\Psi=\delta_{i[\xi,\lambda]}\Psi\end{equation}
Written out this means that\begin{equation}
-i\delta_{\xi}\Lambda_{\lambda}[a]+i\delta_{\lambda}\Lambda_{\xi}[a]+[\Lambda_{\lambda}[a]\stackrel{\star}{,}\Lambda_{\xi}[a]]=i\Lambda_{i[\xi,\lambda]}[a].\label{st_consistency_condition}\end{equation}
 This consistency condition for the the SW-map of the gauge parameter
can be solved order by order. Then the solutions can be used to calculate
the other SW-maps by inserting them into (\ref{SW condition 1}) and
using (\ref{SW condition 2}). 

There are also methods for constructing the SW-maps to all orders
\cite{Jurco:2000fs,Jurco:2001my,Okawa:2001mv}, which we will discuss
later in chapters \ref{swmap_form} and \ref{Inverse SW map}, where
we extend them to more complicated $\star$-products.

\section{\label{canonical observables}Observables}

One characteristic property of noncommutative gauge theory is the
fact that there are no local observables. As the gauge group of noncommutative
gauge theory also comprises translations in space, gauge invariant
quantities (such as observables) cannot be local fields. But nevertheless
observables can be constructed by integrating over special Wilson
lines \cite{Das:2000md,Gross:2000ba,Ishibashi:1999hs}, which can
be interpreted as the Fourier transform of a Wilson line with fixed
momentum. Unlike the commutative case, where closed Wilson lines are
gauge invariant, these noncommutative Wilson lines do not quite close.
The gap between the endpoints is related to the momentum via the parameter
of the noncommutativity. To see this, we will first present finite
expressions for noncommutative gauge theory.

\subsection{Finite gauge transformations}

In a finite version of a noncommutative gauge theory, a scalar field
should transform like\begin{equation}
\phi^{\prime}=g\star\phi,\end{equation}
 where $g$ is a function that is invertible with respect to the $\star$-product\begin{equation}
g\star g^{-1}=g^{-1}\star g=1.\end{equation}
 Again, multiplication with a coordinate function is not covariant
any more\begin{equation}
(x^{i}\star\phi)^{\prime}\neq x^{i}\star\phi^{\prime}.\end{equation}
 Just as in the infinitesimal formulation, covariant coordinates\begin{equation}
X^{i}(x)=x^{i}+\theta^{ij}A_{j}(x)\end{equation}
 can be introduced, transforming in the adjoint representation\begin{equation}
X^{i\prime}=g\star X^{i}\star g^{-1}.\end{equation}
 Now the product of a covariant coordinate with a field is again a
field. In perfect analogy to the commutative case, the gauge field
$A^{i}$ transforms as \begin{equation}
A^{i\prime}=ig\star\dd_{i}g^{-1}+g\star A^{i}\star g^{-1}.\end{equation}
This finite formulation of noncommutative gauge theory is equivalent
to the infinitesimal formulation presented before. For details on
this equivalence, see \cite{Jurco:2001rq}.

\subsection{Wilson lines}

Just think of a field $\phi$ transforming covariantly under a gauge
transformation with gauge parameter $\lambda=l_{i}x^{i}$. The corresponding
finite expression is\begin{equation}
\phi(x)\rightarrow e_{\star}^{il_{i}x^{i}}\star\phi(x^{k})\star e_{\star}^{-il_{i}x^{i}}=\phi(x^{k}-l_{j}\theta^{jk}),\end{equation}
 i.e. a translation by $-l_{j}\theta^{jk}$. This means that noncommutative
gauge transformations in fact contain translations in space! The $\star$
subscript on the exponential means that all the multiplications are
done using the $\star$-product. But it is a special property of the
Moyal-Weyl $\star$-product that the $\star$-exponential actually
is the same as the ordinary one, i. e. we have $e_{\star}^{ix^{i}}=e^{ix^{i}}$,
which is why we will drop the $\star$-subscript in the following.

The fact that translations are gauge transformations can be used to
construct noncommutative analogs of Wilson lines. Such a Wilson line
\begin{equation}
W_{l}=e^{il_{i}X^{i}}\star e^{-il_{i}x^{i}}\label{wilson line canonical}\end{equation}
 has indeed the same transformation properties under a gauge transformation\begin{equation}
W_{l}^{\prime}(x)=g(x)\star W_{l}(x)\star g^{-1}(x-l_{i}\theta^{ij}).\end{equation}
 as a Wilson line starting at $x$ and ending at $x-l\theta$. Here
we only treat straight Wilson lines, but for the canonical case they
can also be generalized to noncommutative Wilson lines with arbitrary
paths \cite{Ishibashi:1999hs,Das:2000md,Gross:2000ba}.

\subsection{Observables}

As space translations are included in the noncommutative gauge transformations,
no local observables can be constructed. One has to integrate over
the whole space to get gauge invariant objects. For this it is useful
to look at the Fourier transform of the Wilson lines (\ref{wilson line canonical})
\begin{equation}
W_{l}(k)=\int d^{n}x\, W_{l}(x)\star e^{ik_{i}x^{i}}\end{equation}
Under a gauge transformation, it transforms as\begin{eqnarray}
W_{l}(k)' & = & \int d^{n}x\, g(x)\star W_{l}(x)\star g^{-1}(x-l_{i}\theta^{ij})\star e^{ik_{i}x^{i}}\\
 & = & \int d^{n}x\, g(x)\star W_{l}(x)\star e^{ik_{i}x^{i}}\star e^{-ik_{i}x^{i}}\star g^{-1}(x-l_{i}\theta^{ij})\star e^{ik_{i}x^{i}}\nonumber \\
 & = & \int d^{n}x\, g(x)\star W_{l}(x)\star e^{ik_{i}x^{i}}\star g^{-1}(x-l_{i}\theta^{ij}+k_{i}\theta^{ij}).\nonumber \end{eqnarray}
This means that the so called open Wilson lines \cite{Ishibashi:1999hs,Das:2000md,Gross:2000ba}
defined as\begin{equation}
U_{l}=W_{l}(l)=\int d^{n}x\, W_{l}(x)\star e^{il_{i}x^{i}},\end{equation}
are gauge invariant. Here, the momentum $k$ of the Wilson line corresponds
to its length $r^{j}=l_{i}\theta^{ij}$ via the parameter of the noncommutativity,
i.e. $r^{j}=k_{i}\theta^{ij}$. Using (\ref{wilson line canonical}),
this is of course even more obvious\begin{equation}
U_{l}=\int d^{n}x\, W_{l}(x)\star e^{il_{i}x^{i}}=\int d^{n}x\, e^{il_{i}X^{i}}\star e^{-il_{i}x^{i}}\star e^{il_{i}x^{i}}=\int d^{n}x\, e^{il_{i}X^{i}}.\end{equation}
These open Wilson lines can even be generalized by inserting an arbitrary
function $f$ of the covariant coordinates as \begin{equation}
\int d^{2n}x\, f(X)\star e^{il_{i}X^{i}}\end{equation}
 without spoiling the gauge invariance.

\chapter{\label{sec: general starproducts}General $\star$-products}

We had introduced the Moyal-Weyl $\star$-product for the canonical
case without explaining how it can be derived. In order to introduce
the notion of $\star$-products in general, we will first have a closer
look at the canonical case again. Suppose we have a two-dimensional
canonical algebra generated by the noncommutative coordinates $\widehat{x}$
and $\widehat{y}$ with relations\begin{equation}
[\widehat{x},\widehat{y}]=-i\theta.\label{can algebra 2 dim}\end{equation}
 To represent this algebra on the space $C_{\infty}(\mathbb{R}^{2})$,
we will define an ordering prescription $\rho$ by mapping monomials
in the commutative variables $x$ and $y$ to the monomials in the
noncommutative variables $\widehat{x}$ and $\widehat{y}$ with all
the $\widehat{x}$ on the left hand side and all the $\widehat{y}$
on the right hand side\begin{equation}
\rho(x^{n}y^{m}):=\widehat{x}^{n}\widehat{y}^{m}.\end{equation}
 This is called normal ordering. If we normal order a monomial, we
get\begin{equation}
\widehat{y}^{m}\widehat{x}^{k}=\sum_{i=0}^{min(m,k)}\frac{(i\theta)^{i}}{i!}\frac{m!}{(m-i)!}\frac{k!}{(k-i)!}\widehat{x}^{k-i}\widehat{y}^{m-i}.\end{equation}
If we multiply two such monomials and normal order the result, we
therefore get\begin{eqnarray}
\rho(x^{n}y^{m})\rho(x^{k}y^{l}) & = & \sum_{i=0}^{min(m,k)}\frac{(i\theta)^{i}}{i!}\frac{m!}{(m-i)!}\frac{k!}{(k-i)!}\widehat{x}^{n+k-i}\widehat{y}^{l+m-i}\\
 & = & \sum_{i=0}^{min(m,k)}\frac{(i\theta)^{i}}{i!}\frac{m!}{(m-i)!}\frac{k!}{(k-i)!}\rho(x^{n+k-i}y^{l+m-i})\nonumber \end{eqnarray}
 As the vector space of the noncommutative polynomials of a certain
degree has the same dimension as the vector space of the commutative
polynomials of the same degree, the ordering $\rho$ can be inverted,
giving\begin{eqnarray}
\rho^{-1}(\rho(x^{n}y^{m})\rho(x^{k}y^{l})) & = & \sum_{i=0}^{min(m,k)}\frac{(i\theta)^{i}}{i!}\frac{m!}{(m-i)!}\frac{k!}{(k-i)!}x^{n+k-i}y^{l+m-i}\\
 & = & \sum_{i=0}^{\infty}\frac{(i\theta)^{i}}{i!}\dd_{y}^{i}(x^{n}y^{m})\dd_{x}^{i}(x^{k}y^{l})\nonumber \\
 & = & m\cdot e^{i\theta\dd_{y}\otimes\dd_{x}}(x^{n}y^{m})\otimes(x^{k}y^{l})\nonumber \\
 & = & (x^{n}y^{m})\star_{n}(x^{k}y^{l}).\nonumber \end{eqnarray}
Therefore get a new $\star$-product\begin{equation}
f\star_{n}g=m\cdot e^{i\theta\dd_{y}\otimes\dd_{x}}f\otimes g=\rho^{-1}(\rho(f)\rho(g))\label{normal ordered star prod}\end{equation}
 for the algebra (\ref{can algebra 2 dim}) by applying the ordering
prescription $\rho$ on polynomial functions! Let's compare it to
the Moyal-Weyl $\star$-product (\ref{moyal-weyl star prod}). For
the algebra (\ref{can algebra 2 dim}) this read\begin{equation}
f\star_{w}g=m\cdot e^{\frac{i}{2}\theta(\dd_{y}\otimes\dd_{x}-\dd_{x}\otimes\dd_{y})}f\otimes g.\end{equation}
If we define a differential operator\begin{equation}
T=e^{-\frac{i}{2}\theta\dd_{x}\dd_{y}},\end{equation}
we can calculate\begin{eqnarray}
f\star_{w}g & = & m\cdot e^{\frac{i}{2}\theta(\dd_{y}\otimes\dd_{x}-\dd_{x}\otimes\dd_{y})}f\otimes g\\
 & = & m\cdot e^{\frac{i}{2}\theta(\dd_{y}\otimes1+1\otimes\dd_{y})(\dd_{x}\otimes1+1\otimes\dd_{x})}\: e^{i\theta\dd_{y}\otimes\dd_{x}}\: e^{-\frac{i}{2}\theta(\dd_{x}\dd_{y}\otimes\dd_{x}\dd_{y})}f\otimes g\nonumber \\
 & = & e^{\frac{i}{2}\theta\dd_{x}\dd_{y}}((e^{-\frac{i}{2}\theta\dd_{x}\dd_{y}}f)\star_{n}(e^{-\frac{i}{2}\theta\dd_{x}\dd_{y}}g))\nonumber \\
 & = & T^{-1}((Tf)\star_{n}(Tg))\nonumber \\
 & = & T^{-1}\rho^{-1}(\rho T(f)\rho T(g)).\nonumber \end{eqnarray}
The two $\star$-products are related by the differential operator
$T$, and the Moyal-Weyl $\star$-product can be expressed by an ordering
prescription $\rho T$. The ordering actually corresponds to symmetric
ordering, e.g. we have\begin{eqnarray}
\rho T(xy) & = & \rho(e^{-\frac{i}{2}\theta\dd_{x}\dd_{y}}xy)=\rho(xy-\frac{i}{2}\theta)=\widehat{x}\widehat{y}-\frac{i}{2}\theta\\
 & = & \widehat{x}\widehat{y}-\frac{1}{2}[\widehat{x},\widehat{y}]=\frac{1}{2}(\widehat{x}\widehat{y}+\widehat{y}\widehat{x}).\nonumber \end{eqnarray}
This method of constructing $\star$-products by applying an ordering
prescription is not limited to the canonical case.

\section{Definition}

But before we have a look at more complicated $\star$-products, we
will fist give an abstract definition. For this we will introduce
a parameter $h$ (which we think of as small) measuring the deformation,
and express everything as formal power series in this parameter. 

A $\star$-product on a manifold $M$ is an associative $\mathbb{C}$-linear
product \begin{equation}
f\star g=fg+\sum_{i=1}^{\infty}h^{i}B_{i}(f,g),\end{equation}
where the $B_{i}$ are bidifferential operators acting on $f,g\,\epsilon\, C_{\infty}(M)[[h]]$.
From the associativity of the $\star$-product follows that the $\star$-commutator
fulfills the Jacobi identity, i.e.\begin{equation}
[f\stackrel{\star}{,}[g\stackrel{\star}{,}h]]+[g\stackrel{\star}{,}[h\stackrel{\star}{,}f]]+[h\stackrel{\star}{,}[f\stackrel{\star}{,}g]]=0.\end{equation}
 For the antisymmetric part $\pi$ of the first order term $B_{1}$,
i.e. \begin{equation}
[f\stackrel{\star}{,}g]=h\,\pi(f,g)+\mc O(2),\end{equation}
 this means that it has to be a Poisson structure. Expressed in some
local coordinates as $\pi=\frac{1}{2}\pi^{ij}\dd_{i}\wedge\dd_{j}$,
this means that it has to fulfill\begin{equation}
\pi^{ij}\dd_{j}\pi^{kl}+\pi^{kj}\dd_{j}\pi^{li}+\pi^{lj}\dd_{j}\pi^{ik}=0.\end{equation}
Therefore for every $\star$-product, there is a Poisson structure
related to it. On the other hand, if we start with some Poisson structure,
we can always construct a corresponding $\star$-product \cite{Kontsevich:1997vb},
the formality $\star$-product we will present in chapter \ref{Formality map}.

\section{$\star$-products by operator ordering}

We will now show how to construct $\star$-products for associative
algebras that are defined by commutator relations\begin{equation}
\mc R\::\;\;\;[\hat{x}^{i},\hat{x}^{j}]=ih\hat{c}^{ij}\end{equation}
in the same way as we constructed the normal ordered $\star$-product
in (\ref{normal ordered star prod}). More abstractly, such an algebra
can be defined as \begin{equation}
\mc A=\mathbb{C}\langle\hat{x}^{1},...,\hat{x}^{n}\rangle[[h]]/\mc R,\end{equation}
where we allow formal power series in the deformation parameter $h$.
As we treat $h$ as a formal parameter, such an algebra always has
the Poincare-Birkoff-Witt property, i.e. the vector space of the noncommutative
polynomials of a certain degree has the same dimension as if the coordinates
were commutative. Especially this means that we can map the basis
of the commutative algebra $\mathbb{C}[x^{1},...,x^{n}][[h]]$ onto
a basis of the noncommutative algebra $\mc A$. Such an ordering prescription
\begin{equation}
\rho:\mathbb{C}[x^{1},...,x^{n}][[h]]\rightarrow\mathbb{C}\langle\hat{x}^{1},...,\hat{x}^{n}\rangle[[h]]/\mc R\end{equation}
 is then an isomorphism of vector spaces. With it, we can introduce
a $\star$-product as\begin{equation}
f\star g=\rho^{-1}(\rho(f)\rho(g)),\end{equation}
making the commutative algebra equipped with the $\star$-product
isomorphic to the noncommutative algebra $\mc A$. Two ordering prescriptions
$\rho$ and $\rho'$ of the same algebra are always related by a similarity
transformation $T$ as $\rho=\rho'T$ with\begin{equation}
T=id+\sum_{i=1}^{\infty}h^{i}T_{i},\end{equation}
where the $T_{i}$ are differential operators. The corresponding $\star$-products
$\star$ and $\star'$ are then related by\begin{equation}
f\star g=\rho^{-1}(\rho(f)\rho(g))=T^{-1}\rho'^{-1}(\rho'(Tf)T\rho'(Tg))=T^{-1}((Tf)\star'(Tg)).\end{equation}

\section{\label{wely_star_construction}The Weyl-ordered $\star$-product}

In this chapter we will construct the Weyl-ordered $\star$-product
of a general noncommutative algebra up to second order. Weyl-ordering
means that we use totally symmetric ordering for the generators. We
start with an algebra generated by $N$ elements $\hat{x}^{i}$ and
relations

\begin{equation}
[\hat{x}^{i},\hat{x}^{j}]=\hat{c}^{ij}(\hat{x}),\end{equation}
where we have suppressed the explicit dependence of $\widehat{c}$
on a formal deformation parameter, but we will always assume that
it is at least of order 1. For such an algebra we will calculate a
$\star$-product up to second order. Let \begin{equation}
f(p)=\int d^{n}x\, f(x)e^{ip_{i}x^{i}}\end{equation}
 be the Fourier transform of $f$. Then the Weyl ordered operator
associated to $f$ is defined by\begin{equation}
W(f)=\int\frac{d^{n}p}{(2\pi)^{n}}f(p)e^{-ip_{i}\hat{x}^{i}}\end{equation}
 (see e. g. \cite{Madore:2000en}) . Every monomial of coordinate
functions is mapped to the corresponding Weyl ordered monomial of
the algebra. We note that \begin{equation}
W(e^{iq_{i}x^{i}})=e^{iq_{i}\hat{x}^{i}}.\end{equation}
 The Weyl ordered $\star$-product is defined by the equation\begin{equation}
W(f\star g)=W(f)W(g).\end{equation}
 If we insert the Fourier transforms of $f$ and $g$ we get\begin{equation}
f\star g=\int\frac{d^{n}k}{(2\pi)^{n}}\int\frac{d^{n}p}{(2\pi)^{n}}f(k)g(p)\, W^{-1}(e^{-ik_{i}\hat{x}^{i}}e^{-ip_{i}\hat{x}^{i}}).\end{equation}
 We are therefore able to write down the $\star$-product of the two
functions if we know the form of the last expression. For this we
expand it in terms of commutators. We use\begin{equation}
e^{\hat{A}}e^{\hat{B}}=e^{\hat{A}+\hat{B}}R(\hat{A},\hat{B})\end{equation}
 with

\begin{eqnarray}
R(\hat{A},\hat{B}) & = & 1+\frac{1}{2}[\hat{A},\hat{B}]\\
 & - & \frac{1}{6}[\hat{A}+2\hat{B},[\hat{A},\hat{B}]]+\frac{1}{8}[\hat{A},\hat{B}][\hat{A},\hat{B}]+\mathcal{O}(3).\nonumber \end{eqnarray}
 If we set $\hat{A}=-ik_{i}\hat{x}^{i}$ and $\hat{B}=-ip_{i}\hat{x}^{i}$,
the above-mentioned expression becomes\begin{eqnarray}
\lefteqn{W^{-1}(e^{-ik_{i}\hat{x}^{i}}e^{-ip_{i}\hat{x}^{i}})=}\\
 &  & e^{-i(k_{i}+p_{i})x^{i}}+\frac{1}{2}(-ik_{i})(-ip_{j})W^{-1}(e^{-i(k_{i}+p_{i})\hat{x}^{i}}[\hat{x}^{i},\hat{x}^{j}])\nonumber \\
 &  & -\frac{1}{6}(-i)(k_{m}+2p_{m})(-ik_{i})(-ip_{j})W^{-1}(e^{-i(k_{i}+p_{i})\hat{x}^{i}}[\hat{x}^{m},[\hat{x}^{i},\hat{x}^{j}]])\nonumber \\
 &  & +\frac{1}{8}(-ik_{m})(-ip_{n})(-ik_{i})(-ip_{j})W^{-1}(e^{-i(k_{i}+p_{i})\hat{x}^{i}}[\hat{x}^{m},\hat{x}^{n}][\hat{x}^{i},\hat{x}^{j}])\nonumber \\
 &  & +\mathcal{O}(3).\nonumber \end{eqnarray}
 If we assume that the commutators of the generators are written in
Weyl ordered form\begin{equation}
\hat{c}^{ij}=W(c^{ij}),\end{equation}
 we see that\begin{equation}
[\hat{x}^{m},[\hat{x}^{i},\hat{x}^{j}]]=W(c^{ml}\partial_{l}c^{ij})+\mathcal{O}(3),\end{equation}
\begin{equation}
[\hat{x}^{m},\hat{x}^{n}][\hat{x}^{i},\hat{x}^{j}]=W(c^{mn}c^{ij})+\mathcal{O}(3).\end{equation}
 Further we can derive\begin{eqnarray}
W^{-1}(e^{-iq_{i}\hat{x}^{i}}W(f(x))) & = & W^{-1}\Big(\int\frac{d^{n}p}{(2\pi)^{n}}f(p)e^{-i(q_{i}+p_{i})\hat{x}^{i}}R(-iq_{i}\hat{x}^{i},-ip_{i}\hat{x}^{i})\Big)\nonumber \\
 & = & W^{-1}\bigg(W\Big(\int\int\frac{d^{n}p}{(2\pi)^{n}}f(p)e^{-i(q_{i}+p_{i})x^{i}}\times\\
 &  & \;\;\;\;\;\;\;\;\;\;\;(1+\frac{1}{2}(-ip_{i})(-iq_{j})[x_{i},x_{j}])\Big)\bigg)+\mc O(2)\nn\nonumber \\
 & = & e^{-iq_{i}x^{i}}f(x)+\frac{1}{2}e^{-iq_{i}x^{i}}(-iq_{i})c^{ij}\partial_{j}f(x)+\mathcal{O}(2),\nn\end{eqnarray}
using\[
\dd_{j}f(x)=\int\frac{d^{n}p}{(2\pi)^{n}}f(p)(-ip_{j})e^{-ip_{i}x^{i}}.\]
Putting all this together yields

\begin{eqnarray}
W^{-1}(e^{-ik_{i}\hat{x}^{i}}e^{-ip_{i}\hat{x}^{i}}) & = & e^{-i(k_{i}+p_{i})x^{i}}\left(1+\frac{1}{2}c^{ij}(-ik_{i})(-ip_{j})\right.\\
 & + & \frac{1}{8}c^{mn}c^{ij}(-ik_{m})(-ip_{n})(-ik_{i})(-ip_{j})\nonumber \\
 & + & \left.\frac{1}{12}c^{ml}\partial_{l}c^{ij}(-i)(k_{m}-p_{m})(-ik_{i})(-ip_{j})\right)\nonumber \\
 & + & \mathcal{O}(3),\nonumber \end{eqnarray}
 and we can write down the Weyl ordered $\star$-product up to second
order for an arbitrary algebra\begin{eqnarray}
f\star g & = & fg+\frac{1}{2}c^{ij}\partial_{i}f\partial_{j}g\label{wo_product}\\
 & + & \frac{1}{8}c^{mn}c^{ij}\partial_{m}\partial_{i}f\partial_{n}\partial_{j}g\nonumber \\
 & + & \frac{1}{12}c^{ml}\partial_{l}c^{ij}(\partial_{m}\partial_{i}f\partial_{j}g-\partial_{i}f\partial_{m}\partial_{j}g)+\mathcal{O}(3).\nonumber \end{eqnarray}

Let us collect some properties of the just calculated $\star$-product.
First\begin{equation}
[x^{i}\stackrel{\star}{,}x^{j}]=c^{ij}\end{equation}
 is the Weyl ordered commutator of the algebra. Further, if there
is a conjugation on the algebra and if we assume that the noncommutative
coordinates are real $\overline{\hat{x}^{i}}=\hat{x}^{i}$, then the
Weyl ordered monomials are real, too. This is also true for the monomials
of the commutative coordinate functions. Therefore this $\star$-product
respects the ordinary complex conjugation\begin{equation}
\overline{f\star g}=\overline{g}\star\overline{f}.\end{equation}
 On the level of the Poisson tensor this means\begin{equation}
\overline{c^{ij}}=-c^{ij}.\end{equation}

\section{\label{Formality map}The formality $\star$-product}

The Weyl-ordered $\star$-product of chapter (\ref{wely_star_construction})
is very useful for explicit calculations, but these can only be done
in a perturbative way order by order. Also, it is only known in general
up to the second order we calculated here. For closed expressions
and questions of existence, Kontsevich's formality $\star$-product
\cite{Kontsevich:1997vb} is the better choice. It is known to all
orders and comes with a strong mathematical framework that can be
used for further constructions.

This mathematical framework, known as Kontsevich's formality map \cite{Kontsevich:1997vb},
is a very useful tool for studying the relations between Poisson tensors
and $\star$-products. To make use of the formality map we first want
to recall some definitions. A polyvector field is a skew-symmetric
tensor in the sense of differential geometry. Every $n$-polyvector
field $\alpha$ may locally be written as\begin{equation}
\alpha=\alpha^{i_{1}\dots i_{n}}\,\partial_{i_{1}}\wedge\dots\wedge\partial_{i_{n}}.\end{equation}
 We see that the space of polyvector fields can be endowed with a
grading $n$. For polyvector fields there is a grading respecting
bracket that in a natural way generalizes the Lie bracket $[\,\cdot\,,\,\cdot\,]_{L}$
of two vector fields, the Schouten-Nijenhuis bracket. For an exact
definition see \ref{Schouten-Nijenhuis bracket}. If $\pi$ is a Poisson
tensor, the Hamiltonian vector field $H_{f}$ for a function $f$
is\begin{equation}
H_{f}={[\pi,f]_{S}}=-\pi^{ij}\partial_{i}f\partial_{j}.\end{equation}
 Note that $[\pi,\pi]_{S}=0$ is the Jacobi identity of a Poisson
tensor. \medskip

On the other hand a $n$-polydifferential operator is a multilinear
map that maps $n$ functions to a function. For example, we may write
a $1$-polydifferential operator $D$ as\begin{equation}
D(f)=D_{0}f+D_{1}^{i}\partial_{i}f+D_{2}^{ij}\partial_{i}\partial_{j}f+\dots.\end{equation}
 The ordinary multiplication $\cdot$ is a $2$-differential operator.
It maps two functions to one function. Again the number $n$ is a
grading on the space of polydifferential operators. Now the Gerstenhaber
bracket $[\:\cdot\:,\:\cdot\:]_{G}$ is natural and respects the grading.
For an exact definition see \ref{Gerstenhaber bracket}.

The formality map is a collection of skew-symmetric multilinear maps
$U_{n}$, $n=0,1,\dots$, that maps $n$ polyvector fields to a $m$-differential
operator. To be more specific let $\alpha_{1},\dots,\alpha_{n}$ be
polyvector fields of grade $k_{1},\dots,k_{n}$. Then $U_{n}(\alpha_{1},\dots,\alpha_{n})$
is a polydifferential operator of grade\begin{equation}
m=2-2n+\sum_{i}k_{i}.\label{relation between grades}\end{equation}
 In particular the map $U_{1}$ is a map from a $k$-vectorfield to
a $k$-differential operator. It is defined by\begin{equation}
U_{1}(\alpha^{i_{1}\dots i_{n}}\partial_{i_{1}}\wedge\dots\wedge\partial_{i_{n}})(f_{1},\dots,f_{n})=\alpha^{i_{1}\dots i_{n}}\partial_{i_{1}}f_{1}\cdot\dots\cdot\partial_{i_{n}}f_{n}.\end{equation}
 The formality maps $U_{n}$ fulfill the formality condition \cite{Kontsevich:1997vb,Arnal:2000hy}

\begin{equation}
Q'_{1}U_{n}(\alpha_{1},\ldots,\alpha_{n})+\frac{1}{2}\sum_{{{I\sqcup J=\{1,\ldots,n\}\atop I,J\neq\emptyset}}}\epsilon(I,J)Q'_{2}(U_{|I|}(\alpha_{I}),U_{|J|}(\alpha_{J}))\label{formality cond}\end{equation}
\[
=\frac{1}{2}\sum_{i\neq j}\epsilon(i,j,\ldots,\hat{i},\ldots,\hat{j},\ldots,n)U_{n-1}(Q_{2}(\alpha_{i},\alpha_{j}),\alpha_{1},\ldots,\widehat{\alpha}_{i},\ldots,\widehat{\alpha}_{j},\ldots,\alpha_{n}).\]

The hats stand for omitted symbols, $Q'_{1}(\Upsilon)=[\Upsilon,\mu]$
with $\mu$ being ordinary multiplication and $Q'_{2}(\Upsilon_{1},\Upsilon_{2})=(-1)^{(|\Upsilon_{1}|-1)|\Upsilon_{2}|}[\Upsilon_{1},\Upsilon_{2}]_{G}$
with $|\Upsilon_{s}|$ being the degree of the polydifferential operator
$\Upsilon_{s}$, i.e. the number of functions it is acting on. For
polyvectorfields $\alpha_{s}^{i_{1}\ldots i_{k_{s}}}\partial_{i_{1}}\wedge\ldots\wedge\partial_{i_{k_{s}}}$
of degree $k_{s}$ we have $Q_{2}(\alpha_{1},\alpha_{2})=-(-1)^{(k_{1}-1)k_{2}}[\alpha_{2},\alpha_{1}]_{S}$.

For a bivectorfield $\pi$ we can now define a bidifferential operator

\begin{equation}
\star=\sum_{n=0}^{\infty}\frac{1}{n!}U_{n}(\pi,\,\ldots,\pi)\end{equation}
 i.e.

\begin{equation}
f\star g=\sum_{n=0}^{\infty}\frac{1}{n!}U_{n}(\pi,\,\ldots,\pi)(f,g).\label{definition formality star product}\end{equation}
To see that the formality $\star$-product is associative, we first
define the special map\begin{equation}
\Phi(\alpha)=\sum_{n=1}^{\infty}\frac{1}{(n-1)!}U_{n}(\alpha,\pi,\,\ldots,\pi).\end{equation}
Using the formality condition (\ref{formality cond}) we calculate
that \begin{equation}
[\star,\star]_{G}=\Phi([\pi,\pi]_{S}),\label{associativity of formality}\end{equation}
where $[\star,\star]_{G}=0$ means that the $\star$-product is associative.
This follows from the fact that $\pi$ is a Poisson tensor, i.e $[\pi\,\pi]_{S}=0$.
Note that the definition (\ref{definition formality star product})
would be equally valid for general bivector fields $\pi'$, but the
resulting product would cease to be associative. Nevertheless the
non-associativity would be controlled by (\ref{associativity of formality}).

\section{The Jambor-Sykora $\star$-product}

The formality $\star$-product of the last chapter is very useful
for abstract proofs (and in fact we will use it for constructing a
SW-map to all orders in chapter \ref{swmap_form}, but it is too complicated
for explicit calculations. But for special cases where the Poisson
structure can be expressed in terms of commuting vector fields, there
is a $\star$-product that is both known to all orders and easy to
handle in calculations, the Jambor-Sykora $\star$-product \cite{Jambor:2004kc}.
For commuting vectorfields $X_{a}=X_{a}^{i}\partial_{i}$ (i.e. $[X_{a},X_{b}]=0$)
and a constant matrix $\sigma$ the Jambor-Sykora $\star$-product
reads\begin{equation}
f\star_{\sigma}g=m\cdot e^{\sigma^{ab}X_{a}\otimes X_{b}}f\otimes g,\label{full starproduct}\end{equation}
where $m\cdot(f\otimes g)=fg$. The constant matrix $\sigma$ can
be written as $\sigma=\sigma_{as}+\sigma_{s}$ with $\sigma_{as}$
antisymmetric and $\sigma_{s}$ symmetric. There is an equivalence
transformation \begin{equation}
\rho=e^{\frac{1}{2}\sigma_{s}^{ab}X_{a}X_{b}}\label{equivalence transformation}\end{equation}
from the antisymmetric $\star$-product \begin{equation}
f\star_{as}g=m\cdot e^{\sigma_{as}^{ab}X_{a}\otimes X_{b}}f\otimes g\label{antisymmetric starproduct}\end{equation}
 to the full one (\ref{full starproduct}) \begin{equation}
f\star_{as}g=\rho^{-1}(\rho(f)\star_{\sigma}\rho(g)).\label{starproduct equivalence}\end{equation}
Note that for real vectorfields $X_{a}$ and $\sigma_{as}$ imaginary,
ordinary complex conjugation is an involution of the antisymmetric
$\star$-product, i.e.\begin{equation}
\overline{f\star_{as}g}=\overline{g}\star_{as}\overline{f}.\end{equation}
 For the full $\star$-product we can pull back this property by $\rho$
and the involution now is $\rho\overline{\rho^{-1}}$. On a function
$f$ this reads $\rho(\overline{\rho^{-1}(f)})$, and we have\begin{eqnarray}
\rho(\overline{\rho^{-1}(f\star_{\sigma}g)}) & = & \rho(\overline{\rho^{-1}f\star_{as}\rho^{-1}g})\\
 & = & \rho(\overline{\rho^{-1}g}\star_{as}\overline{\rho^{-1}f})\nonumber \\
 & = & \rho(\overline{\rho^{-1}g})\star_{\sigma}\rho(\overline{\rho^{-1}f}).\nonumber \end{eqnarray}

\section{\label{sec: traces}Traces}

For the Moyal-Weyl $\star$-product, ordinary integration still had
the trace property, i.e. it was invariant under cyclic permutations
of the elements in the integrand. Unluckily, this is in general no
longer the case for the more complicated $\star$-products of this
chapter. But the cyclicity of integration is crucial for turning gauge-covariant
objects into gauge invariant ones. Therefore, we have to guarantee
the trace property of the integral by introducing a measure function
$\Omega$. For many $\star$-products the trace may then be written
as\begin{equation}
tr\, f=\int d^{2n}x\,\Omega(x)\, f(x).\label{DEF_OMEGA}\end{equation}
 Due to the cyclicity of the trace the measure function $\Omega$
has to fulfill\begin{equation}
\partial_{i}(\Omega\theta^{ij})=0\label{PROP_OMEGA}\end{equation}
 which can easily be seen by using partial integration. If we take
the Poisson structure $\theta^{ij}$ to be invertible, the inverse
of the Pfaffian\begin{equation}
\frac{1}{\Omega}=Pf(\theta)=\sqrt{det(\theta)}=\frac{1}{2^{n}n!}\epsilon_{i_{1}i_{2}\cdots i_{2n}}\theta^{i_{1}i_{2}}\cdots\theta^{i_{2n-1}i_{2n}}\end{equation}
 is a solution to this equation. Unluckily, there is no such formula
for $\star$-products whose Poisson structures are not invertible. 

If equation (\ref{PROP_OMEGA}) is fulfilled, cyclicity is only guaranteed
to first order. In principle we have to calculate higher orders of
$\Omega$ according to the $\star$-product chosen. Nevertheless there
can always be found a $\star$-product so that a measure function
fulfilling (\ref{PROP_OMEGA}) guarantees cyclicity to all orders
\cite{Felder:2000hy}.

\section{\label{example: star products}Example: $\star$-products for the
$\kappa$-deformed plane}

We will exemplify the ideas of the last chapter by applying them to
the algebra generated by $x$ and $y$ with commutation relations\begin{equation}
[x,y]=-iax.\label{kappa}\end{equation}
 This is the 2-dimensional version of what is known as $\kappa$-deformed
spacetime. The generalization to higher dimensions is straightforward.

\subsection{The Weyl-ordered $\star$-product}

The Poisson structure for this algebra quite obviously is \begin{equation}
\{ f,g\}_{p}=-iax\pp xf\pp yg+ia\pp yfx\pp xg,\label{Poisson structure kappa}\end{equation}
 and the Poisson tensor therefore\begin{equation}
c^{ij}=-iax\delta_{x}^{i}\delta_{y}^{j}+iax\delta_{y}^{i}\delta_{x}^{j}.\label{Poisson tensor kappa}\end{equation}
 As $c^{ij}$ is linear in the coordinates, the Weyl-ordering of the
expression doesn't play a role. Inserting (\ref{Poisson tensor kappa})
into (\ref{wo_product}) produces the Weyl-ordered $\star$-product
up to second order for (\ref{kappa})\begin{eqnarray}
f\star g & = & fg-\frac{ia}{2}x(\pp xf\pp yg-\pp yf\pp xg)\label{Weyl ordered kappa}\\
 &  & -\frac{a^{2}}{8}x^{2}(\dd_{x}^{2}f\dd_{y}^{2}g-2\dd_{x}\dd_{y}f\dd_{x}\dd_{y}g+\dd_{y}^{2}f\dd_{x}^{2}g)\nonumber \\
 &  & +\frac{a^{2}}{12}x(\dd_{x}\dd_{y}f\dd_{y}g+\dd_{y}f\dd_{x}\dd_{y}g-\dd_{x}f\dd_{y}^{2}g-\dd_{y}^{2}f\dd_{x}g)+\mc O(3).\nonumber \end{eqnarray}

\subsection{The Jambor-Sykora $\star$-product}

If we choose vectorfields \begin{equation}
X_{1}=x\partial_{x}\;\;\;\;\mbox{and}\;\;\;\; X_{2}=-a\partial_{y}\end{equation}
 and $\sigma=\left(\begin{array}{cc}
0 & i\\
0 & 0\end{array}\right)$ (see also \cite{Jambor:2004kc}), the Jambor-Sykora $\star$-product
(\ref{full starproduct}) will reproduce the algebra (\ref{kappa}).
This $\star$-product corresponds to normal ordering. It reads\begin{equation}
f\star_{\sigma}g=m\cdot e^{-ia\: x\partial_{x}\otimes\partial_{y}}f\otimes g,\label{kappa normal ord star prod}\end{equation}
while the antisymmetric $\star$-product (\ref{antisymmetric starproduct})
reads \begin{equation}
f\star_{as}g=m\cdot e^{-\frac{ia}{2}x\partial_{x}\otimes\partial_{y}+\frac{ia}{2}\partial_{y}\otimes x\partial_{x}}f\otimes g.\label{kappa antisymm starprod}\end{equation}
 Notice that the antisymmetric $\star$-product differs (\ref{kappa antisymm starprod})
from the Weyl-ordered $\star$-product (\ref{Weyl ordered kappa})
at second order and therefore does not correspond to symmetric ordering.
The equivalence transformation (\ref{starproduct equivalence}) between
(\ref{kappa normal ord star prod}) and (\ref{kappa antisymm starprod})
is\begin{equation}
\rho=e^{-\frac{i}{2}ax\partial_{x}\partial_{y}}.\label{kappa equiv trans}\end{equation}

\chapter{\label{sec: derivations}Derivatives and Derivations}

We are now able to represent more complicated algebras on ordinary
functions by using the $\star$-products of the last chapter. But
there is an important element still missing: derivatives. In the canonical
case, we could just use the ordinary derivatives to construct noncommutative
actions. This was unproblematic as the usual derivatives had an undeformed
Leibniz rule, i.e.\begin{equation}
\dd_{i}(f\star g)=\dd_{i}f\star g+f\star\dd_{i}g.\end{equation}
 But with more complicated $\star$-products, this is in general no
longer the case. The derivatives do not only act on the functions,
but also on the $\star$-product, which now depends on the coordinates.
Symbolically we can write\begin{equation}
\dd_{i}(f\star g)=\dd_{i}f\star g+f\star\dd_{i}g+f(\dd_{i}\star)g,\end{equation}
where $\dd_{i}\star$ means that the derivative is acting on the bidifferential
operator $\star$ represents. This additional term can already be
seen at the level of the Poisson structure. Take e.g. the Poisson
structure (\ref{Poisson structure kappa}) of the $\kappa$-deformed
plane. The derivative $\dd_{y}$ in the $y$-direction does not act
on it, so that we still have\begin{equation}
\dd_{y}\{ f,g\}_{p}=\{\dd_{y}f,g\}_{p}+\{ f,\dd_{y}g\}_{p},\label{Leibniz rule poisson s}\end{equation}
 but in the $x$-direction, things are different:\begin{equation}
\dd_{x}\{ f,g\}_{p}=\dd_{x}(-iax\pp xf\pp yg+ia\pp yfx\pp xg)=\{\dd_{x}f,g\}_{p}+\{ f,\dd_{x}g\}_{p}+\frac{\{ f,g\}_{p}}{x}\end{equation}
 for $x\neq0$. The same is true for derivatives acting on $\star$-products:
the usual Leibniz rule is deformed. For the antisymmetric $\star$-product
on the $\kappa$-deformed plane, this deformed Leibniz rule reads\begin{equation}
\pp x(f\star_{as}g)=(\pp xf)\star_{as}(e^{-\frac{i}{2}a\pp y}g)+(e^{\frac{i}{2}a\pp y}f)\star_{as}(\pp xg),\end{equation}
 see (\ref{coproduct pp x}). Such derivatives with a deformed Leibniz
rule can nevertheless be used to construct gauge theory \cite{Dimitrijevic:2003pn,Dimitrijevic:2003wv,Dimitrijevic:2004vv},
but it is far more involved than in the canonical case. Especially,
the gauge fields associated to these derivatives become derivative
valued (see also chapter \ref{sec: general NC gauge theory}). 

Here, we will pursue a different approach. As we saw in (\ref{Leibniz rule poisson s}),
the derivative in the $y$-direction did act on the Poisson structure
as in the canonical case. And on the antisymmetric $\star$-product,
its Leibniz rule is indeed undeformed, i.e\begin{equation}
\pp y(f\star_{as}g)=(\pp yf)\star_{as}g+f\star_{as}(\pp yg),\end{equation}
 see (\ref{coproduct pp y}). Such a derivative can be gauged much
in the same way as in the canonical case, leading to function valued
gauge fields. But before we actually construct gauge theory in chapter
(\ref{sec: gauge theory on}), we will first have a closer look at
objects that behave like $\dd_{y}$ did in our example, i.e we will
be looking for vector fields that commute with the Poisson structure
and how we can use them to get differential operators that have an
undeformed Leibniz rule with the corresponding $\star$-product. These
differential operators with an undeformed Leibniz rule we will call
derivations of the $\star$-product algebra.

\section{\label{sec: derivations and forms}Derivations }

We will be able to identify derivations of $\star$-product algebras
with what we call Poisson vector fields of the Poisson structure associated
with the $\star$-product, i.e. vector fields $X$ with \begin{equation}
X\{ f,g\}_{p}=\{ Xf,g\}_{p}+\{ f,Xg\}_{p}.\end{equation}
 If we locally write \begin{equation}
\{ f,g\}_{p}=\pi(f,g)=\pi^{ij}\dd_{i}f\dd_{j}g\;\;\;\;\mbox{\mbox{and}}\;\;\;\; X=X^{i}\dd_{i},\end{equation}
this is equivalent to saying that the Schouten-Nijenhuis bracket (see
\ref{Schouten-Nijenhuis bracket}) of the vector field $X$ with the
Poisson structure $\pi$ vanishes \begin{equation}
[X,\pi]_{S}=0\;\;\Leftrightarrow\;\; X^{k}\partial_{k}\pi^{ij}-\pi^{ik}\partial_{k}X^{j}+\pi^{jk}\partial_{k}X^{i},\end{equation}
 or that the vector field $X$ commutes with the Poisson structure
$\pi$. If we have such a Poisson vector field $X$, we are looking
for a differential operator $\delta_{X}$ with the following property
\begin{equation}
\delta_{X}(f\star g)=\delta_{X}f\star g+f\star\delta_{X}g.\end{equation}

Such a map $\delta$ from the vectorfields to the differential operators,
which maps the derivations of the Poisson manifold $T_{\pi}M=\{ X\in TM|[X,\pi]_{S}=0\}$
to the derivations of the $\star$-product $D_{\star}M=\{\delta\in D_{poly}|[\delta,\star]_{G}=0\}$,
can be constructed both for the Weyl ordered $\star$-product (see
\ref{weyl derivations}), for the formality $\star$-product (see
\ref{formality derivations} ) and the Jambor-Sykora $\star$-product
(see \ref{Syko derivations}). Here we want to investigate the general
properties of such a map $\delta$. For this we expand it on a local
patch in terms of partial derivatives\begin{equation}
\delta_{X}=\delta_{X}^{i}\partial_{i}+\delta_{X}^{ij}\partial_{i}\partial_{j}+\cdots.\end{equation}
 Due to its property to be a derivation, $\delta_{X}$ is completely
determined by the first term $\delta_{X}^{i}\partial_{i}$. This means
that if the first term is zero, the other terms have to vanish, too.
If further $e$ is an arbitrary derivation of the $\star$-product,
there must exist a vector field $X_{e}$ such that\begin{equation}
\delta_{X_{e}}=e.\end{equation}
 If $X,Y\in T_{\pi}M$, then $[\delta_{X},\delta_{Y}]$ is again a
derivation of the $\star$-product and we can conclude that \begin{equation}
[\delta_{X},\delta_{Y}]=\delta_{[X,Y]_{\star}},\label{lie_delta}\end{equation}
 where $[X,Y]_{\star}$ is a deformation of the ordinary Lie bracket
of vector fields. Obviously it is linear, skew-symmetric and fulfills
the Jacobi identity.

With the help of the map $\delta$ and the deformed bracket $[\,\cdot\,,\,\cdot\,]_{\star}$
it is also possible to construct noncommutative forms over the derivations
of the $\star$-product algebra, a formulation we will present in
appendix \ref{sec:Noncommutative-forms}.

\section{\label{weyl derivations}Derivations for the Weyl-ordered $\star$-product}

We now want to calculate the derivations $\delta_{X}$ of the Weyl-ordered
$\star$-product (\ref{wo_product}) from the derivations $X$ of
the Poisson structure $c^{ij}$ up to second order. We assume that
$\delta_{X}$ can be expanded in the following way\begin{equation}
\delta_{X}=X^{i}\partial_{i}+\delta_{X}^{ij}\partial_{i}\partial_{j}+\delta_{X}^{ijk}\partial_{i}\partial_{j}\partial_{k}+\cdots.\end{equation}
 Expanding the equation \begin{equation}
\delta_{X}(f\star g)=\delta_{X}(f)\star g+f\star\delta_{X}(g)\end{equation}
 order by order and using $[X,c]_{S}=0$ we find that\begin{eqnarray}
\delta_{X} & = & X^{i}\partial_{i}-\frac{1}{12}c^{lk}\partial_{k}c^{im}\partial_{l}\partial_{m}X^{j}\partial_{i}\partial_{j}\\
 &  & +\frac{1}{24}c^{lk}c^{im}\partial_{l}\partial_{i}X^{j}\partial_{k}\partial_{m}\partial_{j}+\mathcal{O}(3).\nonumber \end{eqnarray}
 For $[\,\cdot\,,\,\cdot\,]_{\star}$ we simply calculate $[\delta_{X},\delta_{Y}]$
and get\begin{eqnarray}
[X,Y]_{\star} & = & [X,Y]_{L}\\
 &  & -\frac{1}{12}(c^{lk}\partial_{k}c^{im}\partial_{l}\partial_{m}X^{j}\partial_{i}\partial_{j}Y^{n}-c^{lk}\partial_{k}c^{im}\partial_{l}\partial_{m}Y^{j}\partial_{i}\partial_{j}X^{n})\partial_{n}\nonumber \\
 &  & +\frac{1}{24}(c^{lk}c^{im}\partial_{l}\partial_{i}X^{j}\partial_{k}\partial_{m}\partial_{j}Y^{n}-c^{lk}c^{im}\partial_{l}\partial_{i}Y^{j}\partial_{k}\partial_{m}\partial_{j}X^{n})\partial_{n}\nonumber \\
 &  & +\mathcal{O}(3).\nonumber \end{eqnarray}

\section{\label{formality derivations}Derivations for the formality $\star$-product}

We saw in chapter \ref{Formality map} that the formality $\star$-product
can be constructed from the maps $U_{n}$ from the polyvectorfields
to the polydifferential operators as\begin{equation}
f\star g=\sum_{n=0}^{\infty}\frac{1}{n!}U_{n}(\pi,\,\ldots,\pi)(f,g).\end{equation}
With these maps, we can further define the special polydifferential
operators

\begin{eqnarray}
\Phi(\alpha) & = & \sum_{n=1}^{\infty}\frac{1}{(n-1)!}U_{n}(\alpha,\pi,\,\ldots,\pi),\\
\Psi(\alpha_{1},\alpha_{2}) & = & \sum_{n=2}^{\infty}\frac{1}{(n-2)!}U_{n}(\alpha_{1},\alpha_{2},\pi,\ldots,\pi).\end{eqnarray}
 For $X$ a vectorfield, we define \begin{equation}
\delta_{X}=\Phi(X).\end{equation}
 Using formula (\ref{relation between grades}) we see that it is
indeed a 1-differential operator. We will now use the formality condition
(\ref{formality cond}) to have a closer look at its properties.

For $g$ a function and $X$ and $Y$ vectorfields, we see that $\Psi(X,Y)$
is a function and we go on to calculate\begin{eqnarray}
{[\delta_{X},\star]_{G}} & = & \Phi([X,\pi]_{S}),\label{fromality rel phi}\\
{[\delta_{X},\delta_{Y}]_{G}+[\Psi(X,Y),\star]_{G}} & = & \delta_{[X,Y]_{S}}\label{formality rel Psi}\\
 &  & +\Psi([\pi,Y]_{S},X)-\Psi([\pi,X]_{S},Y).\nn\end{eqnarray}
 If $\pi$ is a Poisson tensor, i. e. $[\pi,\pi]_{S}=0$ and if $X$
and $Y$ are Poisson vector fields, i. e. $[X,\pi]_{S}=[Y,\pi]_{S}=0$,
the relations (\ref{fromality rel phi}) and (\ref{formality rel Psi})
become\begin{eqnarray}
\delta_{X}(f\star g) & = & \delta_{X}(f)\star g+f\star\delta_{X}(g),\\
({[\delta_{X},\delta_{Y}]-\delta_{[X,Y]_{L}}})(g) & = & [\Psi(X,Y)\stackrel{\star}{,}g].\end{eqnarray}
 when evaluated on functions. We see that $\delta$ really is the
map we were looking for, i.e. it maps derivations of the Poisson structure
to derivations of the associated formality $\star$-product.

Additionally the map $\delta$ preserves the bracket up to an inner
derivation. This can be cast into the following form:\begin{equation}
{[\delta_{X},\delta_{Y}]}=\delta_{[X,Y]_{\star}}\end{equation}
 with\begin{equation}
{[X,Y]_{\star}}=[X,Y]_{L}+H_{\Phi^{-1}\Psi(X,Y)}.\end{equation}

\section{\label{Syko derivations}Derivations for the Jambor-Sykora $\star$-product}

While looking for derivatives for the Jambor-Sykora $\star$-products,
we can confine ourselves to the antisymmetric case (\ref{antisymmetric starproduct}),
as all the properties can be pulled back to the full one (\ref{full starproduct})
via the equivalence transformation (\ref{starproduct equivalence}). 

In the framework of Kontsevich's formality $\star$-product, we saw
that we could construct a map from the vectorfields to the differential
operators that maps derivations of the Poisson structure to derivations
of the formality $\star$-product to all orders. 

We will now look for such a map that maps derivations of the Poisson
structure \begin{equation}
\pi=\frac{1}{2}\sigma_{as}^{ab}X_{a}^{i}\partial_{i}\wedge X_{b}^{j}\partial_{j}\label{Poisson structure}\end{equation}
associated with the Jambor-Sykora $\star$-product to derivations
of the antisymmetric Jambor-Sykora $\star$-product (\ref{antisymmetric starproduct}).
As the vectorfields $X_{a}$ commute with each other and $\sigma_{as}$
is antisymmetric, (\ref{antisymmetric starproduct}) can be rewritten
in terms of the Poisson structure as \begin{equation}
f\star_{as}g=m(e^{\sigma_{as}^{ab}X_{a}^{i}\partial_{i}\otimes X_{b}^{j}\partial_{j}}(f\otimes g))=m(e^{\pi}(f\otimes g)).\label{antisymm starprod}\end{equation}
 In this notation it is obvious that vectorfields commuting with the
Poisson structure (\ref{Poisson structure}) will be derivations of
the $\star$-product (\ref{antisymm starprod}) as well. This means
that for vectorfields $Y$ with\[
[Y,\pi]_{S}=0\;\;\Leftrightarrow\;\;[Y\otimes1+1\otimes Y,\pi]=0\;\;\Leftrightarrow\;\; Y\{ f,g\}_{P}=\{ Yf,g\}_{P}+\{ f,Yg\}_{P}\]
 we also have\begin{eqnarray}
Y(f\star_{as}g) & = & m((Y\otimes1+1\otimes Y)e^{\pi}(f\otimes g))\\
 & = & m(e^{\pi}(Y\otimes1+1\otimes Y)(f\otimes g))\nonumber \\
 & = & Yf\star_{as}g+f\star_{as}Yg.\nonumber \end{eqnarray}
We therefore do not get higher order terms, the map from the vectorfields
to the differential operators is the inclusion. This also implies
that the algebra of the vectorfields remains undeformed under quantization. 

For the coproduct of general vectorfields acting on the $\star$-product
(\ref{antisymmetric starproduct}), we get a deformed Leibniz rule

\begin{eqnarray}
\lefteqn{X(f\star_{as}g)=}\label{def coproduct}\\
 &  & \sum_{n=0}^{\infty}\frac{1}{n!}\sigma_{as}^{a_{1}a_{2}}...\sigma_{as}^{a_{n}b_{n}}([...[X,X_{a_{1}}],X_{a_{2}}],...],X_{a_{n}}]f)\star_{as}(X_{b_{1}}...X_{b_{n}}g)\nonumber \\
 &  & +\sum_{n=0}^{\infty}\frac{1}{n!}\sigma_{as}^{a_{1}a_{2}}...\sigma_{as}^{a_{n}b_{n}}(X_{a_{1}}...X_{a_{n}}f)\star_{as}([...[X,X_{b_{1}}],X_{b_{2}}],...],X_{b_{n}}]g).\nonumber \end{eqnarray}

We can use the equivalence transformation (\ref{equivalence transformation})
to pull back these structures to the full $\star$-product (\ref{full starproduct}).
Note that the Poisson structure of the full $\star$-product (\ref{full starproduct})
only depends on the antisymmetric part of $\sigma$ and therefore
is (\ref{Poisson structure}), the same as for the antisymmetric $\star$-product
(\ref{antisymmetric starproduct}).

A vectorfield $Y$ commuting with the Poisson structure of the full
$\star$-product (\ref{full starproduct}) will be a derivation of
the antisymmetric $\star$-product (\ref{antisymmetric starproduct}).
We use (\ref{starproduct equivalence}) to get \begin{equation}
\rho Y\rho^{-1}(f\star_{\sigma}g)=(\rho Y\rho^{-1}f)\star_{\sigma}g+f\star_{\sigma}(\rho Y\rho^{-1}g)\end{equation}
 from\begin{equation}
Y(f\star_{as}g)=Yf\star_{as}g+f\star_{as}Yg.\end{equation}
 The map $\delta_{X}$ from the vectorfields to the differential operators
for the full $\star$-product (\ref{full starproduct}) is therefore
given by\begin{equation}
\delta_{X}=\rho X\rho^{-1}.\label{deformed vectorfields}\end{equation}
The algebra of these deformed vectorfields is isomorphic to the algebra
of the undeformed vectorfields. Also the coalgebra of the deformed
vectorfields with respect to the full $\star$-product (\ref{full starproduct})
is isomorphic to the coalgebra of the undeformed vectorfields with
respect to the antisymmetric $\star$-product (\ref{antisymmetric starproduct}).

\section{\label{example: derivatives}Example: Derivatives and Derivations
for the $\kappa$-deformed plane}

We will now exemplify the ideas of this chapter by applying them to
the $\kappa$-deformed plane we already studied in chapter \ref{example: star products}.
We will now concentrate on the Jambor-Sykora $\star$-products, where
the basic mechanisms at work can be best seen.

\subsection{The antisymmetric case}

For the antisymmetric $\star$-product (\ref{kappa antisymm starprod}),
the vectorfields are undeformed, but they may acquire nontrivial coproducts
(\ref{def coproduct}). The derivative in the $y$-direction commutes
with the Poisson structure and is therefore a derivation of the $\star$-product:\begin{equation}
\pp y(f\star_{as}g)=(\pp yf)\star_{as}g+f\star_{as}(\pp yg).\label{coproduct pp y}\end{equation}
But the derivative in the $x$-direction does not commute with the
Poisson structure and has a deformed Leibniz rule:\begin{equation}
\pp x(f\star_{as}g)=(\pp xf)\star_{as}(e^{-\frac{i}{2}a\pp y}g)+(e^{\frac{i}{2}a\pp y}f)\star_{as}(\pp xg).\label{coproduct pp x}\end{equation}
Multiplication from the left with a function (without $\star$-multiplication)
also acquires a derivative quality:\begin{eqnarray}
y(f\star_{as}g) & = & (yf)\star_{as}g-\frac{i}{2}af\star_{as}(x\pp xg)\\
 & = & f\star_{as}(yg)+\frac{i}{2}a(x\pp xf)\star_{as}g\nonumber \end{eqnarray}
and\begin{eqnarray}
x(f\star_{as}g) & = & (xf)\star_{as}(e^{\frac{i}{2}a\pp y}g)\label{coproduct x}\\
 & = & (e^{-\frac{i}{2}a\pp y}f)\star_{as}(xg).\nonumber \end{eqnarray}
This also implies the following relations\begin{equation}
(xf)\star_{as}g=(e^{-\frac{i}{2}a\pp y}f)\star_{as}(xe^{-\frac{i}{2}a\pp y}g)\end{equation}
and\begin{equation}
(yf)\star_{as}g=f\star_{as}(yg)+\frac{i}{2}ax\pp x(f\star_{as}g).\end{equation}
 If we combine (\ref{coproduct pp x}) and (\ref{coproduct x}) to
calculate the coproduct of $x\pp x$, we see that it is indeed a derivation
of the $\star$-product, as expected.\begin{equation}
x\pp x(f\star_{as}g)=(x\pp xf)\star_{as}g+f\star_{as}(x\pp xg)\end{equation}
 For the other vectorfields linear in the coordinates we get\begin{equation}
x\pp y(f\star_{as}g)=(x\pp yf)\star_{as}(e^{\frac{i}{2}a\pp y}g)+(e^{-\frac{i}{2}a\pp y}f)\star_{as}(x\pp yg),\end{equation}
\begin{eqnarray}
y\pp y(f\star_{as}g) & = & (y\pp yf)\star_{as}g+f\star_{as}(y\pp yg)\\
 &  & -\frac{i}{2}a(\pp yf)\star_{as}(x\pp xg)+\frac{i}{2}a(x\pp xf)\star_{as}(\pp yg)\nonumber \end{eqnarray}
and\begin{eqnarray}
y\pp x(f\star_{as}g) & = & (y\pp xf)\star_{as}(e^{-\frac{i}{2}a\pp y}g)+(e^{\frac{i}{2}a\pp y}f)\star_{as}(y\pp xg)\\
 &  & -\frac{i}{2}a(\pp xf)\star_{as}(x\pp xe^{-\frac{i}{2}a\pp y}g)+\frac{i}{2}a(x\pp xe^{\frac{i}{2}a\pp y}f)\star_{as}(\pp xg).\nonumber \end{eqnarray}

\subsection{The normal ordered case}

Of course we can switch to the $\star$-product (\ref{kappa normal ord star prod})
corresponding to normal ordering by using the transformation (\ref{kappa equiv trans})\begin{equation}
\rho=e^{-\frac{i}{2}ax\partial_{x}\partial_{y}}.\end{equation}
The vectorfields are then mapped to the differential operators by
applying (\ref{deformed vectorfields}). For the coordinates we get
\begin{equation}
\delta_{y}=\rho y\rho^{-1}=y-\frac{i}{2}ax\pp x\;\;\;\;\mbox{and}\;\;\;\;\delta_{x}=\rho x\rho^{-1}=xe^{-\frac{i}{2}a\pp y},\label{delta x and y}\end{equation}
 revealing the derivative nature of multiplication of coordinates
from the left. The derivative in the $y$-direction stays undeformed\begin{equation}
\delta_{\pp y}=\rho\pp y\rho^{-1}=\pp y,\label{delta pp y}\end{equation}
 the derivative in the $x$-direction becomes\begin{equation}
\delta_{\pp x}=\rho\pp x\rho^{-1}=e^{\frac{i}{2}a\pp y}\pp x.\label{delta pp x}\end{equation}
 We can combine (\ref{delta x and y}) and (\ref{delta pp x}) to
give\begin{equation}
\delta_{x\pp x}=\rho x\pp x\rho^{-1}=\rho x\rho^{-1}\rho\pp x\rho^{-1}=x\pp x.\end{equation}
 Note that the deformation $\delta$ acts trivially on $x\dd_{x}$,
as it does commute with $\rho$. For the other vectorfields linear
in the coordinates we get from (\ref{delta x and y},\ref{delta pp y},\ref{delta pp x})\begin{eqnarray}
\delta_{x\pp y} & = & x\pp ye^{-\frac{i}{2}a\pp y},\nonumber \\
\delta_{y\pp y} & = & y\pp y-\frac{i}{2}ax\pp x\pp y,\\
\delta_{y\pp x} & = & y\pp xe^{\frac{i}{2}a\pp y}-\frac{i}{2}ax\pp x^{2}e^{\frac{i}{2}a\pp y}.\nonumber \end{eqnarray}

\chapter{Gauge theory on curved NC spaces\label{sec: gauge theory on}}

One hope associated with the application of noncommutative geometry
in physics is a better description of quantized gravity. At least
it should be possible to construct effective actions where traces
of this unknown theory remain. If one believes that quantum gravity
is in a sense a quantum field theory, then its observables are operators
on a Hilbert space and therefore elements of an algebra. Some properties
of this algebra should be reflected in the noncommutative geometry
the effective actions are constructed on. As the noncommutativity
should be induced by background gravitational fields, the classical
limit of the effective actions should reduce to actions on curved
spacetimes \cite{Madore:1997ta,Connes:1996gi}.

In the canonical case, the gauge theory reduces in the commutative
limit to a theory on flat spacetime. Therefore it is necessary to
develop concepts working with more general algebras, since one would
expect that curved backgrounds are related to algebras with nonconstant
commutation relations. We will use the derivations of $\star$-product
algebras we studied in chapter \ref{sec: derivations} to build covariant
derivatives for noncommutative gauge theory. We will be able to write
down a noncommutative action by linking these derivations to a frame
field induced by a nonconstant metric. In the commutative limit, this
action reduces to gauge theory on a curved manifold. As an example
we will again study $\kappa$-deformed spacetime, where the action
reduces in the commutative limit to scalar electrodynamics on a manifold
with constant curvature.

We will also introduce Seiberg-Witten maps to do noncommutative gauge
theory with arbitrary gauge groups. A proof of the existence of the
Seiberg-Witten-map for an Abelian gauge potential will be given for
the formality $\star$-product. We will also give explicit formulas
for the Weyl ordered $\star$-product up to second order.

\section{The general formalism}

\subsection{\label{sec: general NC gauge theory}Noncommutative gauge theory}

To do gauge theory on the noncommutative spaces equipped with the
more complicated $\star$-products of chapter \ref{sec: general starproducts},
we will try to follow the formalism of the canonical case as much
as possible. 

Fields in the fundamental representation will again transform as\begin{equation}
\delta_{\Lambda}\Psi=i\Lambda\star\Psi.\end{equation}
 The commutator of two such gauge transformations should again be
a gauge transformation, i.e we again want \begin{equation}
(\delta_{\Lambda}\delta_{\Xi}-\delta_{\Xi}\delta_{\Lambda})\Psi=\delta_{i[\Xi\stackrel{\star}{,}\Lambda]}\Psi.\end{equation}
As in the canonical case, this is only possible for gauge groups $U(N)$.
The first difference to the canonical case occurs when we look at
the transformation properties of a derivative\begin{equation}
\delta_{\Lambda}(\dd_{i}\Psi)=\dd_{i}(i\Lambda\star\Psi)=i(\dd_{i}\Lambda)\star\Psi+i\Lambda\star(\dd_{i}\Psi)+i\Lambda(\dd_{i}\star)\Psi.\end{equation}
 The additional term $i\Lambda(\dd_{i}\star)\Psi$ is in general no
longer zero, corresponding to a nontrivial coproduct of the derivative.
If we now want to add a gauge field $A_{i}$ to the derivative to
make it gauge invariant, i.e. \begin{equation}
D_{i}\Psi=\dd_{i}\Psi-iA_{i}\star\Psi,\end{equation}
 the transformation properties of $A_{i}$ also have to offset this
new term to get\begin{equation}
\delta_{\Lambda}(D_{i}\Psi)=i\Lambda\star D_{i}\Psi.\end{equation}
 From this we get\begin{equation}
\delta_{\Lambda}(A_{i})\star\Psi=\dd_{i}\Lambda\star\Psi+i[\Lambda\stackrel{\star}{,}A_{i}]\star\Psi+\Lambda(\dd_{i}\star)\Psi,\end{equation}
 which means that the gauge potential can no longer be a function,
it has to be derivative valued. To see this better, we take as an
example the $\star$-product (\ref{kappa antisymm starprod}) for
the $\kappa$-deformed plane. The above formula then reads \begin{eqnarray}
\delta_{\Lambda}(A_{x})\star_{as}\Psi & = & (\pp x\Lambda)\star_{as}(e^{-\frac{i}{2}a\pp y}\Psi)\\
 &  & +((e^{\frac{i}{2}a\pp y}-1)\Lambda)\star_{as}(\pp x\Psi)+i[\Lambda\stackrel{\star_{as}}{,}A_{x}]\star_{as}\Psi.\nonumber \end{eqnarray}
 To offset the terms coming from the deformed Leibniz rule for $\dd_{x}$
(where additional derivatives act on the right hand side), the gauge
field $A_{x}$ has to become derivative valued. Gauge theory using
such derivative valued gauge fields was constructed in \cite{Dimitrijevic:2003pn,Dimitrijevic:2003wv,Dimitrijevic:2004vv},
but we will try a different approach here.

We saw in chapter \ref{sec: canonical NC gauge theory} that there
is a different formulation for noncommutative gauge theory in terms
of covariant coordinates. So let us see what happens if we try to
gauge the coordinates with a more complicated $\star$-product. We
want to have\begin{equation}
\delta_{\Lambda}(X_{i}\star\Psi)=\delta_{\Lambda}((x_{i}+\widetilde{A}_{i})\star\Psi)=i\Lambda\star X_{i}\star\Psi.\end{equation}
Therefore the gauge field $\widetilde{A}_{i}$ has to transform as\begin{equation}
\delta_{\Lambda}\widetilde{A}_{i}=i[\Lambda\stackrel{\star}{,}x_{i}]+i[\Lambda\stackrel{\star}{,}\widetilde{A}_{i}].\end{equation}
 This means that $\widetilde{A}_{i}$ is still a function, because
the commutator with a coordinate of course has an undeformed Leibniz
rule. But there is a problem with this Ansatz: the gauge field $\widetilde{A}_{i}$
vanishes in the commutative limit. In the canonical case, this could
be solved by defining a new field $(\theta^{-1})^{ij}\widetilde{A}_{j}$,
but this is no longer possible as the now coordinate dependent $\theta^{-1}$
would spoil the transformation properties of the new object. 

This is why we introduced derivations $\delta_{X}$ in chapter \ref{sec: derivations}.
They do have both an undeformed Leibniz rule and a nonvanishing commutative
limit. So we introduce covariant derivations as \begin{equation}
D_{X}=\delta_{X}-iA_{X},\end{equation}
 where $X$ is a Poisson vector field. The gauge field $A_{X}$ will
transform as \begin{equation}
\delta_{\Lambda}A_{X}=\delta_{X}\Lambda+i[\Lambda\stackrel{\star}{,}A_{X}].\end{equation}
 Then, a field strength $F_{X,Y}$ can be defined as \begin{equation}
{}-iF_{X,Y}=[D_{X}\stackrel{\star}{,}D_{Y}]-D_{[X,Y]_{\star}},\end{equation}
the properties of $D$ and $[\,\cdot\,,\,\cdot\,]_{\star}$ making
sure that the field strength is function-valued and transforms covariantly%
\footnote{This can also be expressed in the language of the noncommutative forms
introduced in appendix \ref{sec:Noncommutative-forms}. $A_{X}$ is
the connection one form evaluated on the vector field $X$. It transforms
like\begin{equation}
\delta_{\Lambda}A=\delta\Lambda+i\Lambda\wedge A-iA\wedge\Lambda.\end{equation}
 The covariant derivative of a field is now\begin{equation}
D\Psi=\delta\Psi-iA\wedge\Psi,\end{equation}
 and the field strength becomes\begin{equation}
F=DA=\delta A-iA\wedge A.\end{equation}
 One easily can show that the field strength is a covariant constant\begin{equation}
DF=\delta F-iA\wedge F=0.\end{equation}
}.

\subsection{Seiberg-Witten gauge theory\label{swmap}}

Up to now, we could only do noncommutative gauge theory for gauge
groups $U(n)$, just as in the canonical case. We will now show how
to implement the concept of Seiberg-Witten maps \cite{Seiberg:1999vs,Jurco:2000ja}
into our new setting of covariant derivations to be able to do gauge
theory for general gauge groups. 

Just as in the canonical case, the Seiberg-Witten maps for the fields
will have to be enveloping algebra valued, but they will only depend
on their commutative counterparts, therefore preserving the right
number of degrees of freedom. Again we demand that their noncommutative
transformation properties are determined by the transformation properties
of the commutative fields they depend on.

Therefore the fields again transform as \cite{Jurco:2001rq}\begin{equation}
\delta_{\alpha}\Psi_{\psi}[a]=i\Lambda_{\alpha}[a]\star\Psi_{\psi}[a],\label{sp_gt_field}\end{equation}
leading to the same consistency condition for the gauge parameter\begin{equation}
i\delta_{\alpha}\Lambda_{\beta}-i\delta_{\beta}\Lambda_{\alpha}+[\Lambda_{\alpha}\stackrel{\star}{,}\Lambda_{\beta}]=i\Lambda_{-i[\alpha,\beta]}.\label{sw_consitency_condition}\end{equation}
 The transformation law for the covariantizer is now\begin{equation}
\delta_{\alpha}(D[a](f))=i[\Lambda_{\alpha}[a]\stackrel{\star}{,}D[a](f)].\label{sp_gt_cov}\end{equation}
 The Seiberg-Witten-map can be easily extended to the derivations
$\delta_{X}$ of the $\star$-product. The noncommutative covariant
derivation $D_{X}[a]$ can be written with the help of a noncommutative
gauge potential $A_{X}[a]$ now depending both on the commutative
gauge potential $a$ and the Poisson vectorfield $X$\begin{equation}
D_{X}[a]\Psi_{\psi}[a]=\delta_{X}\Psi_{\psi}[a]-iA_{X}[a]\star\Psi_{\psi}[a].\label{sp_gt_cov_derivative}\end{equation}
 It follows that the gauge potential has to transform like\begin{equation}
\delta_{\alpha}A_{X}[a]=\delta_{X}\Lambda_{\alpha}[a]+i[\Lambda_{\alpha}[a]\stackrel{\star}{,}A_{X}[a]].\label{sp_gt_gauge_pot}\end{equation}
 We will give explicit formulas for the Seiberg-Witten maps in chapters
\ref{swmap_weyl} and \ref{swmap_form}.

\subsection{Commutative actions with the frame formalism}

To link the noncommutative constructions of the last chapters with
commutative gauge theory, we first want to recall some aspects of
classical differential geometry. Suppose we are working on a $n$-dimensional
manifold $M$ with metric $g_{\mu\nu}$. Then there are locally $n$
derivatives $\partial_{\mu}$ which form a basis of the tangent space
$TM$ of the manifold. We can always make a local basis transformation
to a frame (or {}``non-coordinate basis'') \begin{equation}
e_{a}=e_{a}{}^{\mu}(x)\partial_{\mu},\label{eq: frame 1}\end{equation}
 (with $e_{a}{}^{\mu}(x)$ invertible, i.e. $e_{a}{}^{\mu}e^{a}{}_{\nu}=\delta_{\nu}^{\mu}$)
where the metric is constant\begin{equation}
\eta_{ab}=e_{a}{}^{\mu}e_{b}{}^{\nu}g_{\mu\nu}.\end{equation}
 Since forms are dual to vector fields, they may be evaluated on the
frame. For the gauge field we get \begin{equation}
a_{a}=a(e_{a}),\end{equation}
leading to the covariant derivate \begin{equation}
D_{a}\psi=(D\psi)(e_{a})=e_{a}\psi-ia_{a}\psi.\end{equation}
 The field strength becomes\begin{equation}
f_{ab}=i[D_{a},D_{b}]-iD([e_{a},e_{b}])=e_{a}a_{b}-e_{b}a_{a}-a([e_{a},e_{b}])-i[a_{a},a_{b}].\end{equation}
 Locally this means that\begin{equation}
a_{a}=e_{a}{}^{\mu}a_{\mu}\,,\;\;\;\;\; D_{a}\psi=e_{a}{}^{\mu}D_{\mu}\;\;\;\;\;\mbox{and}\;\;\;\;\; f_{ab}=e_{a}{}^{\mu}e_{b}{}^{\nu}f_{\mu\nu}.\end{equation}
Using these definitions, the action for gauge theory on a curved manifold
can be written in the two different bases as\begin{equation}
\mathcal{S}=-\frac{1}{4}\int d^{n}x\,\sqrt{g}\,\eta^{ab}\eta^{cd}f_{ac}f_{bd}=-\frac{1}{4}\int d^{n}x\,\sqrt{g}\, g^{\mu\nu}g^{\rho\sigma}f_{\mu\rho}f_{\nu\sigma},\end{equation}
 where \begin{equation}
\sqrt{g}=\sqrt{\det\,(g_{\mu\nu})}=\sqrt{\det(e^{a}{}_{\mu}e^{b}{}_{\nu}\eta_{ab})}=\det\, e^{a}{}_{\mu}\end{equation}
 is the measure function induced by the metric.

\subsection{\label{Generalization to NC geometry}Gauge theory on curved noncommutative
spacetime}

In order to formulate gauge theory on a curved noncommutative spacetime,
we need a frame $e_{a}$ and a Poisson structure $\{\,\cdot\,,\,\cdot\,\}_{p}=\pi^{\mu\nu}\dd_{\mu}\wedge\dd_{\nu}$
that are compatible with each other. Compatibility means that the
frame $e_{a}$ commutes with the Poisson structure $\{\:\cdot\:,\:\cdot\:\}_{p}$,
i.e.\begin{equation}
e_{a}\{ f,g\}_{p}=\{ e_{a}f,g\}_{p}+\{ f,e_{a}g\}_{p},\end{equation}
and that the measure function $\sqrt{g}$ induced by the metric $g_{\mu\nu}=e^{a}{}_{\mu}e^{b}{}_{\nu}\eta_{ab}$
is also a measure function for the Poisson manifold, i.e. that we
have\[
\dd_{\mu}(\sqrt{g}\pi^{\mu\nu})=0.\]
 We will call the $\star$-product algebra generated by quantizing
such a Poisson structure a curved noncommutative space, as the gauge
theory we will define on it in this chapter will reduce to gauge theory
on a curved manifold in the commutative limit. In appendix \ref{sec: Frames}
we will propose a method how to find frames commuting with the Poisson
structure in the context of quantum spaces. How to find Poisson structures
compatible with a given frame by a construction based on differential
equations can be found in \cite{Sykora:2004pt}.

For the gauge theory, we saw in chapter \ref{sec: general NC gauge theory}
that we can define a covariant derivative of a field by using a derivation
$\delta_{X}$\begin{equation}
D_{X}\Psi_{\psi}=\delta_{X}\Psi_{\psi}-iA_{X}\star\Psi_{\psi}.\label{x_derivative}\end{equation}
 With this, a field strength could be defined as \begin{equation}
-iF_{X,Y}=[D_{X}\stackrel{\star}{,}D_{Y}]-D_{[X,Y]_{\star}}.\label{xy_field_strength}\end{equation}
 The properties of $\delta_{\cdot}$ and $[\,\cdot\,,\,\cdot\,]_{\star}$
ensured that this really is a function and not a polydifferential
operator.

On a curved noncommutative space, we can quantize the frame $e_{a}$
with the map $\delta$ to get derivations of the $\star$-product.
These we can use to define our covariant derivatives. The noncommutative
covariant derivative (\ref{x_derivative}) and field strength (\ref{xy_field_strength})
evaluated on the frame $e_{a}$ then read\begin{equation}
D_{a}\Phi=D_{e_{a}}\Phi=\delta_{e_{a}}\Phi-iA_{e_{a}}\star\Phi,\end{equation}
\begin{equation}
-iF_{ab}=-iF_{e_{a},e_{b}}=[D_{e_{a}}\stackrel{\star}{,}D_{e_{b}}]-D_{[e_{a},e_{b}]_{\star}}.\end{equation}
The field strength will transform covariantly under gauge transformations,
i.e. we have\begin{equation}
\delta_{\Lambda}(F)=i[\Lambda\stackrel{\star}{,}F].\end{equation}
To make the action gauge invariant, the integral has to have the trace
property, i.e. it has to be invariant under cyclic permutations. For
this we need a measure function $\Omega$, which in our case will
be the measure function induced by the metric plus possible higher
orders in the noncommutativity (see also chapter \ref{sec: traces}),
i.e. we will have \begin{equation}
\Omega=\sqrt{g}+\mc O(1).\end{equation}
With this, we have a noncommuative gauge action \begin{equation}
\mathcal{S}=-\frac{1}{4}\int d^{n}x\,\Omega\,\eta^{ab}\eta^{cd}F_{ac}\star F_{bd}.\end{equation}
 that goes in the commutative limit\begin{equation}
\mathcal{S}\rightarrow-\frac{1}{4}\int d^{n}x\,\sqrt{g}\, g^{\mu\nu}g^{\rho\sigma}f_{\mu\rho}f_{\nu\sigma}\end{equation}
 to gauge theory on a curved manifold.

\subsubsection{Scalars}

For the noncommutative version of a scalar Lagrangian \begin{equation}
\eta^{ab}D_{a}\overline{\phi}D_{b}\phi+m^{2}\overline{\phi}\phi,\end{equation}
we also need an involution $\bar{\cdot}$ of the $\star$-product,
i.e.\begin{equation}
\overline{(f\star g)}=\overline{g}\star\overline{f}.\end{equation}
 To make the NC Lagrangian invariant under NC gauge transformations,
the NC gauge parameter $\Lambda$ and the NC gauge field $A_{X}$
have to be invariant under this involution to get\begin{equation}
\delta_{\Lambda}\overline{\phi}=\overline{(\Lambda\star\phi)}=\overline{\phi}\star\overline{\Lambda}=\overline{\Phi}\star\Lambda\end{equation}
and \begin{equation}
\overline{(A_{X}\star\phi)}=\overline{\phi}\star\overline{A_{X}}=\overline{\phi}\star A_{X}.\end{equation}
 For the Weyl-ordered $\star$-product, ordinary complex conjugation
still is an involution, and the hermiticity of the NC gauge parameter
$\Lambda$ and the NC gauge field $A_{X}$ can be checked explicitly
on the formulas of the Seiberg-Witten map in chapter \ref{swmap_weyl}.

Putting everything together, we therefore end up with an action \begin{equation}
\mathcal{S}=\int d^{n}x\,\Omega\,(-\frac{1}{4}\eta^{ab}\eta^{cd}F_{ac}\star F_{bd}+\eta^{ab}D_{a}\overline{\Phi}\star D_{b}\Phi-m^{2}\overline{\Phi}\star\Phi).\label{gen_action}\end{equation}
 that is invariant under noncommutative gauge transformations\begin{equation}
\delta_{\Lambda}S=0\end{equation}
 and reduces in the commutative limit \begin{equation}
\mathcal{S}\rightarrow\int d^{n}x\,\sqrt{g}\,(-\frac{1}{4}g^{\mu\nu}g^{\rho\sigma}f_{\mu\rho}f_{\nu\sigma}+g^{\mu\nu}D_{\mu}\bar{\phi}D_{\nu}\phi-m^{2}\bar{\phi}\phi),\end{equation}
 to scalar electrodynamics on a curved manifold.

\subsubsection{Spinors}

Even though it isn't clear how to define NC spinors on general curved
spacetimes due to the nontrivial spin-connection, it should still
be possible in two dimensions. There, the spin connection vanishes
and the commutative spinor action can be written as\begin{equation}
S_{spinor}=\frac{1}{2}\int d^{2}x\sqrt{g}\overline{\Psi}i\gamma^{a}e_{a}^{\mu}(\partial_{\mu}-iA_{\mu}+m)\Psi.\label{S_spinor}\end{equation}

Note that with the usual gamma-matrices $\{\gamma^{a},\gamma^{b}\}=2\eta^{ab}$
and $\gamma^{\mu}=\gamma^{a}e_{a}^{\mu}$, we get $\{\gamma^{\mu},\gamma^{\nu}\}=2g^{\mu\nu}$.
The noncommutative version of (\ref{S_spinor}) is easily constructed,
and we get\begin{equation}
S_{spinor}=\frac{1}{2}\int d^{2}x\Omega\overline{\Psi}i\gamma^{a}(\delta_{e_{a}}-iA_{e_{a}}+m)\star\Psi\end{equation}
with $e_{a}=e_{a}^{\mu}\partial_{\mu}$, which is invariant under
NC gauge transformations and reduces in the commutative limit to (\ref{S_spinor}).

\subsection{\label{example1}Example: A frame for $\kappa$-deformed spacetime}

In this chapter we will construct a frame for the $n$-dimensional
generalization of the $\kappa$-deformed plane studied in chapters
\ref{example: star products} and \ref{example: derivatives}. The
relations of this quantum space%
\footnote{Compared to the two-dimensional example in (\ref{example: star products})
and (\ref{example: derivatives}), the coordinate $x^{0}$ corresponds
to $x$ and the coordinates $x^{i}$ correspond to $y$. A Jambor-Sykora
$\star$-product for the $n$-dimensional $\kappa$-deformed space
reads e.g. \[
f\star g=m\cdot e^{-ia\, x^{i}\dd_{i}\otimes\dd_{0}}f\otimes g.\]
} are\begin{equation}
[\widehat{x}^{0},\widehat{x}^{i}]=ia\widehat{x}^{i}\;\;\;\mbox{for}\;\;\; i\neq0,\end{equation}
 with $a$ a real number. The Poisson structure for this space is
\begin{equation}
c^{\mu\nu}=iax^{i}\delta_{0}^{\mu}\delta_{i}^{\nu}-iax^{i}\delta_{i}^{\mu}\delta_{0}^{\nu}.\end{equation}
The derivative in the $x^{0}$-direction obviously commutes with this
Poisson structure, and we can use it for the frame, setting $e_{0}{}^{\mu}=\delta_{0}^{\mu}$.
For the other directions, we see that $\rho\dd_{i}$ with $\rho=\sqrt{\sum_{i=1}^{n-1}(x^{i})^{2}}$
commutes with the Poisson structure, as we have\begin{equation}
\rho\dd_{i}c^{\mu\nu}=ia\rho\delta_{0}^{\mu}\delta_{i}^{\nu}-ia\rho\delta_{i}^{\mu}\delta_{0}^{\nu}\;\;\;\;\mbox{and}\;\;\;\; c^{\mu\sigma}\dd_{\sigma}(\rho\delta_{i}^{\nu})=ia\rho\delta_{0}^{\mu}\delta_{i}^{\nu},\end{equation}
giving \begin{equation}
\rho\dd_{i}\c^{\mu\nu}-c^{\mu\sigma}\dd_{\sigma}\rho\delta_{i}^{\nu}+c^{\nu\sigma}\dd_{\sigma}\rho\delta_{i}^{\mu}=0.\end{equation}
For the frame, we can therefore take \begin{eqnarray}
e_{o} & = & \partial_{o},\\
e_{i} & = & \rho\partial_{i},\nonumber \end{eqnarray}
 leading to a commutative metric \begin{equation}
g=(dx^{0})^{2}+\rho^{-2}((dx^{1})^{2}+\cdots+(dx^{n-1})^{2}).\end{equation}
 We know that we can write\begin{equation}
(dx^{1})^{2}+\cdots+(dx^{n-1})^{2}=d\rho^{2}+\rho^{2}d\Omega_{n-2}^{2},\end{equation}
 where $d\Omega_{n-2}^{2}$ is the metric of the $n-2$ dimensional
sphere. Therefore in this new coordinate system \begin{equation}
g=(dx^{0})^{2}+(d\ln\rho)^{2}+d\Omega_{n-2}^{2}\end{equation}
 and we see that the commutative space is a cross product of a two
dimensional Euclidean space and a $n-2$-sphere. Therefore it is a
space of constant non-vanishing curvature. Further \begin{equation}
\sqrt{\det g}=\rho^{-(n-1)}\end{equation}
is both the measure function on this curved space and it fulfills
\begin{equation}
\dd_{\mu}(\sqrt{\det g}c^{\mu\nu})=0,\end{equation}
i.e. it also guarantees the cyclicity of the integral, see chapter
\ref{sec: traces}. 

We have found a frame compatible with the Poisson structure of $\kappa$-deformed
spacetime and can therefore construct noncommutative gauge theory
on this space. We will continue this example in chapter \ref{example2},
where we will also have explicit formulas for the SW-maps, writing
down an explicit action for the gauge theory.

\section{Explicit formulas for the Seiberg-Witten map \label{swmap_weyl}}

We will now present a consistent solution for the Seiberg-Witten-maps
up to second order for the Weyl ordered $\star$-product and non-abelian
gauge transformations. The solutions have been chosen in such a way
that they reproduce the ones obtained in \cite{Jurco:2001rq} for
the canonical case.

For calculating the Seiberg-Witten maps, we will write the Weyl ordered
$\star$-product (\ref{wo_product}) expanded to second order as \begin{equation}
f\star g=fg+f\star_{1}g+f\star_{2}g+\mc O(3)\end{equation}
with\begin{equation}
f\star_{1}g=\frac{1}{2}c^{ij}\partial_{i}f\,\partial_{j}g\end{equation}
 and\begin{equation}
f\star_{2}g=\frac{1}{8}c^{mn}c^{ij}\partial_{m}\partial_{i}f\partial_{n}\partial_{j}g+\frac{1}{12}c^{ml}\partial_{l}c^{ij}(\partial_{m}\partial_{i}f\partial_{j}g-\partial_{i}f\partial_{m}\partial_{j}g).\end{equation}

\subsection{The gauge parameter}

The gauge parameter is equally expanded as\begin{equation}
\Lambda_{\alpha}[a]=\Lambda_{\alpha}^{0}[a]+\Lambda_{\alpha}^{1}[a]+\Lambda_{\alpha}^{2}[a]+\mc O(3)\end{equation}
The solution for the gauge transformations is obtained by solving
the consistency condition (\ref{sw_consitency_condition}) order by
order. To zeroth order, we clearly have $\Lambda_{\alpha}^{0}[a]=\alpha$. 

To first order, the consistency condition reads\begin{eqnarray}
i\delta_{\alpha}\Lambda_{\beta}^{1}-i\delta_{\beta}\Lambda_{\alpha}^{1}+[\alpha,\Lambda_{\beta}^{1}]+[\Lambda_{\alpha}^{1},\beta]-i\Lambda_{-i[\alpha,\beta]}^{1} & = & -[\alpha\stackrel{\star_{1}}{,}\beta]\label{cc 1}\\
 & = & -\frac{1}{2}c^{ij}[\partial_{i}\alpha,\partial_{j}\beta]\nn\end{eqnarray}
 A solution to this equation is\begin{equation}
\Lambda_{\alpha}^{1}[a]=-\frac{i}{4}c^{ij}\{\partial_{i}\alpha,a_{j}\}.\label{lambda 1 ord}\end{equation}
Note that this solution is not unique. Especially, we could always
add terms solving the homogeneous part of (\ref{cc 1}). 

To second order, the consistency condition reads\begin{eqnarray}
\lefteqn{i\delta_{\alpha}\Lambda_{\beta}^{2}-i\delta_{\beta}\Lambda_{\alpha}^{2}+[\alpha,\Lambda_{\beta}^{2}]+[\Lambda_{\alpha}^{2},\beta]-i\Lambda_{-i[\alpha,\beta]}^{2}}\\
 & = & -[\alpha\stackrel{\star_{1}}{,}\Lambda_{\beta}^{1}]-[\Lambda_{\alpha}^{1}\stackrel{\star_{1}}{,}\beta]-[\Lambda_{\alpha}^{1},\Lambda_{\beta}^{1}]-[\alpha\stackrel{\star_{2}}{,}\beta]\nonumber \\
 & = & -\frac{1}{2}c^{ij}[\partial_{i}\alpha,\partial_{j}\Lambda_{\beta}^{1}]-\frac{1}{2}c^{ij}[\partial_{i}\Lambda_{\alpha}^{1},\partial_{j}\beta]-[\Lambda_{\alpha}^{1},\Lambda_{\beta}^{1}]\nonumber \\
 &  & -\frac{1}{8}c^{mn}c^{ij}[\partial_{m}\partial_{i}\alpha,\partial_{n}\partial_{j}\beta]-\frac{1}{12}c^{ml}\partial_{l}c^{ij}([\partial_{m}\partial_{i}\alpha,\partial_{j}\beta]-[\partial_{i}\alpha,\partial_{m}\partial_{j}\beta]).\nonumber \end{eqnarray}
Using the first order term (\ref{lambda 1 ord}), we calculate the
second order term\begin{eqnarray}
\Lambda_{\alpha}^{2}[a] & = & +\frac{1}{32}c^{ij}c^{kl}\Big{(}4\{\partial_{i}\alpha,\{ a_{k},\partial_{l}a_{j}\}\}-2i[\partial_{i}\partial_{k}\alpha,\partial_{j}a_{l}]\\
 &  & \;\;\;\;\;\;+2[\partial_{j}a_{l},[\partial_{i}\alpha,a_{k}]]-2i[[a_{j},a_{l}],[\partial_{i}\alpha,a_{k}]]\nonumber \\
 &  & \;\;\;\;\;\;+i\{\partial_{i}\alpha,\{ a_{k},[a_{j},a_{l}]\}\}+\{ a_{j},\{ a_{l},[\partial_{i}\alpha,a_{k}]\}\}\Big{)}\nonumber \\
 &  & +\frac{1}{24}c^{kl}\partial_{l}c^{ij}\Big{(}\{\partial_{i}\alpha,\{ a_{k},a_{j}\}\}-2i[\partial_{i}\partial_{k}\alpha,a_{j}]\Big{)}.\nonumber \end{eqnarray}

\subsection{Fields in the fundamental representation}

In the same way a solution for the field $\Psi$ in the fundamental
representation is obtained by solving equation (\ref{sp_gt_field}).
We expand it to second order as \begin{equation}
\Psi_{\psi}[a]=\Psi_{\psi}^{0}[a]+\Psi_{\psi}^{1}[a]+\Psi_{\psi}^{2}[a]+\mc O(3).\end{equation}
The zeroth order is the commutative field, i. e. $\Psi_{\psi}^{0}[a]=\psi$.
To first order, the equation (\ref{sp_gt_field}) reads \begin{equation}
\delta_{\alpha}\Psi_{\psi}^{1}-i\alpha\Psi_{\psi}^{1}=i\alpha\star_{1}\psi+i\Lambda_{\alpha}^{1}\psi=\frac{i}{2}c^{ij}\partial_{i}\alpha\partial_{j}\psi+i\Lambda_{\alpha}^{1}\psi,\end{equation}
 which is solved using (\ref{lambda 1 ord}) to give \begin{equation}
\Psi_{\psi}^{1}[a]=\frac{1}{4}c^{ij}\Big{(}2ia_{i}\partial_{j}\psi+a_{i}a_{j}\psi\Big{)}.\label{psi 1 ord}\end{equation}
To second order the equation (\ref{sp_gt_field}) reads\begin{eqnarray}
\delta_{\alpha}\Psi_{\psi}^{2}-i\alpha\Psi_{\psi}^{2} & = & i\alpha\star_{2}\psi+i\alpha\star_{1}\Psi_{\psi}^{1}+i\Lambda_{\alpha}^{1}\star_{1}\psi+i\Lambda_{\alpha}^{1}\Psi_{\psi}^{1}+i\Lambda_{\alpha}^{2}\psi\\
 & = & \frac{i}{8}c^{mn}c^{ij}\partial_{m}\partial_{i}\alpha,\partial_{n}\partial_{j}\psi+\frac{i}{12}c^{ml}\partial_{l}c^{ij}(\partial_{m}\partial_{i}\alpha\partial_{j}\psi-\partial_{i}\alpha\partial_{m}\partial_{j}\psi)\nonumber \\
 &  & +\frac{i}{2}c^{ij}\partial_{i}\alpha\partial_{j}\Psi_{\psi}^{1}+\frac{i}{2}c^{ij}\partial_{i}\Lambda_{\alpha}^{1}\partial_{j}\psi+i\Lambda_{\alpha}^{1}\Psi_{\psi}^{1}+i\Lambda_{\alpha}^{2}\psi.\nonumber \end{eqnarray}
Using the solutions to first order (\ref{lambda 1 ord}) and (\ref{psi 1 ord}),
a solution \begin{eqnarray}
\Psi_{\psi}^{2}[a] & = & +\frac{1}{32}c^{ij}c^{kl}\Big{(}4i\partial_{i}a_{k}\partial_{j}\partial_{l}\psi-4a_{i}a_{k}\partial_{j}\partial_{l}\psi-8a_{i}\partial_{j}a_{k}\partial_{l}\psi\\
 &  & \;\;\;\;\;\;\;\;\;\;\;\;\;\;\;\;+4a_{i}\partial_{k}a_{j}\partial_{l}\psi+4ia_{i}a_{j}a_{k}\partial_{l}\psi-4ia_{k}a_{j}a_{i}\partial_{l}\psi\nonumber \\
 &  & \;\;\;\;\;\;\;\;\;\;\;\;\;\;\;\;+4ia_{j}a_{k}a_{i}\partial_{l}\psi-4\partial_{j}a_{k}a_{i}\partial_{l}\psi+2\partial_{i}a_{k}\partial_{j}a_{l}\psi\nonumber \\
 &  & \;\;\;\;\;\;\;\;\;\;\;\;\;\;\;\;-4ia_{i}a_{l}\partial_{k}a_{j}\psi-4ia_{i}\partial_{k}a_{j}a_{l}\psi+4ia_{i}\partial_{j}a_{k}a_{l}\psi\nonumber \\
 &  & \;\;\;\;\;\;\;\;\;\;\;\;\;\;\;\;-3a_{i}a_{j}a_{l}a_{k}\psi-4a_{i}a_{k}a_{j}a_{l}\psi-2a_{i}a_{l}a_{k}a_{j}\psi\Big{)}\nonumber \\
 &  & +\frac{1}{24}c^{kl}\partial_{l}c^{ij}\Big{(}2ia_{j}\partial_{k}\partial_{i}\psi+2i\partial_{k}a_{i}\partial_{j}\psi+2\partial_{k}a_{i}a_{j}\psi\nonumber \\
 &  & \;\;\;\;\;\;\;\;\;\;\;\;\;\;\;\;\;\;\;\;-a_{k}a_{i}\partial_{j}\psi-3a_{i}a_{k}\partial_{j}\psi-2ia_{j}a_{k}a_{i}\psi\Big{)}\nonumber \end{eqnarray}
can be calculated for the second order term.

\subsection{The covariantizer}

The covariantizer is expanded as well as

\begin{equation}
D[a](f)=D^{0}[a](f)+D^{1}[a](f)+D^{2}[a](f)+\mc O(3).\end{equation}
 We will now solve (\ref{sp_gt_cov}) order to order. To zeroth order,
we take $D$ to be the identity, i.e. $D^{0}[a](f)=f$. To first order
(\ref{sp_gt_cov}) reads \begin{equation}
\delta_{\alpha}D^{1}(f)-i[\alpha,D^{1}(f)]=i[\alpha\stackrel{\star_{1}}{,}f]=\frac{i}{2}c^{ij}[\partial_{i}\alpha,\partial_{j}f],\end{equation}
having a solution\begin{equation}
D^{1}[a](f)=ic^{ij}a_{i}\partial_{j}f.\end{equation}
 To second order we get for (\ref{sp_gt_cov})\begin{eqnarray}
\delta_{\alpha}D^{2}(f)-i[\alpha,D^{2}(f)] & = & i[\alpha\stackrel{\star_{2}}{,}f]+i[\alpha\stackrel{\star_{1}}{,}D^{1}(f)]\\
 &  & +i[\Lambda_{\alpha}^{1}\stackrel{\star_{1}}{,}f]+i[\Lambda_{\alpha}^{1},D^{1}(f)]\nonumber \\
 & = & \frac{i}{8}c^{mn}c^{ij}[\partial_{m}\partial_{i}\alpha,\partial_{n}\partial_{j}f]\nonumber \\
 &  & +\frac{i}{12}c^{ml}\partial_{l}c^{ij}([\partial_{m}\partial_{i}\alpha,\partial_{j}f]-[\partial_{i}\alpha,\partial_{m}\partial_{j}f])\nonumber \\
 &  & +\frac{i}{2}c^{ij}[\partial_{i}\alpha,\partial_{j}D^{1}(f)]+\frac{i}{2}c^{ij}[\partial_{i}\Lambda_{\alpha}^{1},\partial_{j}f]+i[\Lambda_{\alpha}^{1},D^{1}(f)],\nonumber \end{eqnarray}
with a solution\begin{eqnarray}
D^{2}[a](f) & = & +\frac{1}{4}c^{ij}c^{kl}\Big{(}-2\{ a_{i},\partial_{j}a_{k}\}\partial_{l}f+\{ a_{i},\partial_{k}a_{j}\}\partial_{l}f\\
 &  & \;\;\;\;\;\;\;\;\;\;\,\;\;\;\;\;+i\{ a_{i},[a_{j},a_{k}]\}\partial_{l}f-\{ a_{i},a_{k}\}\partial_{j}\partial_{l}f\Big{)}\nonumber \\
 &  & +\frac{1}{4}c^{il}\partial_{l}c^{jk}\{ a_{i},a_{k}\}\partial_{j}f.\nonumber \end{eqnarray}

\subsection{The gauge field}

Finally we want to calculate a SW-map for the gauge potential $A_{X}$
evaluated on a Poisson vector field $X$. Again we expand it as \begin{equation}
A_{X}[a]=A_{X}^{0}[a]+A_{X}^{1}[a]+A_{X}^{2}[a]+\mc O(3).\end{equation}
 Expanding the equation (\ref{sp_gt_gauge_pot}) as well, we see that
to zeroth order it is solved by $A_{X}^{0}[a]=X^{n}a_{n}$. To first
order it reads\begin{eqnarray}
\delta_{\alpha}A_{X}^{1}-i[\alpha,A_{X}^{1}] & = & X^{i}\partial_{i}\Lambda_{\alpha}^{1}+\delta_{X}^{1}\alpha+i[\alpha\stackrel{\star_{1}}{,}X^{n}a_{n}]+i[\Lambda_{\alpha}^{1},X^{n}a_{n}]\\
 & = & X^{i}\partial_{i}\Lambda_{\alpha}^{1}+\frac{i}{2}c^{ij}[\partial_{i}\alpha,\partial_{j}(X^{n}a_{n})]+i[\Lambda_{\alpha}^{1},X^{n}a_{n}].\nonumber \end{eqnarray}
 Using $[X,c]_{S}=0$, we can calculate a solution\begin{equation}
A_{X}^{1}[a]=\frac{i}{4}c^{kl}X^{n}\{ a_{k},\partial_{l}a_{n}+f_{ln}\}+\frac{i}{4}c^{kl}\partial_{l}X^{n}\{ a_{k},a_{n}\},\end{equation}
where $f_{ij}=\dd_{i}a_{j}-\dd_{j}a_{i}-i[a_{i},a_{j}]$ is the commutative
field strength. 

The equation to second order is\begin{eqnarray}
\delta_{\alpha}A_{X}^{2}-i[\alpha,A_{X}^{2}] & = & X^{i}\partial_{i}\Lambda_{\alpha}^{2}+\delta_{X}^{1}\Lambda_{\alpha}^{1}+\delta_{X}^{2}\alpha+i[\Lambda_{\alpha}^{1},A_{X}^{1}]+i[\Lambda_{\alpha}^{2},X^{n}a_{n}]\\
 &  & +i[\alpha\stackrel{\star_{2}}{,}X^{n}a_{n}]+i[\alpha\stackrel{\star_{1}}{,}A_{X}^{1}]+i[\Lambda_{\alpha}^{1}\stackrel{\star_{1}}{,}X^{n}a_{n}]\nonumber \\
 & = & X^{i}\partial_{i}\Lambda_{\alpha}^{2}+\delta_{X}^{1}\Lambda_{\alpha}^{1}+i[\Lambda_{\alpha}^{1},A_{X}^{1}]+i[\Lambda_{\alpha}^{2},X^{n}a_{n}].\nonumber \\
 &  & +\frac{i}{8}c^{mn}c^{ij}[\partial_{m}\partial_{i}\alpha,\partial_{n}\partial_{j}(X^{n}a_{n})]\nonumber \\
 &  & +\frac{i}{12}c^{ml}\partial_{l}c^{ij}([\partial_{m}\partial_{i}\alpha,\partial_{j}(X^{n}a_{n})]-[\partial_{i}\alpha,\partial_{m}\partial_{j}(X^{n}a_{n})])\nonumber \\
 &  & +\frac{i}{2}c^{ij}[\partial_{i}\alpha,\partial_{j}A_{X}^{1}]+\frac{i}{2}c^{ij}[\partial_{i}\Lambda_{\alpha}^{1},\partial_{j}(X^{n}a_{n})]\nonumber \\
 &  & -\frac{1}{12}c^{lk}\partial_{k}c^{im}\partial_{l}\partial_{m}X^{j}\partial_{i}\partial_{j}\alpha+\frac{1}{24}c^{lk}c^{im}\partial_{l}\partial_{i}X^{j}\partial_{k}\partial_{m}\partial_{j}\alpha\nonumber \end{eqnarray}

The second order solution for the NC gauge potential is \begin{eqnarray}
A_{X}^{2}[a] & = & +\frac{1}{32}c^{kl}c^{ij}X^{n}\Big{(}-4i[\partial_{k}\partial_{i}a_{n},\partial_{l}a_{j}]+2i[\partial_{k}\partial_{n}a_{i},\partial_{l}a_{j}]\\
 &  & \;\;\;\;\;\;\;\;\;-4\{ a_{k},\{ a_{i},\partial_{j}f_{ln}\}\}-2[[\partial_{k}a_{i},a_{n}],\partial_{l}a_{j}]+4\{\partial_{l}a_{n},\{\partial_{i}a_{k},a_{j}\}\}\nonumber \\
 &  & \;\;\;\;\;\;\;\;\;-4\{ a_{k},\{ f_{li},f_{jn}\}\}+i\{\partial_{n}a_{j},\{ a_{l},[a_{i},a_{k}]\}\}\nn\\
 &  & \;\;\;\;\;\;\;\;\;+i\{ a_{i},\{ a_{k},[\partial_{n}a_{j},a_{l}]\}\}-4i[[a_{i},a_{l}],[a_{k},\partial_{j}a_{n}]]\nonumber \\
 &  & \;\;\;\;\;\;\;\;\;+2i[[a_{i},a_{l}],[a_{k},\partial_{n}a_{j}]]+\{ a_{i},\{ a_{k},[a_{l},[a_{j},a_{n}]]\}\}\nonumber \\
 &  & \;\;\;\;\;\;\;\;\;-\{ a_{k},\{[a_{l},a_{i}],[a_{j},a_{n}]\}\}-[[a_{i},a_{l}],[a_{k},[a_{j},a_{n}]]]\Big{)}\nonumber \\
 &  & +\frac{1}{32}c^{kl}c^{ij}\partial_{j}X^{n}\Big{(}2i[\partial_{k}a_{i},\partial_{l}a_{n}]+2i[\partial_{i}a_{k},\partial_{l}a_{n}]\nonumber \\
 &  & \;\;\;\;\;\;\;\;\;+2i[\partial_{i}a_{k},\partial_{l}a_{n}-\partial_{n}a_{l}]+4\{ a_{n},\{ a_{l},\partial_{k}a_{i}\}\}\nonumber \\
 &  & \;\;\;\;\;\;\;\;\;+4\{ a_{k},\{ a_{i},\partial_{n}a_{l}-\partial_{l}a_{n}\}\}-2i\{ a_{k},\{ a_{i},[a_{n},a_{l}]\}\}\nonumber \\
 &  & \;\;\;\;\;\;\;\;\;+i\{ a_{i},\{ a_{l},[a_{n},a_{k}]\}\}+i\{ a_{n},\{ a_{l},[a_{i},a_{k}]\}\}\Big{)}\nonumber \\
 &  & +\frac{1}{24}c^{kl}c^{ij}\partial_{l}\partial_{j}X^{n}\Big{(}\partial_{i}\partial_{k}a_{n}-2i[a_{i},\partial_{k}a_{n}]-\{ a_{n},\{ a_{k},a_{i}\}\}\Big{)}\nonumber \\
 &  & +\frac{1}{24}c^{kl}\partial_{l}c^{ij}X^{n}\Big{(}2i[a_{j},\partial_{k}\partial_{i}a_{n}]+2i[\partial_{k}a_{i},f_{jn}]\nonumber \\
 &  & \;\;\;\;\;\;\;\;\;-\{\partial_{j}a_{n},\{ a_{k},a_{i}\}\}+2\{ a_{i},\{ a_{k},f_{nj}\}\}\Big{)}\nonumber \\
 &  & +\frac{1}{24}c^{kl}\partial_{l}c^{ij}\partial_{j}X^{n}\Big{(}-4i[a_{i},\partial_{k}a_{n}]+2i[a_{k},\partial_{i}a_{n}]-\{ a_{n},\{ a_{k},a_{i}\}\}\Big{)}\nonumber \\
 &  & -\frac{1}{12}c^{kl}\partial_{l}c^{ij}\partial_{j}\partial_{k}X^{n}\partial_{i}a_{n}+\mathcal{O}(3).\nonumber \end{eqnarray}

\subsection{Field strength, covariant derivative and action}

We will now use the Seiberg-Witten maps of the preceding chapters
to calculate actions for noncommutative gauge theory to first order.
We start with calculating the field strength (\ref{xy_field_strength}).
It is\begin{eqnarray}
F_{ab}=F(X_{a},X_{b}) & = & [D_{X_{a}}\stackrel{\star}{,}D_{X_{b}}]-D_{[X_{a},X_{b}]_{\star}}\\
 & = & X_{a}^{k}X_{b}^{l}f_{kl}+\frac{i}{2}c^{ij}\{ a_{i},\partial_{j}(X_{a}^{k}X_{b}^{l}f_{kl})\}\nn\\
 &  & +\frac{i}{2}c^{ij}X_{a}^{k}X_{b}^{l}\{ f_{jl},f_{ki}\}+\frac{1}{4}c^{ij}X_{a}^{k}X_{b}^{l}\{ a_{i},[a_{j},f_{kl}]\}+\mathcal{O}(2).\nonumber \end{eqnarray}
 The covariant derivative is\begin{eqnarray}
D_{a}\Phi=D_{X_{a}}\Phi & = & \delta_{X_{a}}\Phi-iA_{X_{a}}\star\Phi\\
 & = & X_{a}^{k}D_{k}\phi+\frac{i}{2}X_{a}^{k}f_{ki}c^{ij}D_{j}\phi\nonumber \\
 &  & +\frac{i}{2}c^{ij}a_{i}\partial_{j}(X_{a}^{k}D_{k}\phi)+\frac{1}{4}c^{ij}a_{i}a_{j}X_{a}^{k}D_{k}\phi+\mathcal{O}(2).\nonumber \end{eqnarray}
 Using partial integration and the trace property of the integral,
i.e. $\partial_{\mu}(\Omega c^{\mu\nu})=0$, we can calculate

\begin{eqnarray}
\widetilde{S}_{gauge} & = & \int d^{n}x\,\Omega\,\eta^{ab}\eta^{cd}F_{ac}\star F_{bd}\\
 & = & \int d^{n}x\,\Omega\,\eta^{ab}\eta^{cd}X_{a}^{\mu}X_{c}^{\nu}X_{b}^{\rho}X_{d}^{\sigma}f_{\mu\nu}f_{\rho\sigma}\nonumber \\
 &  & +\int d^{4}x\,\Omega\,\eta^{ab}\eta^{cd}\Big{(}\frac{i}{4}c^{ij}[a_{i},[\partial_{j}(X_{a}^{\mu}X_{c}^{\nu}f_{\mu\nu}),X_{b}^{\rho}X_{d}^{\sigma}f_{\rho\sigma}]]\nonumber \\
 &  & +\frac{i}{8}c^{ij}X_{a}^{\mu}X_{c}^{\nu}X_{b}^{\rho}X_{d}^{\sigma}\{ f_{\mu\nu},\{ f_{ij},f_{\rho\sigma}\}\}-\frac{1}{4}c^{ij}X_{a}^{\mu}X_{c}^{\nu}X_{b}^{\rho}X_{d}^{\sigma}[f_{\mu\nu}a_{i},f_{\rho\sigma}a_{j}]\nonumber \\
 &  & -\frac{1}{4}c^{ij}X_{a}^{\mu}X_{c}^{\nu}X_{b}^{\rho}X_{d}^{\sigma}[a_{i}f_{\mu\nu},a_{j}f_{\rho\sigma}]+\frac{i}{2}c^{ij}X_{a}^{\mu}X_{c}^{\nu}X_{b}^{\rho}X_{d}^{\sigma}\{ f_{\mu\nu},\{ f_{j\sigma},f_{\rho i}\}\}\Big{)}\nonumber \\
 &  & +\mathcal{O}(2),\nn\end{eqnarray}
 where we haven't done the trace over the gauge representation jet.
Doing this now, the action for the gauge particles is\begin{eqnarray}
S_{gauge} & = & -\frac{1}{4}tr(\widetilde{S}_{gauge})\\
 & = & \int d^{n}x\,\Omega\,\eta^{ab}\eta^{cd}X_{a}^{\mu}X_{c}^{\nu}X_{b}^{\rho}X_{d}^{\sigma}\Big{(}-\frac{1}{4}tr(f_{\mu\nu}f_{\rho\sigma})\nonumber \\
 &  & -\frac{i}{8}c^{ij}tr(f_{ij}f_{\mu\nu}f_{\rho\sigma})-\frac{i}{2}c^{ij}tr(f_{\mu\nu}f_{j\sigma}f_{\rho i})\Big{)}+\mathcal{O}(2).\nonumber \end{eqnarray}
 With $\overline{c}^{ij}=-c^{ij}$ we get 

\begin{eqnarray}
S_{scalar} & = & \int d^{n}x\,\Omega\,\eta^{ab}\overline{D_{a}\Phi}\star D_{b}\Phi\\
 & = & \int d^{n}x\,\Omega\,\eta^{ab}\Big{(}\overline{X}_{a}^{\mu}X_{b}^{\nu}\overline{D_{\mu}\phi}D_{\nu}\phi\nonumber \\
 &  & +\frac{i}{2}c^{ij}\overline{X}_{a}^{\mu}X_{b}^{\nu}\overline{D_{\mu}\phi}f_{\nu i}D_{j}\phi+\frac{i}{2}c^{ij}\overline{X}_{a}^{\mu}X_{b}^{\nu}\overline{D_{j}\phi}f_{\nu i}D_{\mu}\phi\nonumber \\
 &  & +\frac{i}{2}c^{ij}\overline{X}_{a}^{\mu}X_{b}^{\nu}\overline{D_{\mu}\phi}f_{ij}D_{\nu}\phi\Big{)}+\mathcal{O}(2)\nonumber \end{eqnarray}
 for the scalar fields.

\subsection{\label{example2}Example: A NC action on $\kappa$-deformed spacetime}

Now we continue our example from chapter \ref{example1}. There, we
had already constructed a frame \begin{eqnarray}
e_{0}^{\mu} & = & \delta_{0}^{\mu},\label{eq:frame0}\\
e_{i}^{\mu} & = & \rho\delta_{i}^{\mu}\label{eq:framei}\end{eqnarray}
 with $\rho=\sqrt{x_{i}x^{i}}$ compatible with the Poisson structure\begin{equation}
c^{\mu\nu}=ia\delta_{0}^{\mu}\delta_{i}^{\nu}x^{i}-ia\delta_{0}^{\nu}\delta_{i}^{\mu}x^{i}.\end{equation}
 These we can plug into our solution of the Seiberg-Witten map and
get\begin{eqnarray}
\Lambda_{\lambda}[a] & = & \lambda+\frac{a}{4}x^{i}\{\partial_{0}\lambda,a_{i}\}-\frac{a}{4}x^{i}\{\partial_{i}\lambda,a_{0}\}+\mathcal{O}(a^{2}),\nonumber \\
\Phi_{\phi}[a] & = & \phi-\frac{a}{2}x^{i}a_{0}\partial_{i}\phi+\frac{a}{2}x^{i}a_{i}\partial_{0}\phi+\frac{ia}{4}x^{i}[a_{0},a_{i}]\phi+\mathcal{O}(a^{2}),\nonumber \\
A_{X_{0}} & = & a_{0}-\frac{a}{4}x^{i}\{ a_{0},\partial_{i}a_{0}+f_{i0}\}+\frac{a}{4}x^{i}\{ a_{i},\partial_{0}a_{0}\}+\mathcal{O}(a^{2}),\\
A_{X_{j}} & = & \rho a_{j}-\frac{a}{4}\rho\{ a_{j},a_{0}\}-\frac{a}{4}\rho x^{i}\{ a_{0},\partial_{i}a_{j}+f_{ij}\}\nonumber \\
 &  & +\frac{a}{4}\rho x^{i}\{ a_{i},\partial_{0}a_{j}+f_{0j}\}+\mathcal{O}(a^{2}),\nn\\
\delta_{X_{\mu}} & = & X_{\mu}^{\nu}\partial_{\nu}+\mathcal{O}(a^{2}).\nonumber \end{eqnarray}
 The measure function induced by the frame (\ref{eq:frame0},\ref{eq:framei})
was\begin{equation}
\Omega=\rho^{-(n-1)}=\sqrt{g},\end{equation}
 also guaranteeing the cyclicity of the integral. With this measure
function the actions become\begin{eqnarray}
S_{gauge} & = & -\frac{1}{2}\int d^{n}x\,\rho^{3-n}\eta^{00}\eta^{ij}Tr(f_{0i}f_{0j})\\
 &  & -\frac{1}{4}\int d^{4}x\,\rho^{5-n}\eta^{kl}\eta^{ij}Tr(f_{ki}f_{lj})\nn\\
 &  & -\frac{a}{2}\int d^{n}x\,\rho^{3-n}\eta^{00}\eta^{ij}x^{p}Tr(f_{0p}f_{0i}f_{0j})\nonumber \\
 &  & +\frac{a}{4}\int d^{4}x\,\rho^{5-n}\eta^{kl}\eta^{ij}x^{p}Tr(f_{0p}f_{ki}f_{lj})\nn\\
 &  & -\frac{a}{2}\int d^{n}x\,\rho^{5-n}\eta^{kl}\eta^{ij}x^{p}Tr(f_{jp}\{ f_{ki},f_{l0}\})+\mathcal{O}(a^{2})\nonumber \end{eqnarray}
 and\begin{eqnarray}
S_{scalar} & = & \int d^{n}x\,\rho^{1-n}\eta^{00}\overline{D_{0}\phi}D_{0}\phi+\int d^{n}x\,\rho^{3-n}\eta^{kl}\overline{D_{k}\phi}D_{l}\phi\\
 &  & -\frac{a}{2}\int d^{n}x\,\rho^{3-n}\eta^{kl}x^{i}\overline{D_{k}\phi}f_{l0}D_{i}\phi+\frac{a}{2}\int d^{n}x\,\rho^{3-n}\eta^{kl}x^{i}\overline{D_{k}\phi}f_{li}D_{0}\phi\nonumber \\
 &  & -\frac{a}{2}\int d^{n}x\,\rho^{3-n}\eta^{kl}x^{i}\overline{D_{i}\phi}f_{l0}D_{k}\phi+\frac{a}{2}\int d^{n}x\,\rho^{3-n}\eta^{kl}x^{i}\overline{D_{0}\phi}f_{li}D_{k}\phi\nonumber \\
 &  & -a\int d^{n}x\,\rho^{3-n}\eta^{kl}x^{i}\overline{D_{k}\phi}f_{0i}D_{l}\phi+\mathcal{O}(a^{2}).\nonumber \end{eqnarray}
 In the commutative limit $a\rightarrow0$ the action reduces to scalar
electrodynamics on a manifold with constant curvature.

\section{Construction of the Seiberg-Witten maps to all orders\label{swmap_form}}

For explicit calculations, the Weyl ordered $\star$-product is the
best choice, but it is only known to second order. For calculations
to all orders, we can use the formality $\star$-product, which also
comes with strong mathematical tools we can use for the construction
of the Seiberg-Witten maps. We already saw how to construct derivations
for the formality $\star$-product in chapter (\ref{formality derivations}).
We can use them to formulate NC gauge theory on any Poisson manifold.
To relate the NC theory to commutative gauge theory, we need the Seiberg-Witten
maps for the formality $\star$-product. In \cite{Jurco:2000fs} and
\cite{Jurco:2001my} the SW maps for the NC gauge parameter and the
covariantizer were already constructed to all orders in $\theta$.
We will extend the method developed there to the SW map for covariant
derivations.

\subsection{Formality}

We saw in chapter \ref{Formality map} that the formality $\star$-product
can be constructed using the maps $U_{n}$ from the polyvectorfields
to the polydifferential operators as\begin{equation}
f\star g=\sum_{n=0}^{\infty}\frac{1}{n!}U_{n}(\pi,\,\ldots,\pi)(f,g).\end{equation}
With these maps, we already introduced the special polydifferential
operators

\begin{eqnarray}
\Phi(\alpha) & = & \sum_{n=1}^{\infty}\frac{1}{(n-1)!}U_{n}(\alpha,\pi,\,\ldots,\pi),\\
\Psi(\alpha_{1},\alpha_{2}) & = & \sum_{n=2}^{\infty}\frac{1}{(n-2)!}U_{n}(\alpha_{1},\alpha_{2},\pi,\ldots,\pi)\end{eqnarray}
in chapter \ref{formality derivations}. For the construction of the
Seiberg-Witten maps, we will need some additional relations, which
we calculate using the formality condition (\ref{formality cond}). 

For $g$ a function, $X$ and $Y$ vectorfields and $\pi$ and $\sigma$
bivectorfields we see that both $\Phi(g)$ and $\Psi(X,Y)$ are functions
and we go on to calculate\begin{eqnarray}
{[\star,\star]_{G}} & = & \Phi([\pi,\pi]_{S}),\label{comm star star}\\
{[\Phi(f),\star]_{G}} & = & -\Phi([f,\pi]_{S}),\label{comm phi f star}\\
{[\delta_{X},\star]_{G}} & = & \Phi([X,\pi]_{S}),\label{comm delta X star}\\
{[\delta_{X},\delta_{Y}]_{G}+[\Psi(X,Y),\star]_{G}} & = & \delta_{[X,Y]_{S}}\label{comm delta X delta Y}\\
 &  & +\Psi([\pi,Y]_{S},X)-\Psi([\pi,X]_{S},Y)\nn,\\
{}[\Phi(\sigma),\Phi(g)]_{G}+[\Psi(\sigma,g),\star]_{G} & = & -\delta_{[\sigma,g]_{S}}\\
 &  & -\Psi([\pi,g]_{S},\sigma)-\Psi([\pi,\sigma]_{S},g)\nn,\\
{}[\delta_{X},\Phi(g)]_{G} & = & \phi([X,g]_{S})\label{comm delta X Phi g}\\
 &  & -\Psi([\pi,g]_{S},X)-\Psi([\pi,X]_{S},g)\nn.\end{eqnarray}
 If $\pi$ is a Poisson tensor, i. e. $[\pi,\pi]_{S}=0$ and if $X$
and $Y$ are Poisson vector fields, i. e. $[X,\pi]_{S}=[Y,\pi]_{S}=0$,
the relations (\ref{comm star star}) to (\ref{comm delta X delta Y})
become \begin{eqnarray}
f\star(g\star h) & = & (f\star g)\star h,\\
\delta_{H_{f}}(g) & = & -[\Phi(f)\stackrel{\star}{,}g],\\
\delta_{X}(f\star g) & = & \delta_{X}(f)\star g+f\star\delta_{X}(g),\label{formality star derivation}\\
({[\delta_{X},\delta_{Y}]-\delta_{[X,Y]_{L}}})(g) & = & [\Psi(X,Y)\stackrel{\star}{,}g]\end{eqnarray}
 when evaluated on functions. $[\,\cdot\,,\,\cdot\,]$ are now ordinary
commutator brackets. $\star$ defines an associative product, the
Hamiltonian vector fields are mapped to inner derivations and Poisson
vector fields are mapped to outer derivations of the $\star$-product.

\subsection{Semi-classical construction}

We will first do the construction in the semi-classical limit, where
the star commutator is replaced by the Poisson bracket. As in \cite{Jurco:2000fs}
and \cite{Jurco:2001my}, we define, with the help of the Poisson
tensor $\theta=\frac{1}{2}\theta^{kl}\partial_{k}\wedge\partial_{l}$\begin{equation}
d_{\theta}=-[\cdot,\theta]\end{equation}
 and (locally) \begin{equation}
a_{\theta}=\theta^{ij}a_{j}\partial_{i}.\end{equation}
 Note that the bracket used in the definition of $d_{\theta}$ is
not the Schouten-Nijenhuis bracket (\ref{Schouten-Nijenhuis bracket}).
For polyvectorfields $\pi_{1}$ and $\pi_{2}$ it is \begin{equation}
{}[\pi_{1},\pi_{2}]=-[\pi_{2},\pi_{1}]_{S},\end{equation}
 giving an extra minus sign for $\pi_{1}$ and $\pi_{2}$ both even
(see \ref{Quantum commutators}). Especially, we get for $d_{\theta}$
acting on a function $g$

\begin{equation}
d_{\theta}g=-[g,\theta]=[g,\theta]_{S}=\theta^{kl}\partial_{l}g\partial_{k}.\end{equation}
 Now a parameter $t$ and $t$-dependent $\theta_{t}=\frac{1}{2}\theta_{t}^{kl}\partial_{k}\wedge\partial_{l}$
and $X_{t}=X_{t}^{k}\partial_{k}$ are introduced, fulfilling

\begin{equation}
\partial_{t}\theta_{t}=f_{\theta}=-\theta_{t}f\theta_{t}\;\;\;\;\mbox{and}\;\;\;\;\partial_{t}X_{t}=-X_{t}f\theta_{t},\end{equation}
 where the multiplication is understood as ordinary matrix multiplication,
e.g. $(\theta_{t}f\theta_{t})^{ij}=\theta_{t}^{ik}f_{kl}\theta_{t}^{kj}$.
Given the Poisson tensor $\theta$ and the Poisson vectorfield $X$,
the formal solutions are\begin{equation}
\theta_{t}=\theta\sum_{n=0}^{\infty}(-t\; f\theta)^{n}=\frac{1}{2}(\theta^{kl}-t\theta^{ki}f_{ij}\theta^{jl}+\ldots)\partial_{k}\wedge\partial_{l}\end{equation}
 and

\begin{equation}
X_{t}=X\sum_{n=0}^{\infty}(-t\; f\theta)^{n}=X^{k}\partial_{k}-tX^{i}f_{ij}\theta^{jk}\partial_{k}+\ldots.\end{equation}
 $\theta_{t}$ is still a Poisson tensor and $X_{t}$ is still a Poisson
vectorfield, i.e.

\begin{equation}
[\theta_{t},\theta_{t}]=0\;\;\;\;\;\;\;\mbox{and}\;\;\;\;\;\;\;\;[X_{t},\theta_{t}]=0.\end{equation}
 For the proof see \ref{Commutator theta theta and theta X}.

With this we calculate\begin{equation}
f_{\theta}=\partial_{t}\theta_{t}=-\theta_{t}f\theta_{t}=-[a_{\theta},\theta]=d_{\theta}a_{\theta}.\label{f theta}\end{equation}
 We now get the following commutation relations

\begin{eqnarray}
{[}a_{\theta_{t}}+\partial_{t},d_{\theta_{t}}(g)] & = & d_{\theta_{t}}((a_{\theta_{t}}+\partial_{t})(g)),\label{comm rel 1}\\
{[}a_{\theta_{t}}+\partial_{t},X_{t}] & = & -d_{\theta_{t}}(X_{t}^{k}a_{k}),\label{comm rel}\end{eqnarray}
 where $g$ is some function which might also depend on $t$ (see
\ref{Semi-classical commutators}).

To construct the Seiberg-Witten map for the gauge potential $A_{X}$,
we first define

\begin{equation}
K_{t}=\sum_{n=0}^{\infty}\frac{1}{(n+1)!}(a_{\theta_{t}}+\partial_{t})^{n}.\end{equation}
 With this, the semi-classical gauge parameter reads \cite{Jurco:2000fs,Jurco:2001my}\begin{equation}
\Lambda_{\lambda}[a]=K_{t}(\lambda)\Big{|}_{t=0}.\end{equation}
 To see that this has indeed the right transformation properties under
gauge transformations, we first note that the transformation properties
of $a_{\theta_{t}}$ and $X_{t}^{k}a_{k}$ are\begin{equation}
\delta_{\lambda}a_{\theta_{t}}=\theta_{t}^{kl}\partial_{l}\lambda\partial_{k}=d_{\theta_{t}}\lambda\label{delta a theta}\end{equation}
 and\begin{equation}
\delta_{\lambda}(X_{t}^{k}a_{k})=X_{t}^{k}\partial_{k}\lambda=[X_{t},\lambda].\label{delta X}\end{equation}
 Using (\ref{delta a theta},\ref{delta X}) and the commutation relations
(\ref{comm rel 1},\ref{comm rel}), a rather tedious calculation
(see \ref{Transformation properties of K}) shows that \begin{equation}
\delta_{\lambda}K_{t}(X_{t}^{k}a_{k})=X_{t}^{k}\partial_{k}K_{t}(\lambda)+d_{\theta_{t}}(K_{t}(\lambda))K_{t}(X_{t}^{k}a_{k}).\end{equation}
 Therefore, the semi-classical gauge potential is

\begin{equation}
A_{X}[a]=K_{t}(X_{t}^{k}a_{k})\Big{|}_{t=0}.\end{equation}

\subsection{Quantum construction}

We can now use the Kontsevich formality map to quantize the semi-classical
construction. All the semi-classical expressions can be mapped to
their counterparts in the $\star$-product formalism without loosing
the properties necessary for the construction. One higher order term
will appear, fixing the transformation properties for the quantum
objects.

The $\star$-product we will use is \begin{equation}
\star=\sum_{n=0}^{\infty}\frac{1}{n!}U_{n}(\theta_{t},\,\ldots,\theta_{t}).\end{equation}
 We define

\begin{equation}
d_{\star}=-[\cdot,\star]_{G\:\:},\end{equation}
 which for functions $f$ and $g$ reads

\begin{equation}
d_{\star}(g)\: f=[f\stackrel{\star}{,}g].\end{equation}
 The bracket used in the definition of $d_{\star}$ is the Gerstenhaber
bracket (\ref{Gerstenhaber bracket}). We now calculate the commutators
(\ref{comm rel 1}) and (\ref{comm rel}) in the new setting (see
\ref{Quantum commutators}). We get

\begin{eqnarray}
{[}\Phi(a_{\theta_{t}})+\partial_{t},d_{\star}(\Phi(g))] & = & d_{\star}((\Phi(a_{\theta_{t}})+\partial_{t})\Phi(g)),\\
{[}\Phi(a_{\theta_{t}})+\partial_{t},\Phi(X_{t})] & = & -d_{\star}(\Phi(X_{t}^{k}a_{k})-\Psi(a_{\theta_{t}},X_{t})).\end{eqnarray}
 The higher order term $\Psi(a_{\theta_{t}},X_{t})$ has appeared,
but looking at the gauge transformation properties of the quantum
objects we see that it is actually necessary. We get \begin{equation}
\delta_{\lambda}\Phi(a_{\theta_{t}})=\Phi(d_{\theta_{t}}\lambda)=d_{\star}\Phi(\lambda)\end{equation}
 with (\ref{formality star derivation}) and (\ref{delta a theta})
and\begin{eqnarray}
\delta_{\lambda}(\Phi(X_{t}^{k}a_{k})-\Psi(a_{\theta},X_{t})) & = & \Phi([X_{t},\lambda])-\Psi(d_{\theta}\lambda,X_{t})\\
 & = & [\Phi(X_{t}),\Phi(\lambda)]-\Psi([\theta_{t},\lambda],X_{t})\nonumber \\
 &  & +\Psi([\theta_{t},X_{t}],\lambda)-\Psi(d_{\theta}\lambda,X_{t})\nonumber \\
 & = & [\Phi(X_{t}),\Phi(\lambda)]\nonumber \\
 & = & \delta_{X_{t}}\Phi(\lambda),\nonumber \end{eqnarray}
 where the addition of the new term preserves the correct transformation
property. With

\begin{equation}
K_{t}^{\star}=\sum_{n=0}^{\infty}\frac{1}{(n+1)!}(\Phi(a_{\theta_{t}})+\partial_{t})^{n},\end{equation}
 a calculation analogous to the semi-classical case gives \begin{eqnarray}
\delta_{\lambda}(K_{t}^{\star}(\Phi(X_{t}^{k}a_{k})-\Psi(a_{\theta_{t}},X_{t}))) & = & \delta_{X_{t}}K_{t}^{\star}(\Phi(\lambda))\\
 &  & +d_{\star}(K_{t}^{\star}(\Phi(\lambda)))K_{t}^{\star}(\Phi(X_{t}^{k}a_{k})-\Psi(a_{\theta_{t}},X_{t})).\nonumber \end{eqnarray}
 As in \cite{Jurco:2000fs,Jurco:2001my}, the NC gauge parameter is

\begin{equation}
\Lambda_{\lambda}[a]=K_{t}^{\star}(\Phi(\lambda))\Big{|}_{t=0},\end{equation}
 and we therefore get for the NC gauge potential

\begin{equation}
A_{X}[a]=K_{t}^{\star}(\Phi(X_{t}^{k}a_{k})-\Psi(a_{\theta_{t}},X_{t}))\Big{|}_{t=0},\end{equation}
 transforming with

\begin{equation}
\delta_{\lambda}A_{X}=\delta_{X}\Lambda_{\lambda}-[\Lambda_{\lambda}\stackrel{\star}{,}A_{X}].\end{equation}

\chapter{\label{sec: covariant coordinates general}Covariant coordinates}

While we can only construct actions for noncommutative gauge theory
if we have a frame commuting with the Poisson structure, covariant
coordinates can always be defined. Therefore we can still extract
information from the noncommutative gauge theory, even if we do not
have the complete picture. We will use these covariant coordinates
to generalize the open Wilson lines of chapter \ref{canonical observables}.
In \cite{Okawa:2001mv} these were used to give an exact formula for
the inverse Seiberg-Witten map. We will generalize this construction
for general $\star$-products with invertible Poisson structure $\theta^{ij}$.

\section{Wilson lines and observables}

As we saw in chapter \ref{sec: general NC gauge theory}, multiplication
with a coordinate from the left is not a covariant operation. For
this, we can define covariant coordinates\begin{equation}
X^{i}=x^{i}+A^{i}\end{equation}
for which we want\begin{equation}
\delta_{\Lambda}(X^{i}\star\Psi)=\delta_{\Lambda}((x^{i}+A^{i})\star\Psi)=i\Lambda\star X^{i}\star\Psi.\end{equation}
Therefore the gauge field $A^{i}$ has to transform as\begin{equation}
\delta_{\Lambda}A^{i}=i[\Lambda\stackrel{\star}{,}x^{i}]+i[\Lambda\stackrel{\star}{,}A^{i}].\end{equation}
 Even though the gauge field $A^{i}$ vanishes in the commutative
limit, its Seiberg-Witten map can nevertheless be calculated \cite{Jurco:2000fs}.
It starts with\begin{equation}
A^{i}=\theta^{ij}a_{j}+\mc O(2).\end{equation}
We can use the covariant coordinates to construct noncommutative Wilson
lines. As in the canonical case we can start with \begin{equation}
W_{l}=e_{\star}^{il_{i}X^{i}}\star e_{\star}^{-il_{i}x^{i}},\end{equation}
 where $\star$ is now an arbitrary $\star$-product. The transformation
property of $W_{l}$ is now\begin{equation}
W_{l}^{\prime}(x)=g(x)\star W_{l}(x)\star g^{-1}(T_{l}x),\end{equation}
 where \begin{equation}
T_{l}x^{j}=e_{\star}^{il_{i}x^{i}}\star x^{j}\star e_{\star}^{-il_{i}x^{i}}\end{equation}
 is an inner automorphism of the algebra, which can be interpreted
as a quantized coordinate transformation. Note that the $e_{\star}^{-l_{i}X^{i}}$
do not close to a group for $\theta^{ij}(x)$ at least quadratic in
the $x$'s. Therefore it is not clear how to generalize NC Wilson
lines for arbitrary curves as in \cite{Ishibashi:1999hs}. If we replace
commutators by Poisson brackets, the semi-classical limit of these
coordinate transformations may be calculated\begin{equation}
T_{l}x^{k}=e_{\star}^{il_{i}[x^{i}\stackrel{\star}{,}\cdot]}x^{k}\approx e^{-l_{i}\{ x^{i},\cdot\}}x^{j}=e^{-l_{i}\theta^{ij}\partial_{j}}x^{k},\end{equation}
 the formula becoming exact for $\theta^{ij}$ constant or linear
in $x$. We see that the semi-classical coordinate transformation
is the flow induced by the Hamiltonian vector field $-l_{i}\theta^{ij}\partial_{j}$.
At the end we may expand $W_{l}$ in terms of $\theta$ and get\begin{equation}
W_{l}=e^{-il_{i}\theta^{ij}a_{j}}+\mathcal{O}(\theta^{2}),\end{equation}
 where we have replaced $A^{i}$ by its Seiberg-Witten expansion.
We see that for $l$ small this really is a Wilson line starting at
$x$ and ending at $x-l\theta$. For a given $\star$-product, the
higher order corrections to this expression can in principle be calculated.
Note that this expression would also depend on the specific choice
of the Seiberg-Witten-map of the covariant coordinates.

If we have a measure function $\Omega(x)$ for our $\star$-product
with $\dd_{i}(\Omega\theta^{ij})=0$, we can use the trace property
of the integral (see chapter \ref{sec: traces}) to generalize the
open Wilson lines of chapter \ref{canonical observables}. They read

\begin{equation}
U_{l}=\int d^{2n}x\,\Omega(x)\, W_{l}(x)\star e_{\star}^{il_{i}x^{i}}=\int d^{2n}x\,\Omega(x)\, e_{\star}^{il_{i}X^{i}(x)}\end{equation}
 and are again gauge invariant objects. Of course, we can again insert
a function $f$ depending only on the covariant coordinates \begin{equation}
f_{l}=\int d^{2n}x\,\Omega(x)\, f(X^{i})\star e_{\star}^{il_{i}X^{i}(x)}\end{equation}
without spoiling the gauge invariance.

\section{\label{Inverse SW map}Inverse Seiberg-Witten-map}

As an application of the above constructed observables we generalize
\cite{Okawa:2001mv} to arbitrary $\star$-products, i. e. we give
a formula for the inverse Seiberg-Witten map for $\star$-products
with invertible Poisson structure. In order to map noncommutative
gauge theory to its commutative counterpart, we need a functional
$f_{ij}[X]$ fulfilling

\begin{equation}
f_{ij}[g\star X\star g^{-1}]=f_{ij}[X],\end{equation}
 \begin{equation}
df=0\end{equation}
 and\begin{equation}
f_{ij}=\partial_{i}a_{j}-\partial_{j}a_{i}+\mathcal{O}(\theta).\end{equation}
 $f$ is a gauge covariant field strength and reduces in the limit
$\theta\rightarrow0$ to the correct expression.

To prove the first and the second property we will only use the algebra
properties of the $\star$-product and the cyclicity of the trace.
All quantities with a hat will be elements of an algebra. With this
let $\hat{X}^{i}$ be covariant coordinates in an algebra, transforming
under gauge transformations like\begin{equation}
\hat{X}^{i\prime}=\hat{g}\hat{X}^{i}\hat{g}^{-1}\end{equation}
 with $\hat{g}$ an invertible element of the algebra. Now define
\begin{equation}
\hat{F}^{ij}=-i[\hat{X}^{i},\hat{X}^{j}]\end{equation}
 and \begin{equation}
(\hat{F}^{n-1})_{ij}=\frac{1}{2^{n-1}(n-1)!}\epsilon_{iji_{1}i_{2}\cdots i_{2n-2}}\hat{F}^{i_{1}i_{2}}\cdots\hat{F}^{i_{2n-3}i_{2n-2}}.\end{equation}
 Note that the space is $2n$ dimensional. Using the the symmetrized
trace $str$, i.e. 

\begin{flushright}\begin{eqnarray}
str_{\hat{F},\hat{X}}\Big{(}\hat{F}^{q}\hat{X}^{r}\Big{)}=\frac{q!r!}{(q+r)!}tr\Big{(}\hat{F}^{q}\hat{X}^{r}+\end{eqnarray}
\mbox{ all other possible permutations of $q$ $\hat{F}'$s and $r$ $\hat{X}'s$} \Big{)}\end{flushright}

\vspace{0,5cm}

see also \cite{Okawa:2001mv}, the expression\begin{equation}
\mathcal{F}_{ij}(k)=str_{\hat{F},\hat{X}}\left((\hat{F}^{n-1})_{ij}e^{ik_{j}\hat{X}^{j}}\right)\label{DEF_F}\end{equation}
 clearly fulfills the first property due to the properties of the
trace. Note that symmetrization is only necessary for space dimension
bigger than 4 due to the cyclicity of the trace. In dimensions 2 and
4 we may replace $str$ by the ordinary trace $tr$. $\mathcal{F}_{ij}(k)$
is the Fourier transform of a closed form if \begin{equation}
k_{[i}\mathcal{F}_{jk]}=0\end{equation}
 or if the current \begin{equation}
J^{i_{1}\cdots i_{2n-2}}=str_{\hat{F},X}\left(\hat{F}^{[i_{1}i_{2}}\cdots\hat{F}^{i_{2n-3}i_{2n-2}]}e^{ik_{j}\hat{F}^{j}}\right)\end{equation}
 is conserved, respectively\begin{equation}
k_{i}J^{i\cdots}=0.\end{equation}
 This is easy to show, if one uses \begin{eqnarray}
\lefteqn{str_{\hat{F},\hat{X}}\left([k_{i}\hat{X}^{i},\hat{X}^{l}]e^{ik_{j}\hat{X}^{j}}\cdots\right)}\\
 &  & =str_{\hat{F},\hat{X}}\left([\hat{X}^{l},e^{ik_{j}\hat{X}^{j}}]\cdots\right)=str_{\hat{F},\hat{X}}\left(e^{ik_{j}\hat{X}^{j}}[\hat{X}^{l},\cdots]\right),\nonumber \end{eqnarray}
 which can be calculated by simple algebra.

To prove that $\mc F$ has the right commutative limit, we have to
switch to the $\star$-product formalism and expand the formula in
$\theta^{ij}$. The expression (\ref{DEF_F}) now becomes \begin{equation}
\mathcal{F}[X]_{ij}(k)=\int\frac{d^{2n}x}{Pf(\theta)}\left((F_{\star}^{n-1})_{ij}\star e_{\star}^{ik_{j}X^{j}}\right)_{sym\, F,X}.\end{equation}
 The expression in brackets has to be symmetrized in $F^{ij}$ and
$X^{i}$ for $n>2$. Up to second order in $\theta^{ij}$, the commutator
$F^{ij}$ of two covariant coordinates is \begin{equation}
F^{ij}=-i[X^{i}\stackrel{\star}{,}X^{j}]=\theta^{ij}-\theta^{ik}f_{kl}\theta^{lj}-\theta^{kl}\partial_{l}\theta^{ij}a_{k}+\mathcal{O}(3)\end{equation}
 with $f_{ij}=\partial_{i}a_{j}-\partial_{j}a_{i}$ the ordinary field
strength. Furthermore we have\begin{equation}
e_{\star}^{ik_{i}X^{i}}=e^{ik_{i}x^{i}}(1+ik_{i}\theta^{ij}a_{j})+\mathcal{O}(2).\end{equation}
 If we choose an antisymmetric $\star$-product, the symmetrization
will annihilate all the first order terms of the $\star$-products
between the $F^{ij}$ and $X^{i}$, and therefore we get

\begin{eqnarray}
\lefteqn{-\mathcal{F}[X]_{ij}(k)}\\
 &  & =-2n\int\frac{d^{2n}x}{\epsilon\theta^{n}}\Big{(}\epsilon_{ij}\theta^{n-1}-(n-1)\epsilon_{ij}\theta^{n-2}\theta f\theta\nonumber \\
 &  & \hspace{70pt}-\theta^{kl}\partial_{l}(\epsilon_{ij}\theta^{n-1})a_{k}\Big{)}e^{ik_{i}x^{i}}+\mathcal{O}(1)\nonumber \\
 &  & =-2n\int\frac{d^{2n}x}{\epsilon\theta^{n}}\Big{(}\epsilon_{ij}\theta^{n-1}-(n-1)\epsilon_{ij}\theta^{n-2}\theta f\theta\nonumber \\
 &  & \hspace{75pt}-\frac{1}{2}\epsilon_{ij}\theta^{n-1}f_{kl}\theta^{kl}\Big{)}e^{ik_{i}x^{i}}+\mathcal{O}(1)\nonumber \\
 &  & =d^{2n}x\,\Big{(}\theta_{ij}^{-1}+2n(n-1)\frac{\epsilon_{ij}\theta^{n-2}\theta f\theta}{\epsilon\theta^{n}}\nonumber \\
 &  & \hspace{80pt}-\frac{1}{2}\theta_{ij}^{-1}f_{kl}\theta^{kl}\Big{)}e^{ik_{i}x^{i}}+\mathcal{O}(1),\nonumber \end{eqnarray}
 using partial integration and $\partial_{i}(\epsilon\theta^{n}\theta^{ij})=0$.
To simplify notation we introduced \begin{equation}
\epsilon_{ij}\theta^{n-1}=\epsilon_{iji_{1}j_{1}\cdots i_{n-1}j_{n-1}}\theta^{i_{1}j_{1}}\cdots\theta^{i_{n-1}j_{n-1}}\mbox{etc.}\end{equation}
 In the last line we have used\begin{equation}
\theta_{ij}^{-1}=-\frac{(\theta^{n-1})_{ij}}{Pf(\theta)}=-2n\frac{\epsilon_{ij}\theta^{n-1}}{\epsilon\theta^{n}}.\end{equation}
 We will now have a closer look at the second term, noting that\begin{equation}
\theta^{ij}\frac{\epsilon_{ij}\theta^{n-2}\theta f\theta}{\epsilon\theta^{n}}=-\frac{1}{2n}\theta_{kl}^{-1}\theta^{kr}f_{rs}\theta^{sl}=-\frac{1}{2n}f_{rs}\theta^{rs}\end{equation}
 and therefore\begin{equation}
\frac{\epsilon_{ij}\theta^{n-2}\theta f\theta}{\epsilon\theta^{n}}=a\frac{\epsilon_{ij}\theta^{n-1}}{\epsilon\theta^{n}}f_{rs}\theta^{rs}+bf_{ij}\label{defining formula for a b}\end{equation}
 with $a+b=-\frac{1}{2n}$. Taking e. g. $i=1,j=2$ we see that \begin{eqnarray}
\epsilon_{12\cdots kl}\theta^{n-2}\theta^{kr}f_{rs}\theta^{sl} & = & \epsilon_{12\cdots kl}\theta^{n-2}(\theta^{k1}\theta^{2l}-\theta^{k2}\theta^{1l})f_{12}\\
 &  & +\mbox{terms\, without}\: f_{12}.\nonumber \end{eqnarray}
 Especially there are no terms involving $f_{12}\theta^{12}$ and
we get for the two terms on the right hand side of (\ref{defining formula for a b})\begin{equation}
2a\epsilon_{12}\theta^{n-1}f_{12}\theta^{12}=-2nb\epsilon_{12}\theta^{12}\theta^{n-1}f_{12}\end{equation}
 and therefore $b=-\frac{a}{n}$. This has the solution \begin{equation}
a=-\frac{1}{2(n-1)}\:\:\:\mbox{and}\:\:\: b=\frac{1}{2n(n-1)}.\end{equation}
 With the resulting \begin{equation}
2n(n-1)\frac{\epsilon_{ij}\theta^{n-2}\theta f\theta}{\epsilon\theta^{n}}=\frac{1}{2}\theta_{ij}^{-1}f_{kl}\theta^{kl}+f_{ij}\end{equation}
 we finally get \begin{equation}
-\mathcal{F}[X]_{ij}(k)=\int d^{2n}x\,\left(\theta_{ij}^{-1}+f_{ij}\right)e^{ik_{i}x^{i}}+\mathcal{O}(1).\end{equation}
 Therefore\begin{equation}
f[X]_{ij}=\mathcal{F}[X]_{ij}(k)-\mathcal{F}[x]_{ij}(k)\end{equation}
 is a closed form that reduces in the commutative limit to the commutative
Abelian field strength. We have found an expression for the inverse
Seiberg-Witten map.

\cleardoublepage

\part{Matrix model approach}

$\mbox{}$

\vspace{40pt}Though $\star$-products are a convenient tool for studying noncommutativity,
their strength lies mainly in the perturbative regime. For other purposes,
especially nonperturbative ones, a different approach using a different
representation of the algebra of functions on noncommutative space
is better suited. 

If we take the simple example of a noncommutative plane with canonical
noncommutativity\begin{equation}
[x,y]=i\theta,\end{equation}
we see immediately that this is nothing but the Heisenberg algebra,
for which we can use the well known Fock-space representation. In
$2n$ dimensions, we can use $n$ such pairs of coordinates which
upon complexification become creation and annihilation operators on
the Fock-space. Using this approach, it was possible to study many
nonperturbative features of noncommutative field theory such as solitons
and instantons (see e.g. \cite{Douglas:2001ba} for references). 

We will call this approach matrix model approach, as the gauge theory
can be described as a matrix model having the noncommutative space
as its ground state, the fluctuations creating the gauge theory. But
noncommutative spacetime with canonical commutation relations has
to be represented on an infinite-dimensional vectorspace, leading
to a number of problems. First of all, there are the well known divergencies
of noncommutative gauge theory. Then, the rank of the gauge group
can't be fixed in this model \cite{Douglas:2001ba}. Therefore we
are looking for spaces that can be represented as finite-dimensional
matrix algebras, where everything is well defined. The space on which
we will base our constructions will be the fuzzy sphere \cite{Madore:1991bw},
an $N$-dimensional matrix algebra corresponding to a truncation of
the spherical harmonics on the sphere at angular momentum $N-1$.
To go to 4 dimensions, we will use the product of two such fuzzy spheres
$S_{N}^{2}\times S_{N}^{2}$, generated by $N^{2}$-dimensional matrices.
In one limit, this fuzzy space goes over to the product of two commutative
spheres, but in a different limit, it also goes to noncommutative
$\mathbb{R}^{4}$ with canonical commutation relations. Our interest
will therefore be twofold: On one hand we will study gauge theory
on this fuzzy space as the deformation of commutative gauge theory,
on the other hand as a regularization of gauge theory on $\mathbb{R}_{\theta}^{4}.$ 

\newpage

\thispagestyle{empty}

\chapter{\label{fuzzy canonical case}The canonical case}

Before we study gauge theory on a finite-dimensional fuzzy space,
we first want to present the usual matrix model approach to noncommutative
gauge theory on $\mathbb{R}_{\theta}^{4}$. After a quick look at
the infinite-dimensional Fock-space representation of $\mathbb{R}_{\theta}^{4}$,
we will show how gauge theory can be formulated as a matrix model
with ground state $\mathbb{R}_{\theta}^{4}$. The fluctuations around
this ground state will create the gauge theory. Finally we will have
a look at a certain class of instantons, the so called fluxon solutions.

\section{The Heisenberg algebra}

In two dimensions, the coordinate algebra with canonical deformation\begin{equation}
[x,y]=i\theta\label{Heisenberg algebra}\end{equation}
is nothing but the well known Heisenberg algebra. But now the noncommutativity
isn't between the coordinates and momenta, but between the coordinates
themselves. Of course we can use the usual Fock space representation
for this algebra by first defining\begin{equation}
x_{\pm}:=x\pm iy\end{equation}
 with\begin{equation}
[x_{+},x_{-}]=2\theta.\end{equation}
 The Fock space is given by\begin{equation}
\mc H=\{|n\rangle,\; n\epsilon\mathbb{N}_{0}\},\end{equation}
 where the creator and annihilator operators act as \begin{equation}
x_{-}|n\rangle=\sqrt{2\theta}\sqrt{n+1}|n+1\rangle,\;\;\;\;\;\;\; x_{+}|n\rangle=\sqrt{2\theta}\sqrt{n}|n-1\rangle.\end{equation}
 This can be generalized to higher dimensions. Any $2n$-dimensional
algebra with canonical commutation relations can by suitable rotations
be brought into a form where it consists of $n$ pairs of noncommuting
variables (\ref{Heisenberg algebra}). As we will mostly be concerned
with the 4-dimensional case in the following, we will present it here
in more detail.

The most general noncommutative $\mathbb{R}_{\theta}^{4}$ is generated
by coordinates subject to the commutation relations \begin{equation}
[x_{\mu},x_{\nu}]=i\theta_{\mu\nu}\,,\label{x-CR}\end{equation}
 where $\mu,\nu\in\{1,\dots,4\}$. Using suitable rotations, $\theta_{\mu\nu}$
can always be cast into the form \begin{equation}
\theta_{\mu\nu}=\left(\begin{array}{cccc}
0 & \theta_{12} & 0 & 0\\
-\theta_{12} & 0 & 0 & 0\\
0 & 0 & 0 & \theta_{34}\\
0 & 0 & -\theta_{34} & 0\end{array}\right)\:.\end{equation}
To simplify the following formulas, we restrict our discussion from
now on to the selfdual case \begin{equation}
\theta_{\mu\nu}=\frac{1}{2}\varepsilon_{\mu\nu\rho\sigma}\theta_{\rho\sigma}\end{equation}
 and denote \begin{equation}
\theta:=\theta_{12}=\theta_{34};\end{equation}
 the generalizations to the antiselfdual and the general case are
obvious. In terms of the complex coordinates \begin{eqnarray}
x_{\pm L}:=x_{1}\pm ix_{2} & , & x_{\pm R}:=x_{3}\pm ix_{4},\end{eqnarray}
 the commutation relations (\ref{x-CR}) take the form \begin{equation}
[x_{+a},x_{-b}]=2\theta\delta_{ab},\,\,\,[x_{+a},x_{+b}]=[x_{-a},x_{-b}]=0,\label{algebra}\end{equation}
 where $a,b\in\{ L,R\}$. The Fock-space representation $\mc H$ of
(\ref{algebra}) has the standard basis \begin{equation}
\mc H=\{|n_{1},n_{2}\rangle,\quad n_{1},n_{2}\in\mathbb{N}_{0}\,\},\label{Fock space}\end{equation}
 with\begin{equation}
\begin{array}{c}
x_{-L}|n_{1},n_{2}\rangle=\sqrt{2\theta}\sqrt{n_{1}\!+\!1}|n_{1}\!+\!1,n_{2}\rangle,\;\;\; x_{+L}|n_{1},n_{2}\rangle=\sqrt{2\theta}\sqrt{n_{1}}|n_{1}\!-\!1,n_{2}\rangle\;\;\\
x_{-R}|n_{1},n_{2}\rangle=\sqrt{2\theta}\sqrt{n_{2}\!+\!1}|n_{1},n_{2}\!+\!1\rangle,\;\;\; x_{+R}|n_{1},n_{2}\rangle=\sqrt{2\theta}\sqrt{n_{2}}|n_{1},n_{2}\!-\!1\rangle\,.\end{array}\end{equation}
 The derivatives on this space are inner, i.e. they are produced by
the commutator with a coordinate\begin{equation}
-i\theta^{\mu\nu}\dd_{\nu}\;\;\widehat{=}\;\;[\,\cdot\,,x^{\mu}],\end{equation}
 just as in the $\star$-product formalism.

\section{\label{sec: matrix canonical case}Noncommutative gauge theory}

We can introduce gauge theory by using a matrix action \begin{equation}
S=-\frac{(2\pi)^{2}}{2g^{2}\theta^{2}}\;\mathrm{tr}\;([X_{\mu},X_{\nu}]-i\theta_{\mu\nu})^{2},\label{eq: action on R4}\end{equation}
where the $X_{\mu}$ are infinite-dimensional matrices, and the trace
is over the Fock space (\ref{Fock space}). The action is obviously
constructed in such a way as to have the Fock-space representation
of $\mathbb{R}_{\theta}^{4}$ as its ground state. As we want the
action to be invariant under unitary transformations\begin{equation}
X_{\mu}\rightarrow U^{\dagger}X_{\mu}U,\end{equation}
 we get fluctuations $A_{\mu}$ around the ground state $x_{\mu}$
as \begin{equation}
X_{\mu}=x_{\mu}+A_{\mu}.\end{equation}
The fluctuations $A_{\mu}$ are understood as infinite-dimensional
matrices acting on the Fock space (\ref{Fock space}) as well. They
have to transform as \begin{equation}
A_{\mu}\rightarrow U^{\dagger}[x_{\mu},U]+U^{\dagger}A_{\mu}U\end{equation}
to make the $X_{\mu}$ gauge covariant. Remembering that the commutator
with a coordinate produces a derivative, we recognize the correct
transformation behavior for the gauge field. The gauge covariant field
strength then reads\begin{equation}
iF_{\mu\nu}=([X_{\mu},X_{\nu}]-i\theta_{\mu\nu})=[x_{\mu},A_{\nu}]-[x_{\nu},A_{\mu}]+[A_{\mu},A_{\nu}]\end{equation}
and the action (\ref{eq: action on R4}) reads \begin{equation}
S=\frac{(2\pi)^{2}}{2g^{2}\theta^{2}}\;\mathrm{tr}(F_{\mu\nu}F_{\mu\nu}).\end{equation}
 To bring the action into a form where it resembles more the creator
and annihilator representation, we can also use the complex covariant
coordinates $X_{\pm a}$\begin{eqnarray}
X_{\pm L}=X_{1}\pm iX_{2} & , & X_{\pm R}=X_{3}\pm iX_{4}\end{eqnarray}
 and the corresponding field strength \begin{equation}
F_{\alpha a,\beta b}=[X_{\alpha a},X_{\beta b}]-2\theta\varepsilon_{\alpha\beta}\delta_{ab}\,\end{equation}
 with $a,b\in\{ L,R\}$ and $\alpha,\beta\in\{+,-\}$. The action
(\ref{eq: action on R4}) can now be written in the form \begin{equation}
S=\frac{\pi^{2}}{g^{2}\theta^{2}}\;\mathrm{tr}(\sum_{a}F_{+a,-a}F_{+a,-a}-\sum_{a,b}F_{+a,+b}F_{-a,-b}),\end{equation}
 and the equations of motion are given by \begin{equation}
\sum_{a,\alpha}[X_{\alpha a},(F_{\alpha a,\beta b})^{\dagger}]=0\,.\label{eq:  EOM of plane}\end{equation}
We now want to discuss a peculiar feature of this formulation of noncommutative
gauge theory. Even though we did construct our action for $U(1)$,
it actually contains sectors for every rank of the gauge group $U(n)$!
This is related to the fact that in noncommutative gauge theories,
the gauge group also contains transformations acting on spacetime
itself. As the size of the matrices $X_{\mu}$ isn't fixed (they are
infinite-dimensional operators), we can't seperate the gauge part
of the unitary transformations from the spacetime part. This can be
seen as follows: If we have\begin{equation}
X_{\mu}=x_{\mu}\label{ground state 1}\end{equation}
as a ground state of the theory, then of course \begin{equation}
X'_{\mu}=\left(\begin{array}{cc}
x_{\mu} & 0\\
0 & x_{\mu}\end{array}\right)\end{equation}
is equally a ground state. In fact, the direct sum of $n$ solutions
$x_{\mu}$ of the equations of motion will again be a solution. As
the covariant coordinates $X_{\mu}=x_{\mu}+A_{\mu}$ corresponding
to the ground state (\ref{ground state 1}) produce a $U(1)$ theory,
any such ground state $X_{\mu}'=x_{\mu}\otimes1_{n\times n}$ can
be viewed as the ground state of a $U(n)$ gauge theory, where the
gauge degrees of freedom act on the right hand side of the tensor
product. The corresponding covariant coordinate can then be written
as\begin{equation}
X'_{\m}=x_{\mu}\otimes1_{n\times n}+A_{\mu,a}T^{a},\end{equation}
with the $T^{a}$ are generalized Gellman matrices for $U(n)$, producing
a $U(n)$ gauge theory. So the matrix action (\ref{eq: action on R4})
cannot be restricted to one gauge group, it contains sectors with
all $U(n).$ As we will see in chapter \ref{sec:scaling}, this problem
can be fixed in a regularized theory.

\section{$U(1)$ instantons on $\mathbb{R}_{\theta}^{4}$}

\label{se: instantons}

We will for the moment stick to the $U(1)$-sector of the theory and
look for solutions of the equations of motion (\ref{eq:  EOM of plane})
which can be understood as instantons of the gauge theory.

On noncommutative $\mathbb{R}_{\theta}^{2}$, all $U(1)$-instantons
were constructed and classified in \cite{Gross:2000ss}. They can
be interpreted as localized flux solutions, sometimes called fluxons.

The situation on $\R_{\theta}^{4}$ is more complicated, and there
are different types of non-trivial $U(1)$ instanton solutions on
$\mathbb{R}_{\theta}^{4}$. Assuming that $\theta_{\mu\nu}$ is self-dual,
there are two types of instantons: first, there exist straightforward
generalizations of the two-dimensional localized fluxon solutions
with self-dual field strength. As in the two-dimensional case, we
will refer to these 4-dimensional solutions as fluxons. 

There are other types of $U(1)$ instantons on $\mathbb{R}_{\theta}^{4}$,
which were found through a noncommutative version of the ADHM equations
\cite{Nekrasov:1998ss,Furuuchi:1999kv,Chu:2001cx,Hamanaka:2001dr,Ivanova:2005fh,Wimmer:2005bz},
in particular anti-selfdual instantons which are much less localized
than the fluxon solutions. However, we will concentrate on the generalizations
of \cite{Gross:2000ss}, as they will become important for us in chapter
\ref{sec:scaling}.

For the construction of the fluxons, let us consider a finite dimensional
subvectorspace $V_{n}$ of the Fock-space $\mc\mathcal{H}$ of dimension
$n$ spanned by a finite set of vectors $|n_{1},n_{2}\rangle\in\mathcal{H}$,
\begin{equation}
V_{n}=\langle\{|i_{k},j_{k}\rangle;\;\; k=1,...,n\}\rangle\,.\label{eq: subspace of  H}\end{equation}
 We introduce a partial isometry%
\footnote{If we index the basis of $\mc H$ as $|i_{k},j_{k}\rangle$ with $k\in\mathbb{N}$
and assume that $V$ is spanned by the first $n$ vectors (which we
can always get by using a suitable unitary transformation), $S$ can
be given by $S:|i_{k},j_{k}\rangle\rightarrow|i_{k+n},j_{k+n}\rangle$.%
} $S$ mapping $\mc H$ to $\mc H\setminus V_{n}$ , which has \begin{eqnarray}
S^{\dagger}S & = & \one,\\
SS^{\dagger} & = & \one-P_{_{V_{n}}}\end{eqnarray}
with the projection operator onto the subspace $V_{n}$\begin{equation}
P_{_{V_{n}}}:=\sum_{k=1}^{n}|i_{k},j_{k}\rangle\langle i_{k},j_{k}|.\end{equation}
Following \cite{Gross:2000ss} one then finds solutions to the equations
of motion given by%
\footnote{Note that $[X_{+L}^{(n)},X_{+R}^{(n)}]=[X_{+L}^{(n)},X_{-R}^{(n)}]=[X_{-L}^{(n)},X_{+R}^{(n)}]=[X_{-L}^{(n)},X_{-R}^{(n)}]=0$
.%
}\begin{eqnarray}
X_{+L}^{(n)} & := & Sx_{+L}S^{\dagger}+\sum_{k=1}^{n}\gamma_{k}^{L}|i_{k},j_{k}\rangle\langle i_{k},j_{k}|\label{eq: sol for  instantons on R4 L}\\
X_{+R}^{(n)} & := & Sx_{+R}S^{\dagger}+\sum_{k=1}^{n}\gamma_{k}^{R}|i_{k},j_{k}\rangle\langle i_{k},j_{k}|\,,\label{eq: sol for instantons on R4 R}\end{eqnarray}
and $X_{-a}^{(n)}=(X_{+a}^{(n)})^{\dagger}$. Here $\gamma_{k}^{L,R}\in\C$
determine the position of the fluxons. The field strength $F_{\mu\nu}$
for this solution is \begin{equation}
F_{\mu\nu}=P_{_{V_{n}}}\theta_{\mu\nu}.\end{equation}
 In particular, the action corresponding to the instanton solution
(\ref{eq: sol for instantons on R4 L},\ref{eq: sol for instantons on R4 R})
is proportional to the dimension of the subspace $V_{n}$ \begin{equation}
S[X_{\pm a}^{(n)}]=\frac{8\pi^{2}}{g^{2}}\;\mathrm{tr}(P_{_{V_{n}}})=\frac{8\pi^{2}}{g^{2}}\; n.\end{equation}
 Since they can be interpreted as localized flux, these $U(1)$-instanton
solutions for $\mathbb{R}_{\theta}^{4}$ are called fluxons. The localization
can be seen as follows: recall \cite{Furuuchi:2000vc} that the above
projection operators can be represented on the space of commutative
functions (using a normal-ordering prescription) as \begin{equation}
|k^{1},k^{2}\rangle\langle k^{1},k^{2}|\cong\frac{1}{k^{1}!k^{2}!}(\frac{x^{-L}}{\sqrt{2\theta}})^{k^{1}}(\frac{x^{+L}}{\sqrt{2\theta}})^{k^{1}}(\frac{x^{-R}}{\sqrt{2\theta}})^{k^{2}}(\frac{x^{+R}}{\sqrt{2\theta}})^{k^{2}}e^{-\frac{x^{+L}x^{-L}}{2\theta}-\frac{x^{+R}x^{-R}}{2\theta}}\,.\end{equation}
 Hence the above field strengths $F_{\mu\nu}=P_{_{V_{n}}}\theta_{\mu\nu}$
are superpositions of Gauss-functions which are localized in a region
in space of size $\sqrt{\theta}$.

\chapter{Fuzzy spaces }

In this chapter we will present a 4-dimensional noncommutative space
that has the advantage of having finite dimensional representations.
Therefore, the gauge theory we will construct on it in chapter \ref{sec: gauge theory on S2S2}
will be well defined and all calculations will become finite. Using
this space we will be able to regularize both $\mathbb{R}_{\theta}^{4}$
itself in chapter \ref{planelimit} and gauge theory on $\mathbb{R}_{\theta}^{4}$
in chapter \ref{sec:scaling}.

\section{The fuzzy sphere $S_{N}^{2}$}

We start by recalling the definition of a 2-dimensional space, the
fuzzy sphere introduced in \cite{Madore:1991bw}. The algebra of functions
on the fuzzy sphere is the finite algebra $S_{N}^{2}$ generated by
Hermitian operators $x_{i}=(x_{1},x_{2},x_{3})$ satisfying the defining
relations \begin{eqnarray}
[x_{i},x_{j}]=i\Lambda_{N}\e_{ijk}x_{k},\label{def1}\\
x_{1}^{2}+x_{2}^{2}+x_{3}^{2}=R^{2}.\label{def2}\end{eqnarray}
 They are obtained from the $N$-dimensional representation of $su(2)$
with generators $\lambda_{i}\;(i=1,2,3)$ and commutation relations
\begin{equation}
{}[\lambda_{i},\lambda_{j}]=i\epsilon_{ijk}\lambda_{k},\qquad\sum_{i=1}^{3}\lambda_{i}\lambda_{i}=\frac{N^{2}-1}{4}\label{lambda-algebra}\end{equation}
 (see Appendix \ref{sec:useful}) by identifying \begin{equation}
x_{i}=\Lambda_{N}\;\lambda_{i},\qquad\Lambda_{N}=\frac{2R}{\sqrt{N^{2}-1}}.\end{equation}
 The noncommutativity parameter $\Lambda_{N}$ is of dimension length.
The algebra of functions $S_{N}^{2}$ therefore coincides with the
simple matrix algebra $Mat(N,\C)$. The normalized integral of a function
$f\in S_{N}^{2}$ is given by the trace \begin{equation}
\int\limits _{S_{N}^{2}}f=\frac{4\pi R^{2}}{N}\mathrm{tr}(f).\end{equation}
 The functions on the fuzzy sphere can be mapped to functions on the
commutative sphere $S^{2}$ using the decomposition into harmonics
under the action \begin{equation}
J_{i}f=[\lambda_{i},f]\end{equation}
 of the rotation group $SU(2)$. One obtains analogs of the spherical
harmonics up to a maximal angular momentum $N-1$. Therefore $S_{N}^{2}$
is a regularization of $S^{2}$ with a UV cutoff, and the commutative
sphere $S^{2}$ is recovered in the limit $N\rightarrow\infty$. Note
also that for the standard representation (\ref{standard-rep}), entries
in the upper-left block of the matrices correspond to functions localized
at $x_{3}=R$. In particular, the fuzzy delta-function at the {}``north
pole'' is given by a suitably normalized projector of rank 1, \begin{equation}
\d{^{(2)}}_{NP}(x)=\frac{N}{4\pi R^{2}}\;|\frac{N-1}{2}\rangle\langle\frac{N-1}{2}|\label{fuzzydelta}\end{equation}
 where $|\frac{N-1}{2}\rangle$ is the highest weight state with maximal
eigenvalue of $\la_{3}$. Delta-functions with arbitrary localization
are obtained by rotating (\ref{fuzzydelta}).

\section{$S_{N_{L}}^{2}\times S_{N_{R}}^{2}$}

The simplest 4-dimensional generalization of the above is the product
$S_{N_{L}}^{2}\times S_{N_{R}}^{2}$ of 2 such fuzzy spheres, with
generally independent parameters $N_{L,R}$. It is generated by a
double set of representations of $su(2)$ commuting with each other,
i. e. by $\lambda_{i}^{L},\lambda_{i}^{R}$ satisfying \begin{eqnarray}
{}[\lambda_{i}^{L},\lambda_{j}^{L}] & = & i\epsilon_{ijk}\lambda_{k}^{L},\qquad{}[\lambda_{i}^{R},\lambda_{j}^{R}]=i\epsilon_{ijk}\lambda_{k}^{R},\\
{}[\lambda_{i}^{L},\lambda_{j}^{R}] & = & 0\nonumber \end{eqnarray}
 for $i,j=1,2,3$, and Casimirs \begin{equation}
\sum_{i=1}^{3}\lambda_{i}^{L}\lambda_{i}^{L}=\frac{N_{L}^{2}-1}{4},\qquad\sum_{i=1}^{3}\lambda_{i}^{R}\lambda_{i}^{R}=\frac{N_{R}^{2}-1}{4}.\end{equation}
 This can be realized as a tensor product of 2 fuzzy sphere algebras
\begin{eqnarray}
\lambda_{i}^{L} & = & \lambda_{i}\otimes1_{N_{R}\times N_{R},}\label{Lambda L}\\
\lambda_{i}^{R} & = & 1_{N_{L}\times N_{L}}\otimes\lambda_{i},\label{Lambda R}\end{eqnarray}
 hence as algebra we have $S_{N_{L}}^{2}\times S_{N_{R}}^{2}\cong\mathrm{Mat}(\cN,\C)$
where \begin{equation}
\cN=N_{L}N_{R}.\end{equation}
 The normalized coordinate functions are given by \begin{equation}
x_{i}^{L,R}=\frac{2R}{\sqrt{(N^{L,R})^{2}-1}}\;\lambda_{i}^{L,R},\qquad\sum(x_{i}^{L})^{2}=R^{2}=\sum(x_{i}^{R})^{2}.\end{equation}
 This space%
\footnote{In principle one could also introduce different radii $R^{L,R}$ for
the 2 spheres, but for simplicity we will keep only one scale parameter
$R$ (and sometimes we will set $R=1$).%
} can be viewed as regularization of $S^{2}\times S^{2}\subset\R^{6}$,
and admits the symmetry group $SU(2)_{L}\times SU(2)_{R}\subset SO(6)$.
The generators $x_{i}^{L,R}$ should be viewed as coordinates in an
embedding space $\R^{6}$. The normalized integral of a function $f\in S_{N_{L}}^{2}\times S_{N_{R}}^{2}$
is now given by \begin{equation}
\int\limits _{S_{N_{L}}^{2}\times S_{N_{R}}^{2}}f=\frac{16\pi^{2}R^{4}}{\cN}\mathrm{tr}(f)=\frac{V}{\cN}\mathrm{tr}(f),\label{int-4}\end{equation}
 where we define the volume $V:=16\pi^{2}R^{4}$. We will mainly consider
$N_{L}=N_{R}$ in the following.

\section{\label{planelimit}The limit to the canonical case $\R_{\theta}^{4}$}

It is well-known \cite{Chu:2001xi} that if a fuzzy sphere is blown
up near a given point, it can be used to obtain a (compactified) noncommutative
plane with canonical commutation relations: Consider the tangential
coordinates $x_{1,2}$ near the north pole $x_{3}=R$. Setting \begin{equation}
R^{2}=N\theta/2,\label{R-theta}\end{equation}
 they satisfy the commutation relations \begin{equation}
[x_{1},x_{2}]\;=\; i\frac{2R}{N}x_{3}\;=\; i\frac{2R}{N}\sqrt{R^{2}-x_{1}^{2}-x_{2}^{2}}\;=\; i\theta+O(1/N).\end{equation}
 Therefore in the double scaling limit with $N,\, R\rightarrow\infty$
keeping $\theta$ fixed, we recover%
\footnote{One could be more sophisticated and use the stereographic projections
as in \cite{Chu:2001xi}, which leads essentially to the same results.%
} the commutation relation of the canonical case, \begin{equation}
[x_{1},x_{2}]=i\theta\end{equation}
 up to corrections of order $\frac{1}{N}$. Similarly, starting with
$S_{N_{L}}^{2}\times S_{N_{R}}^{2}$ and setting \begin{equation}
R^{2}=N_{L,R}\theta_{L,R}/2,\label{R-thetaLR}\end{equation}
 we obtain in the large $N_{L},\, N_{R}$ limit \begin{eqnarray}
[x_{i}^{L},x_{j}^{L}] & = & i\epsilon_{ij}\theta^{L},\qquad[x_{i}^{R},x_{j}^{R}]=i\epsilon_{ij}\theta^{R},\qquad\\
{}[x_{i}^{L},x_{j}^{R}] & = & 0.\nonumber \end{eqnarray}
 This is the most general form of $\R_{\theta}^{4}$ with coordinates
$(x_{1},...,x_{4})\equiv(x_{1}^{L},x_{2}^{L},x_{1}^{R},x_{2}^{R})$
(after a suitable orthogonal transformation). The integral of a function
$f(x)$ then becomes \begin{equation}
\int\limits _{S_{N_{L}}^{2}\times S_{N_{R}}^{2}}f(x)\rightarrow4\pi^{2}\theta_{L}\theta_{R}\mathrm{tr}(f(x))=:\int\limits _{\R_{\theta}^{4}}f(x),\end{equation}
 which has indeed the standard normalization, giving each {}``Planck
cell'' the appropriate volume.

\chapter{\label{sec: gauge theory on S2S2}Gauge theory on fuzzy $S^{2}\times S^{2}$}

Now that we have the fuzzy space $S_{N_{L}}^{2}\times S_{N_{R}}^{2}$
corresponding to $\mc N^{2}$-dimensional matrices, we want to construct
a matrix model having $S_{N_{L}}^{2}\times S_{N_{R}}^{2}$ as its
ground state. As in the canonical case, the fluctuations around this
ground state will produce a gauge theory. But as the matrices are
now finite-dimensional, the model will be well defined and finite. 

We will start with the most obvious formulation, gauging every coordinate
seperately. But there is also a more elegant formulation using collective
matrices. This will be especially usefull to introduce fermions, which
can be embedded very naturally in this framework.

We will also study non-trivial solutions of the EOMs, identifying
some of them as the monopoles on the commutative $S^{2}\times S^{2}$,
while others will become important in the limit to $\mathbb{R}_{\theta}^{4}$
in the following chapter.

\section{Gauge theory}

In the fuzzy case, it is natural to construct $S_{L}^{2}\times S_{R}^{2}$
as a submanifold of $\R^{6}$. We therefore consider a multi-matrix
model with 6 dynamical fields (covariant coordinates) $B_{i}^{L}$
and $B_{i}^{R}$ $(i=1,2,3)$, which are $\cN\times\cN$ Hermitian
matrices. As action we choose the following generalization of the
action in \cite{Steinacker:2003sd}, \begin{equation}
S=\frac{1}{g^{2}}\int\frac{1}{2}F_{ia\: jb}F_{ia\: jb}+\varphi_{L}^{2}+\varphi_{R}^{2}\label{action}\end{equation}
 with $a,b=L,R$ and $i,j=1,2,3$; summation over repeated indices
is implied. Here $\varphi_{L,R}$ are defined as \begin{equation}
\varphi_{L}:=\frac{1}{R^{2}}(B_{i}^{L}B_{i}^{L}-\frac{N_{L}^{2}-1}{4}),\qquad\varphi_{R}:=\frac{1}{R^{2}}(B_{i}^{R}B_{i}^{R}-\frac{N_{R}^{2}-1}{4}),\label{phi-def}\end{equation}
 and $R$ denotes the radius of the two spheres, which we keep explicitly
to have the correct dimensions. The field strength is defined by \begin{eqnarray}
F_{iL\: jL} & = & \frac{1}{R^{2}}(i[B_{i}^{L},B_{j}^{L}]+\epsilon_{ijk}B_{k}^{L}),\label{F-def}\\
F_{iR\: jR} & = & \frac{1}{R^{2}}(i[B_{i}^{R},B_{j}^{R}]+\epsilon_{ijk}B_{k}^{R}),\nonumber \\
F_{iL\: jR} & = & \frac{1}{R^{2}}(i[B_{i}^{L},B_{j}^{R}]).\nonumber \end{eqnarray}
 This model (\ref{action}) is manifestly invariant under $SU(2)_{L}\times SU(2)_{R}$
rotations acting in the obvious way, and $U(\cN)$ gauge transformations
acting as $B_{i}^{L,R}\rightarrow UB_{i}^{L,R}U^{-1}$. We will see
below that this reduces indeed to the $U(1)$ Yang-Mills action on
$S^{2}\times S^{2}$ in the commutative limit. Note that if the action
(\ref{action}) is considered as a matrix model, the radius drops
out using (\ref{int-4}). The equations of motion for $B_{i}^{L}$
are \begin{eqnarray}
 &  & {}\{ B_{i}^{L},B_{j}^{L}B_{j}^{L}-\frac{N_{L}^{2}-1}{4}\}+(B_{i}^{L}+i\epsilon_{ijk}B_{j}^{L}B_{k}^{L})\label{EOM}\\
 &  & +i\epsilon_{ijk}[B_{j}^{L},(B_{k}^{L}+i\epsilon_{krs}B_{r}^{L}B_{s}^{L})]+[B_{j}^{R},[B_{j}^{R},B_{i}^{L}]]=0,\nonumber \end{eqnarray}
 and those for $B_{i}^{R}$ are obtained by exchanging $L\leftrightarrow R$.
By construction, the minimum or ground state of the action is given
by $F=\varphi=0$, hence $B_{i}^{L,R}=\lambda_{i}^{L,R}$ as in (\ref{Lambda L},\ref{Lambda R})
up to gauge transformations; cp. \cite{Grosse:2004wm} for a similar
approach on $\C P^{2}$. We can therefore expand the covariant coordinates
$B_{i}^{L}$ and $B_{i}^{R}$ around the ground state \begin{equation}
B_{i}^{a}=\lambda_{i}^{a}+R{A}_{i}^{a},\label{B-A-relation}\end{equation}
 where $a\in\{ L,R\}$ and ${A}_{i}^{a}$ is small. Then ${A}_{i}^{L,R}$
transforms under gauge transformations as \begin{equation}
{A}_{i}^{L,R}\rightarrow{A'}_{i}^{L,R}=U{A}_{i}^{L,R}U^{-1}+U[{\la}_{i}^{L,R},U^{-1}],\label{gaugetrafo-A}\end{equation}
 and the field strength takes a more familiar form%
\footnote{We do not distinguish between upper and lower indices $L,R$.%
}, \begin{eqnarray}
F_{iL\: jL} & = & i([\frac{\lambda_{i}^{L}}{R},A_{j}^{L}]-[\frac{\lambda_{j}^{L}}{R},A_{i}^{L}]+[A_{i}^{L},A_{j}^{L}]),\label{F-A}\\
F_{iR\: jR} & = & i([\frac{\lambda_{i}^{R}}{R},A_{j}^{R}]-[\frac{\lambda_{j}^{R}}{R},A_{i}^{R}]+[A_{i}^{R},A_{j}^{R}]),\nonumber \\
F_{iL\: jR} & = & i([\frac{\lambda_{i}^{L}}{R},A_{j}^{R}]-[\frac{\lambda_{j}^{R}}{R},A_{i}^{L}]+[A_{i}^{L},A_{j}^{R}]).\nonumber \end{eqnarray}
 So far, the spheres are described in terms of 3 Cartesian covariant
coordinates each. In the commutative limit, we can separate the radial
and tangential degrees of freedom. There are many ways to do this;
perhaps the most elegant for the present purpose is to note that the
terms $\int\varphi_{L}^{2}+\varphi_{R}^{2}$ in the action imply that
$\varphi_{L,R}$ is bounded for configurations with finite action.
Using \begin{equation}
\varphi_{L}=\;\frac{\lambda_{i}^{L}}{R}{A}_{i}^{L}+{A}_{i}^{L}\frac{\lambda_{i}^{L}}{R}+{A}_{i}^{L}{A}_{i}^{L},\label{phi-2}\end{equation}
 and similarly for $\varphi_{R}$ it follows that \begin{equation}
x_{i}{A}_{i}^{a}+{A}_{i}^{a}x_{i}=O(\frac{\varphi}{N})\label{A-constraint}\end{equation}
 for finite $A_{i}^{a}$. This means that ${A}_{i}^{a}$ is tangential
in the (commutative) large $N$ limit. Alternatively, one could consider
$\phi_{L}=N\varphi_{L}$, which would acquire a mass of order $N$
and decouple from the other fields%
\footnote{The constraints $\varphi_{L}=0=\varphi_{R}$ could also be imposed
by hand; however the suppression through the above terms in the action
is more flexible, as we will see in chapter \ref{se: monopoles}.%
}. The commutative limit of (\ref{action}) therefore gives the standard
action for electrodynamics on $S^{2}\times S^{2}$, \begin{equation}
S=\frac{1}{2g^{2}}\int\limits _{S^{2}\times S^{2}}F_{ia\: jb}^{t}F_{ia\: jb}^{t}\end{equation}
 with $a,b=L,R$ and $i,j=1,2,3$. Here $F_{iL\: jR}^{t}$ denotes
the usual tangential field strength. This can be seen most easily
by noting that e.g. at the north pole $x_{3}^{L,R}=R$, one can replace
\begin{equation}
i[\frac{\lambda_{i}^{L,R}}{R},\cdot]\;\;\rightarrow\;\;-\varepsilon_{ij}\frac{\partial}{\partial x_{j}^{L,R}}\end{equation}
 in the commutative limit, so that upon identifying the commutative
gauge fields $A_{i}^{(cl)}$ via \begin{equation}
A_{i}^{(cl)L,R}=-\varepsilon_{ij}A_{i}^{L,R}\end{equation}
 the field strength is given by the standard expression $F_{iL\: jR}^{t}=\pp i^{L}A_{j}^{(cl)R}-\pp j^{R}A_{i}^{(cl)L}$
etc.

\section*{$U(k)$ gauge theory}

The above action generalizes immediately to the nonabelian case, keeping
precisely the same action (\ref{action}), (\ref{phi-def}), but replacing
the matrices $B_{i}^{L,R}$ by $k\cN\times k\cN$ matrices, cp. \cite{Steinacker:2003sd}.
The constraint term will then impose as ground state $\lambda_{i}^{L/R}\otimes1_{k\times k}$.
Expanding the covariant coordinates $B_{i}^{L,R}=\lambda_{i}^{L/R}\otimes1_{k\times k}+A_{i,a}^{L/R}T^{a}$
in terms of the Gellman matrices $T^{a}$, the action (\ref{action})
is the fuzzy version of nonabelian $U(k)$ Yang-Mills on $S^{2}\times S^{2}$.

\section{A formulation based on $SO(6)$}

\label{so6 formulation}

The above action can be cast into a nicer form by assembling the matrices
$B_{i}^{L,R}$ into bigger collective matrices, following \cite{Steinacker:2003sd}.
Since it is natural from the fuzzy point of view to embed $S^{2}\times S^{2}\subset\mathbb{R}^{6}$
with corresponding embedding of the symmetry group $SO(3)_{L}\times SO(3)_{R}\subset SO(6)$,
we consider \begin{equation}
B_{\mu}=(B_{i}^{L},B_{i}^{R})\end{equation}
 to be the $6$ -dimensional irrep of $so(6)\cong su(4)$. Since $(4)\tens(4)=(6)\oplus(10)$,
it is natural to introduce the intertwiners \begin{equation}
\g_{\mu}=(\g_{i}^{L},\g_{i}^{R})=(\g_{\mu})^{\a,\b{}}\end{equation}
 where $\a,\b{}$ denote indices of $(4)$. We could then assemble
our dynamical fields into a single $4\cN\times4\cN$ matrix \begin{equation}
B=B_{\mu}\g_{\mu}\;\;+\mathrm{const}\cdot\one.\label{B-def}\end{equation}
 Of course the most general such $4\cN\times4\cN$ matrix contains
far too many degrees of freedom, and we have to constrain these $B$
further. Since $SU(4)$ acts on $B$ as $B\rightarrow U^{T}BU$, the
$\g_{\mu}$ can be chosen as totally anti-symmetric matrices, which
precisely singles out the $(6)\subset(4)\tens(4)$. One can moreover
impose \begin{equation}
(\g_{i}^{L})^{\dagger}=\g_{i}^{L},\qquad(\g_{i}^{R})^{\dagger}=-\g_{i}^{R},\label{g-conj}\end{equation}
 and \begin{eqnarray}
\gamma_{i}^{L}\gamma_{j}^{L} & = & \delta_{ij}+i\epsilon_{ijk}\gamma_{k}^{L},\label{g-rel-L}\\
\gamma_{i}^{R}\gamma_{j}^{R} & = & -\delta_{ij}-\epsilon_{ijk}\gamma_{k}^{R},\label{g-rel-R}\\
{}[\gamma_{i}^{L},\gamma_{j}^{R}] & = & 0,\label{g-L-R}\end{eqnarray}
 which will be assumed from now on; we will give two explicit such
representations in (\ref{gamma-rep-1}), (\ref{gamma-wolfg}). This
would suggest to constrain $B$ to be antisymmetric. However, the
component fields $B_{\mu}$ are naturally considered as Hermitian
rather than symmetric matrices. Furthermore, since the $\g_{\mu}=(\g_{\mu})^{\a,\b{}}$
have two upper indices, they do not form an algebra. There are two
ways to proceed. We can either separate them again by introducing
two $4\mc N\times4\mc N$ matrices, \begin{equation}
B^{L}=\frac{1}{2}+B_{i}^{L}\g_{i}^{L},\qquad B^{R}=\frac{i}{2}+B_{i}^{R}\g_{i}^{R},\label{B-L-R}\end{equation}
 breaking $SO(6)\rightarrow SO(3)\times SO(3)$. This will be pursued
in Appendix \ref{sec:4Nmatrices}. Alternatively, we can use the $\g_{\mu}$
with the above properties to construct the $8\times8$ Gamma-matrices
\begin{equation}
\Gamma^{\mu}=\left(\begin{array}{cc}
0 & \gamma^{\mu}\\
\gamma^{\mu\dagger} & 0\end{array}\right),\label{Gamma}\end{equation}
 which generate the $SO(6)$-Clifford algebra \begin{equation}
\{\Gamma^{\mu},\Gamma^{\nu}\}=\left(\begin{array}{cc}
\gamma^{\mu}\gamma^{\nu\dagger}+\gamma^{\nu}\gamma^{\mu\dagger} & 0\\
0 & \gamma^{\mu\dagger}\gamma^{\nu}+\gamma^{\nu\dagger}\gamma^{\mu}\end{array}\right)=2\delta^{\mu\nu}.\label{clifford-so6}\end{equation}
 This suggests to consider the single Hermitian $8\cN\times8\cN$
matrix \begin{equation}
C=\Gamma^{\mu}B_{\mu}+C_{0}=\left(\begin{array}{cc}
0 & B^{L}\\
B^{L} & 0\end{array}\right)+\left(\begin{array}{cc}
0 & B^{R}\\
-B^{R} & 0\end{array}\right)=:C^{L}+C^{R},\label{C-def}\end{equation}
 where $C_{0}=C_{0}^{L}+C_{0}^{R}$ denote the constant $8\times8$-matrices
\begin{eqnarray}
C_{0}^{L} & = & -\frac{i}{2}\G_{1}^{L}\G_{2}^{L}\G_{3}^{L}=\frac{1}{2}\left(\begin{array}{cc}
0 & 1\\
1 & 0\end{array}\right),\label{B0-l}\\
C_{0}^{R} & = & -\frac{i}{2}\G_{1}^{R}\G_{2}^{R}\G_{3}^{R}=\frac{i}{2}\left(\begin{array}{cc}
0 & 1\\
-1 & 0\end{array}\right)\label{B0-R}\end{eqnarray}
 in the above basis. Using the Clifford algebra and the above definitions
one then finds \begin{equation}
C^{2}=B_{\mu}B_{\mu}+\frac{1}{2}+\Sigma_{8}^{\mu\nu}F_{\mu\nu}.\label{c-2}\end{equation}
 Here $\Sigma_{8}^{\mu\nu}=-\frac{i}{4}[\G_{\mu},\G_{\nu}]$, and
the field strength $F_{\mu\nu}$ coincides with the definition in
(\ref{F-def}) if written in the $L-R$ notation, \begin{equation}
F_{ia\: jb}=i[B_{ia},B_{jb}]+\delta_{ab}\epsilon_{ijk}B_{ka}.\end{equation}
 Therefore the action \begin{equation}
S_{6}=\mathrm{Tr}((C^{2}-\frac{N^{2}}{2})^{2})=8\mathrm{tr}(B_{\mu}B_{\mu}-\frac{N^{2}-1}{2})^{2}+4\mathrm{tr}F_{\mu\nu}F_{\mu\nu}\label{action-so6}\end{equation}
 is quite close to what we want. The only difference is the term $(B_{\mu}B_{\mu}-\frac{N^{2}-1}{2})^{2}$
instead of $(B_{iL}B_{iL}-\frac{N_{L}^{2}-1}{4})^{2}+(B_{iR}B_{iR}-\frac{N_{R}^{2}-1}{4})^{2}$,
for $2N^{2}=N_{L}^{2}+N_{R}^{2}$. This difference is easy to understand:
since (\ref{action-so6}) is $SO(6)$-invariant, the ground state
should be some $S^{5}$. We therefore have to break this $SO(6)$-
invariance explicitly, which will be done in the next chapter. However
before doing that, let us try to understand action (\ref{action-so6})
better and see whether it leads to a meaningful 4-dimensional field
theory. We show in Appendix \ref{app:stability} by carefully integrating
out the scalar components of $B_{i}^{L,R}$ that the $SO(6)$- invariant
constraint term in (\ref{action-so6}) induces the second term in
the following effective action \begin{equation}
S_{6}^{\mathrm{eff}}\sim4\mathrm{tr}\left(F_{\mu\nu}F_{\mu\nu}-(F_{iL}x_{iL}-F_{iR}x_{iR})\frac{1}{4(\frac{1}{2}-\dd_{\mu}\dd_{\mu})}(F_{iL}x_{iL}-F_{iR}x_{iR})\right)\end{equation}
 in the commutative limit, where $F_{iL}=\frac{1}{2}\epsilon_{ijk}F_{jL\, kL}$
etc. Comparing the second term with $F_{\mu\nu}F_{\mu\nu}$, we see
that the zero mode of the Laplace operator $\dd_{\mu}\dd_{\mu}$ can
produce a contribution that cancels the corresponding contribution
from $F_{\mu\nu}F_{\mu\nu}$, but that all higher modes are smaller
by at least a factor of $2(\frac{1}{2}-\dd_{\mu}\dd_{\mu})$. Therefore,
the action (\ref{action-so6}) is positive definite except for the
obvious zero mode $\d{B}_{i}^{L}=\epsilon,\;\;\d{B}_{i}^{R}=-\epsilon$.
This means that the geometry of $S_{L}^{2}\times S_{R}^{2}$ is locally
stable even with the $SO(6)$-symmetry unbroken, except for opposite
fluctuations of the radii.

\subsection{Breaking $SO(6)\rightarrow SO(3)\times SO(3)$}

To obtain the original action (\ref{action}) for $S^{2}\times S^{2}$,
we have to break the $SO(6)$-symmetry down to $SO(3)\times SO(3)$.
We can do this by using the left and right gauge fields $C^{L}$ and
$C^{R}$ introduced in (\ref{C-def}) separately. Their squares are
\begin{eqnarray}
C_{L}^{2} & = & B_{iL}B_{iL}+\frac{1}{4}+\left(\begin{array}{cc}
\gamma_{L}^{i} & 0\\
0 & \gamma_{L}^{i}\end{array}\right)(B_{iL}+i\epsilon_{ijk}B_{jL}B_{kL}),\\
C_{R}^{2} & = & B_{iR}B_{iR}+\frac{1}{4}-i\:\left(\begin{array}{cc}
\gamma_{R}^{i} & 0\\
0 & \gamma_{R}^{i}\end{array}\right)(B_{iR}+i\epsilon_{ijk}B_{jR}B_{kR}).\nonumber \end{eqnarray}
 As both $\gamma_{L}^{i},\:\gamma_{R}^{i}$ and $\gamma_{L}^{i}\gamma_{R}^{j}$
are traceless, we have\begin{eqnarray}
S_{\mathrm{break}} & := & 2\mathrm{Tr}((C_{L}^{2}-\frac{N_{L}^{2}}{4})(C_{R}^{2}-\frac{N_{R}^{2}}{4}))\\
 & = & 16\mathrm{Tr}((B_{iL}B_{iL}-\frac{N_{L}^{2}-1}{4})(B_{iR}B_{iR}-\frac{N_{R}^{2}-1}{4})).\nonumber \end{eqnarray}
 With these terms we can recover our action as \begin{eqnarray}
S & = & S_{6}-S_{\mathrm{break}}=\mathrm{Tr}\big((C^{2}-\frac{N^{2}}{2})^{2}-2(C_{L}^{2}-\frac{N_{L}^{2}}{4})(C_{R}^{2}-\frac{N_{R}^{2}}{4})\big)\label{eq: action S_6 - S_break}\\
 & = & 8\,\mathrm{tr}\,\big((B_{iL}B_{iL}-\frac{N_{L}^{2}-1}{4})^{2}+(B_{iR}B_{iR}-\frac{N_{R}^{2}-1}{4})^{2}+\frac{1}{2}F_{\mu\nu}F_{\mu\nu}\big),\nonumber \end{eqnarray}
 which is precisely the action (\ref{action}) for gauge theory on
$S_{N_{L}}^{2}\times S_{N_{R}}^{2}$ omitting the overall constants.
Hence the action is formulated as a 2-matrix model, however with highly
constrained matrices $C_{L},C_{R}$. This formulation using the Gamma-matrices
is very natural and useful if one wants to couple the gauge fields
to fermions, as discussed in chapter \ref{sec:fermions}.

For simplicity, we will only consider $N_{L}=N_{R}=N$ from now on.

\section{Quantization}

The quantization of the gauge theory defined by (\ref{action}) or
its reformulation (\ref{eq: action S_6 - S_break}) is straightforward
in principle, by a path integral over the Hermitian matrices \begin{equation}
Z[J]=\int dB_{\mu}e^{-S[B_{\mu}]+\mathrm{tr}B_{\mu}J_{\mu}}.\label{ZJ}\end{equation}
 Note that there is no need to fix the gauge since the gauge group
$U(\cN)$ is compact. The above path integral is well-defined and
finite for any fixed $\cN$. To see this, it is enough to show that
the integral\begin{equation}
\int dB_{\mu}e^{-(B_{i}^{L}B_{i}^{L}-\frac{N^{2}-1}{4})^{2}-(B_{i}^{R}B_{i}^{R}-\frac{N^{2}-1}{4})^{2}}\end{equation}
 converges, since the contributions from the field strength further
suppress the integrand. This integral is obviously convergent for
any fixed $N$.

For perturbative computations it is necessary to fix the gauge, and
to substitute gauge invariance by BRST-invariance. Such a gauge-fixed
action will be presented next.

\subsection{BRST Symmetry}

To construct a gauge-fixed BRST-invariant action, we have to introduce
ghost fields $c$ and anti-ghost fields $\bar{c}$. These are fermionic
fields, more precisely $\cN\times\cN-$ matrices with entries which
are Grassman variables.

The full gauge-fixed action reads: \begin{equation}
S_{\mathrm{BRST}}=S+\frac{1}{\cN}\mathrm{tr}\big(\bar{c}[\lambda_{\mu},[B_{\mu},c]]-(\frac{\alpha}{2}b-[\lambda_{\mu},B_{\mu}])b\big)\,,\end{equation}
 where $b$ is an auxiliary (Nakanishi-Lautrup) field. This action
is invariant with respect to the following BRST-transformations: \begin{eqnarray}
sB_{\mu}=[B_{\mu},c] &  & sc=cc\label{eq: BRST-trafos}\\
s\bar{c}=b &  & sb=0\nonumber \end{eqnarray}
 (matrix product is understood), where the BRST-differential $s$
acts on a product of fields as follows: \begin{equation}
s(XY)=X(sY)+(-1)^{\varepsilon_{Y}}(sX)Y\,.\end{equation}
 Here $\varepsilon_{Y}$ denotes the Grassman-parity of $Y$ \begin{equation}
\varepsilon_{Y}=\left\{ \begin{array}{cc}
0 & Y\,\textrm{bosonic}\\
1 & Y\,\textrm{fermionic}\,.\end{array}\right.\end{equation}
 It is not difficult to check that these BRST-transformations are
indeed nilpotent, i.e. \begin{equation}
s^{2}=0\,.\end{equation}
 Integrating out the auxiliary field $b$ leads to the following action
\begin{equation}
S'_{\mathrm{BRST}}=S+\frac{1}{\cN}\mathrm{tr}\big(\bar{c}[\lambda_{\mu},[B_{\mu},c]]-\frac{1}{2\alpha}[\lambda_{\mu},B_{\mu}][\lambda_{\nu},B_{\nu}]\big)\,.\end{equation}
 Setting $\alpha=1$ corresponds to the Feynman gauge. This is indeed
what one would obtain by the Faddeev-Popov procedure. The action $S'$
is invariant with respect to the following operations: \begin{eqnarray}
s'B_{\mu} & = & [B_{\mu},c]\\
s'c & = & cc\nonumber \\
s'\bar{c} & = & [\lambda_{\mu},B_{\mu}]\,.\nonumber \end{eqnarray}
 Since we have used the equations of motion of $b$, the BRST-differential
$s'$ is \emph{not} nilpotent off-shell anymore, but we still have
\begin{equation}
s'^{2}|_{\textrm{on}-\textrm{shell}}=0\,.\end{equation}

\section{Fermions}

\label{sec:fermions}To introduce spinors on fuzzy $S^{2}\times S^{2}$,
we will first have to have a look at the commutative case. There,
we will calculate the Dirac operator and bring it into a form which
is more suitable for the fuzzy case. The formulation of fuzzy gauge
theory using the $SO(6)$-Clifford algebra will proove very usefull,
and the fuzzy Dirac operator will be a simple generalization of the
commutative one. But this Dirac operator (because it is based on $SO(6)$
instead of $SO(3)\times SO(3)$) will be reducible, which is why we
will have to introduce projectors onto the physical Dirac fermions.
Chirality can be introduced either using the chirality operator inherited
from $SO(6)$ or using a Ginsparg-Wilson system.

\subsection{The commutative Dirac operator on $S^{2}\times S^{2}$}

To find a form of the commutative Dirac operator on $S^{2}\times S^{2}$
which is suitable for the fuzzy case, one can generalize the approach
of \cite{Grosse:1995pr} for $S^{2}$, which is carried out in detail
in Appendix \ref{sec:so6-dirac-4}: One can write the flat $SO(6)$
Dirac operator $D_{6}$ in 2 different forms, using the usual flat
Euclidean coordinates and also using the spherical coordinates of
the spheres. Then one can relate $D_{6}$ with the curved four-dimensional
Dirac operator $D_{4}$ on $S^{2}\times S^{2}$ in the same spherical
coordinates. This leads to an explicit expression for $D_{4}$ involving
only the angular momentum generators, which is easy to generalize
to the fuzzy case. In terms of these tangential derivatives $J_{\mu},$
the result becomes the simple expression\begin{equation}
D_{4}=\Gamma^{\mu}J_{\mu}+\left(\begin{array}{cc}
0 & 1\\
1 & 0\end{array}\right)+i\left(\begin{array}{cc}
0 & 1\\
-1 & 0\end{array}\right)=\Gamma^{\mu}J_{\mu}+2C_{0},\label{Dirac operator}\end{equation}
 which is clearly a $SO(3)\times SO(3)$-covariant Hermitian first-oder
differential operator. Here $\Gamma^{\mu}$ generate the $SO(6)$
Clifford algebra (\ref{clifford-so6}), $C_{0}$ is defined in (\ref{B0-R}),
and we put $R=1$ for simplicity here. However this Dirac operator
is reducible, acting on 8-dimensional spinors $\Psi_{8}$ corresponding
to the $SO(6)$ Clifford algebra. Hence $\Psi_{8}$ should be a combination
of two independent 4-component Dirac spinors on the 4-dimensional
space $S^{2}\times S^{2}$. To see this, we will construct explicit
projectors projecting onto these 4-dimensional spinors, and identify
the appropriate 4-dimensional chirality operators. This will provide
us with the desired physical Dirac or Weyl fermions.

\subsection{Chirality and projections for the spinors}

\label{chapter on projections}

There are 3 obvious operators which anti-commute with $D_{4}$. One
is the usual 6-dimensional chirality operator \begin{equation}
\G:=i\G_{1}^{L}\G_{2}^{L}\G_{3}^{L}\G_{1}^{R}\G_{2}^{R}\G_{3}^{R}=\left(\begin{array}{cc}
-1 & 0\\
0 & 1\end{array}\right),\label{chiral-G-op}\end{equation}
 which satisfies \begin{equation}
\{ D_{4},\G\}=0,\qquad\G^{\dagger}=\G,\qquad\G^{2}=1.\label{G-chiral}\end{equation}
 The 8-component spinors $\Psi_{8}$ split accordingly into two 4-component
spinors $\Psi_{8}=\left(\begin{array}{c}
\psi_{\a}\\
\obar\psi_{\obar{\b{}}}\end{array}\right)$, which transform as $4$ resp. $\obar{4}$ under $so(6)\cong su(4)$;
recall the related discussion in chapter \ref{so6 formulation}. The
other operators of interest are \begin{equation}
\chi_{L}=\Gamma^{iL}x_{iL}\;\;\;\mbox{and}\;\;\;\chi_{R}=\Gamma^{iR}x_{iR}.\end{equation}
 They preserve $SO(3)\times SO(3)\subset SO(6)$, and satisfy \begin{equation}
\{ D_{4},\chi_{L,R}\}=0=\{\chi_{L},\chi_{R}\}\end{equation}
 as well as \begin{equation}
\chi_{L,R}^{2}=1.\end{equation}
 We will also use \begin{equation}
\chi=\frac{1}{\sqrt{2}}\;\Gamma^{\mu}x_{\mu}=\frac{1}{\sqrt{2}}\;(\chi_{L}+\chi_{R})\label{chiralityop}\end{equation}
 which satisfies similar relations. This means that \begin{equation}
P_{\pm}=\frac{1}{2}(1\pm i\chi_{L}\chi_{R})\label{projector}\end{equation}
 with \begin{equation}
P_{\pm}^{2}=P_{\pm},\;\;\;\; P_{+}+P_{-}=1\;\;\;\;\mbox{and}\;\;\;\; P_{+}P_{-}=0\label{properties of p+-}\end{equation}
 are Hermitian projectors commuting with the Dirac operator on $S^{2}\times S^{2}$
as well as with $\G$, \begin{equation}
P_{\pm}^{\dagger}=P_{\pm}\;\;\;\;\mbox{and}\;\;\;\;[P_{\pm},D_{4}]=[P_{\pm},\G]=0.\label{P-properties}\end{equation}
 Therefore they project onto subspaces which are preserved by $D_{4}$
and $\G$. Hence the spinor Lagrangian can be written as \begin{equation}
\Psi_{8}^{\dagger}D_{4}\Psi_{8}=\Psi_{+}^{\dagger}D_{4}\Psi_{+}+\Psi_{-}^{\dagger}D_{4}\Psi_{-}\end{equation}
 involving two Dirac spinors $\Psi_{\pm}=P_{\pm}\Psi_{8}$. In order
to get one 4-component Dirac spinor, we can e.g. impose the constraint
\begin{equation}
P_{+}\Psi_{8}=\Psi_{8},\label{constraint}\end{equation}
 or equivalently give one of the two components a large mass, by adding
a term \begin{equation}
M\,\Psi_{8}^{\dagger}P_{-}\Psi_{8}\end{equation}
 to the action with $M\rightarrow\infty$. The physical chirality
operator is now identified using (\ref{P-properties}) and (\ref{G-chiral})
as $\G$ acting on $\Psi_{+}$. It can be used to define 2-component
Weyl spinors on $S^{2}\times S^{2}$.

To make the above more explicit, consider the a pole of the spheres,
i.e. \begin{equation}
x_{L}=\left(\begin{array}{c}
1\\
0\\
0\end{array}\right)\;\;\;\;\mbox{and}\;\;\;\; x_{R}=\left(\begin{array}{c}
1\\
0\\
0\end{array}\right).\end{equation}
 In the basis (\ref{Gamma}) for the Clifford algebra we then get
explicitly \begin{equation}
P_{\pm}=\frac{1}{2}(1\pm i\left(\begin{array}{cc}
-\gamma_{L}^{1}\gamma_{R}^{1} & 0\\
0 & \gamma_{L}^{1}\gamma_{R}^{1}\end{array}\right))=\frac{1}{2}(1\pm\sigma_{3}\otimes\sigma_{3}\otimes\sigma_{3}).\end{equation}
 This means that \begin{equation}
P_{+}=\mathrm{diag}(1,0,0,1,0,1,1,0)\end{equation}
 projects onto a 4-dimensional subspace exactly as expected.

\subsection{Gauged fuzzy Dirac and chirality operators}

To find a fuzzy analogue of the Dirac operator (\ref{Dirac operator})
coupled to the gauge fields, we recall the connection between the
gauge theory on $S^{2}\times S^{2}$ and the $SO(6)$ Gamma matrices
established in chapter \ref{so6 formulation}. In the spirit of that
chapter a natural fuzzy spinor action would involve \begin{equation}
\Psi^{\dagger}C\Psi,\label{try}\end{equation}
 where $\Psi$ is now a $8\cN\times\cN$-matrix (with Grassman entries).
Of course, (\ref{try}) does not have the appropriate commutative
limit, but we can split $C$ into a fuzzy Dirac operator $\widehat{D}$
and the operator $\widehat{\chi}$ defined by \begin{equation}
\widehat{\chi}\Psi=\frac{\sqrt{2}}{N}\;(\Gamma^{\mu}\Psi\lambda_{\mu}-C_{0}\Psi),\label{fuzzy chirality operator}\end{equation}
 which generalizes (\ref{chiralityop}); here we used the definition
(\ref{B0-l},\ref{B0-R}) of $C_{0}$. This operator satisfies \begin{equation}
\widehat{\chi}^{2}=1,\end{equation}
 and reduces to (\ref{chiralityop}) in the commutative limit. Note
also that $\widehat{\chi}$ commutes with gauge transformations, since
the coordinates $\lambda_{\mu}$ are acting from the right in (\ref{fuzzy chirality operator}).
Setting \begin{equation}
\widehat{J}_{\mu}\Psi=[\lambda_{\mu},\Psi],\end{equation}
 we get for the fuzzy Dirac operator \begin{equation}
\widehat{D}=C-\frac{N}{\sqrt{2}}\;\widehat{\chi}=\Gamma^{\mu}(\widehat{J}_{\mu}+A_{\mu})+2C_{0}=\Gamma^{\mu}{\mathcal{D}}_{\mu}+2C_{0}.\label{fuzzy dirac operator}\end{equation}
 Here%
\footnote{We set $R=1$ in this chapter for simplicity.%
}\begin{equation}
\widehat{\mathcal{D}}_{\mu}:=\widehat{J}_{\mu}+A_{\mu}\end{equation}
 is a covariant derivative operator, i.e. $\widehat{\mathcal{D}}{}_{\mu}\psi\rightarrow U\widehat{\mathcal{D}}_{\mu}\psi$
which is easily verified using (\ref{gaugetrafo-A}). This $\widehat{D}$
clearly has the correct commutative limit (\ref{Dirac operator})
for vanishing $A$, and the gauge fields are coupled correctly. In
particular, this definition of $\widehat{D}$ applies also to the
topologically non-trivial solutions of chapter \ref{se: monopoles}
without any modifications. Moreover, the chirality operator $\G$
as defined in (\ref{chiral-G-op}) anti-commutes with $\widehat{D}$
also in the fuzzy case, \begin{equation}
\{\widehat{D},\G\}=0.\end{equation}
 Furthermore, using some identities given at the beginning of chapter
\ref{so6 formulation} we obtain for $\widehat{D}^{2}\psi$: \begin{eqnarray}
\widehat{D}^{2}\psi & = & (\Sigma^{\mu\nu}F_{\mu\nu}+\widehat{\mathcal{D}}_{\mu}\widehat{\mathcal{D}}_{\mu}+\{\Gamma^{\mu},C_{0}\}\widehat{\mathcal{D}}_{\mu}+2)\psi\label{Dirac-squared}\\
 & =: & (\Sigma^{\mu\nu}F_{\mu\nu}+\widehat{\square}+2)\psi,\nonumber \end{eqnarray}
 defining the covariant 4-dimensional Laplacian $\widehat{\square}$
acting on the spinors. This corresponds to the usual expression for
$\widehat{D}^{2}$ on curved spaces, and the constant 2 is due to
the curvature scalar. Since $\widehat{D}^{2}$ and $\Sigma^{\mu\nu}F_{\mu\nu}$
are both Hermitian and commute with $\G$ and $\widehat{P}_{\pm}$
as defined in (\ref{projection operator 1}) in the large $N$ limit,
it follows that $\widehat{\square}$ satisfies these properties as
well.

\subsection{Projections for the fuzzy spinors}

\label{chapter on fuzzy projections}

For the fuzzy case, we can again consider the following operators
\begin{eqnarray}
\widehat{\chi}_{L}\Psi & = & \frac{2}{N}(\Gamma^{iL}\Psi\lambda_{iL}+C_{0}^{L}\Psi),\\
\widehat{\chi}_{R}\Psi & = & \frac{2}{N}(\Gamma^{iR}\Psi\lambda_{iR}+C_{0}^{R}\Psi)\nonumber \end{eqnarray}
 which satisfy \begin{equation}
\widehat{\chi}_{L,R}^{2}=1,\qquad\{\widehat{\chi}_{L},\widehat{\chi}_{R}\}=0\,.\end{equation}
 This implies $(\widehat{\chi}_{L}\widehat{\chi}_{R})^{2}=-1$, and
we can write down the projection operators \begin{equation}
\widehat{P}_{\pm}=\frac{1}{2}(1\pm i\widehat{\chi}_{L}\widehat{\chi}_{R})\label{projection operator 1}\end{equation}
 which have the commutative limit (\ref{projector}) and the properties
(\ref{properties of p+-}). However, the projector no longer commutes
with the fuzzy Dirac operator (\ref{fuzzy dirac operator}): \begin{eqnarray}
{}[\widehat{D},\widehat{\chi}_{L}\widehat{\chi}_{R}] & = & \{\widehat{D},\widehat{\chi}_{L}\}\widehat{\chi}_{R}-\widehat{\chi}_{L}\{\widehat{D},\widehat{\chi}_{R}\}\\
 & = & -\frac{2}{N}\Big(\big(2(\lambda_{iL}+A_{iL})\widehat{J}_{iL}-2A_{iL}\lambda_{iL}+2C_{0}^{L}\;\Gamma^{iL}\widehat{\mathcal{D}}_{iL}+1\big)\widehat{\chi}_{R}\nonumber \\
 &  & -\widehat{\chi}_{L}\big(2(\lambda_{iR}+A_{iR})\widehat{J}_{iR}-2A_{iR}\lambda_{iR}+2C_{0}^{R}\;\Gamma^{iR}\widehat{\mathcal{D}}_{iR}+1\big)\Big)\,,\nonumber \end{eqnarray}
 which only vanishes for $N\rightarrow\infty$ and tangential $\mc A_{\mu}$
(\ref{A-constraint}). To reduce the degrees of freedom to one Dirac
4-spinor, we should therefore add a mass term \begin{equation}
M\,\Psi_{8}^{\dagger}\widehat{P}_{-}\Psi_{8}\label{M-constraint-fuzzy}\end{equation}
 which for $M\rightarrow\infty$ suppresses one of the spinors, rather
than impose an exact constraint as in (\ref{constraint}). This is
gauge invariant since $\widehat{P}_{\pm}$ commutes with gauge transformations,
\begin{equation}
\widehat{P}_{\pm}\psi\rightarrow U\widehat{P}_{\pm}\psi\,.\end{equation}
 The complete action for a Dirac fermion on fuzzy $S_{N}^{2}\times S_{N}^{2}$
is therefore given by \begin{equation}
S_{\mathrm{Dirac}}=\int\Psi_{8}^{\dagger}(\widehat{D}+m)\Psi_{8}+M\,\Psi_{8}^{\dagger}\widehat{P}_{-}\Psi_{8}\end{equation}
 with $M\rightarrow\infty$. The physical chirality operator is given
by $\G$ (\ref{chiral-G-op}), which allows to consider Weyl spinors
as well.

\subsection{The Ginsparg-Wilson relations}

There is an alternative approach to introduce chirality on fuzzy spaces,
using the Ginsparg-Wilson relations. These were initially designed
to study chiral fermions on the lattice \cite{Ginsparg:1981bj}, but
they proved to be applicable to fuzzy fermions as well \cite{Balachandran:2000du,Balachandran:2003ay}.
On the fuzzy sphere, the Dirac and the chirality operator can be cast
into a form in which they fulfill these relations. This makes it possible
to study issues such as topological properties and index theory \cite{Aoki:2002fq,Ydri:2002nt}.
We will see that the same relations can be formulated for our model,
too.

A Ginsparg-Wilson system consists of two involutions $\Gamma$ and
$\Gamma'$, i.e.\begin{equation}
\Gamma^{2}=1\;;\;\Gamma^{\dagger}=\Gamma\;\;\;\;\mbox{and}\;\;\;\;\Gamma'^{2}=1\;;\;\Gamma'^{\dagger}=\Gamma'.\end{equation}
In our case, these two involutions are defined as two different noncommutative
versions of chirality, one acting from the left, the other one acting
from the right\begin{eqnarray}
\Gamma\Psi & = & \frac{\sqrt{2}}{N}(\Gamma^{\mu}\lambda_{\mu}+C_{0})\Psi,\\
\Gamma'\Psi & = & \frac{\sqrt{2}}{N}(\Gamma^{\mu}\Psi\lambda_{\mu}-C_{0}\Psi).\end{eqnarray}
We recognize $\Gamma'$ as the fuzzy operator (\ref{fuzzy chirality operator}).
But also $\Gamma$ has the commutative operator (\ref{chiralityop})
as its limit.

In the Ginsparg-Wilson system, the Dirac operator was initially defined
to be\begin{equation}
d=\frac{1}{a}\Gamma(\Gamma-\Gamma'),\end{equation}
 where $a$ is the lattice spacing, but here we will choose\begin{equation}
D=\frac{N}{2}\sqrt{2}(\Gamma-\Gamma'),\end{equation}
as this reproduces our fuzzy Dirac operator (\ref{fuzzy dirac operator})
(with gauge fields switched off). We can now define an alternative
chirality operator\begin{equation}
\chi=\frac{1}{2}(\Gamma+\Gamma').\end{equation}
It fulfills\begin{eqnarray}
\{ D,\chi\} & = & 0,\\
2N^{2}\chi^{2}+D^{2} & = & 2N^{2}.\nonumber \end{eqnarray}
 Therefore $\chi$ exactly anticommutes with $D$, but it vanishes
on the top modes of $D$, i.e. for $|D|=\sqrt{2}N$. But at least
for every eigenstate $\Psi_{E}$ with positive eigenvalue $E<\sqrt{2}N$\begin{equation}
D\Psi_{E}=E\Psi,\end{equation}
the ungauged fuzzy Dirac operator has also an eigenstate $\Psi_{-E}=\chi\Psi_{E}$
with the negative eigenvalue $-E$ because of\begin{equation}
D\Psi_{-E}=D\chi\Psi=-\chi D\Psi=-\chi E\Psi=-E\Psi_{-E}.\end{equation}
This can be used \cite{Aoki:2002fq} to derive the following index
theorem for $D$ \begin{equation}
Ind(D)=n_{+}-n_{-}=Tr(\chi).\end{equation}
To include gauge fields, we can write\begin{equation}
\Gamma_{A}=\frac{\sqrt{2}}{N}(\Gamma^{\mu}(\lambda_{\mu}+A_{\mu})+C_{0})=\frac{\sqrt{2}}{N}C.\end{equation}
 With \begin{eqnarray}
D_{A} & = & \frac{N}{2}\sqrt{2}(\Gamma_{A}-\Gamma'),\\
\chi_{A} & = & \frac{1}{2}(\Gamma_{A}+\Gamma')\end{eqnarray}
we now get\begin{eqnarray}
\{ D,\chi\} & = & \frac{N}{2}\sqrt{2}(\Gamma_{\mc A}^{2}-1)\\
 & = & \frac{N}{2}\sqrt{2}(\frac{2}{N^{2}}(B_{\mu}B_{\mu}+\frac{1}{2}+\Sigma_{8}^{\mu\nu}F_{\mu\nu})-1)\nonumber \\
 & = & \frac{\sqrt{2}}{N}(B_{\mu}B_{\mu}+\frac{N^{2}-1}{2}+\Sigma_{8}^{\mu\nu}F_{\mu\nu}),\nonumber \end{eqnarray}
which corresponds exactly to the result of \cite{Ydri:2002nt} for
the fuzzy sphere. Other results of \cite{Ydri:2002nt} are therefore
expected to hold in our case, too.

Alternatively, the gauge fields could also be introduced in a way
that is closer to the Ginsparg-Wilson setting by normalizing $\Gamma_{\mc A}$.

\section{Topologically non-trivial solutions on $S_{N}^{2}\times S_{N}^{2}$}

\label{se: monopoles} We will now go back to pure gauge theory on
$S_{N}^{2}\times S_{N}^{2}$, looking for non-trivial solutions of
the equations of motion (\ref{EOM}). We will find that the theory
is rich in topological solutions, some corresponding to monopoles
on the commutative limit $S^{2}\times S^{2}$, others corresponding
to the fluxon solutions found on the second limit $\mathbb{R}_{\theta}^{4}$. 

In order to understand better the non-trivial solutions found below,
we first note that the classical space $S^{2}\times S^{2}$ is symplectic
with symplectic form \begin{equation}
\omega=\omega^{L}+\omega^{R},\end{equation}
 where \begin{equation}
\omega^{L}=\frac{1}{4\pi R^{3}}\epsilon_{ijk}x_{i}^{L}dx_{j}^{L}dx_{k}^{L}\end{equation}
 and similarly $\omega^{R}$. The normalization is chosen such that
\begin{equation}
\int_{S_{L,R}^{2}}\omega^{L,R}=1=\int_{S^{2}\times S^{2}}\omega^{L}\wedge\omega^{R}\end{equation}
 so that $\omega^{L},\omega^{R}$ generate the integer cohomology
$H^{*}(S^{2}\times S^{2},\Z)$. Noting that $\omega$ is self-dual
while $\tilde{\omega}:=\omega^{L}-\omega^{R}$ is anti-selfdual, it
follows immediately that both $F=2\pi\omega$ and $F=2\pi\tilde{\omega}$
are solutions of the Abelian field equations. More generally, any
\begin{equation}
F^{(m_{L},m_{R})}=2\pi m_{L}\omega^{L}+2\pi m_{R}\omega^{R}\end{equation}
 for any integers $m_{L},m_{R}$ is a solution. In bundle language,
they correspond to products of 2 monopole bundles with connections
and monopole number $m_{L,R}$ over $S_{L,R}^{2}$. Following the
literature we will denote any such non-trivial solution as instanton.

\subsection{Instantons and fluxons}

We are interested in similar non-trivial solutions of the EOMs (\ref{EOM})
in the fuzzy case. The monopole solutions on the fuzzy sphere $S_{N}^{2}$
are given by \reps $\lambda_{i}^{N-m}$ of $su(2)$ of size $N-m$
\cite{Karabali:2001te}, which lead to the classical monopole gauge
fields in the commutative limit as shown in \cite{Steinacker:2003sd}.
It is hence easy to guess that we will obtain solutions on $S_{N}^{2}\times S_{N}^{2}$
by taking products of these: \begin{eqnarray}
B_{i}^{L} & = & \alpha^{L}\;\lambda_{i}^{N-m_{L}}\otimes\one_{N-m_{R}},\label{monopoles L}\\
B_{i}^{R} & = & \alpha^{R}\;\one_{N-m_{L}}\otimes\lambda_{i}^{N-m_{R}}\label{monopole-ansatz}\end{eqnarray}
 where $\lambda_{i}^{N-m_{L,R}}$ are the $N-m_{L,R}$ dimensional
generators of $su(2)$. It is not difficult to verify that these are
solutions of (\ref{EOM}) with $\alpha^{L,R}=1+\frac{m_{L,R}}{N}$
for $m_{L,R}\ll N$, with field strength \begin{equation}
F_{iLjL}=-\frac{m^{L}}{2R^{3}}\epsilon_{ijk}x_{k}^{L},\quad F_{iRjR}=-\frac{m^{R}}{2R^{3}}\epsilon_{ijk}x_{k}^{R},\quad F_{iLjR}=0,\label{monopole-field}\end{equation}
 while $B\cdot B-\frac{N^{2}-1}{4}\rightarrow0$ as $N\rightarrow\infty$.
This means that $F=-2\pi m^{L}\omega^{L}-2\pi m^{R}\omega^{R}$ in
the commutative limit, so that indeed \begin{equation}
\int\limits _{S_{2}^{L,R}}\frac{F}{2\pi}=-m^{L,R}.\label{monopole-flux}\end{equation}
 Notice that the Ansatz (\ref{monopole-ansatz}) implies that all
matrices have size $\cN=(N-m_{L})(N-m_{R})$, which is inconsistent
if we require that $\cN=N^{2}$ in order to have the original $S_{N}^{2}\times S_{N}^{2}$
vacuum. Therefore it appears that these solutions live in a different
configuration space, similar as the commutative monopoles which live
on different bundles. However, the situation is in fact more interesting:
the above solutions can be embedded in the \emph{same} configuration
spaces of $N^{2}\times N^{2}$ matrices as the vacuum solution if
we combine them with other solutions, which have finite action in
four dimensions%
\footnote{as opposed to 2 dimensions, where their action goes to infinity for
$N\rightarrow\infty$.%
}. They are in fact crucial to recover some of the known $U(1)$ instantons
in the limit $S_{N}^{2}\rightarrow\mathbb{R}_{\theta}^{2}$ resp.
$S_{N}^{2}\times S_{N}^{2}\rightarrow\mathbb{R}_{\theta}^{4}$, as
we will see. Consider the following Ansatz \begin{equation}
B_{i}^{L,R}=\mathrm{diag}(d_{i,1}^{L,R},...,d_{i,n}^{L,R})\end{equation}
 in terms of diagonal matrices (ignoring the size of the matrices
for the moment). These are solutions of (\ref{EOM}) in two cases,
\begin{equation}
\sum_{i}d_{i,k}^{L,R}d_{i,k}^{L,R}=\left\{ \begin{array}{ll}
\frac{N^{2}-3}{4}, & \mbox{typeA}\\
0, & \mbox{typeB}\end{array}\right.\label{singular-solutionsAB}\end{equation}
 (i.e. $d_{i,k}^{L,R}=0$ in type B). The associated field strength
is \begin{equation}
F_{iLjL}=\frac{\epsilon_{ijk}}{R^{2}}\;\mathrm{diag}(d_{k,1}^{L},...,d_{k,n}^{L}),\quad F_{LR}=0,\end{equation}
 and a similar formula for $F_{iRjR}$. The constraint term is then
$(B\cdot B-\frac{N^{2}-1}{4})\rightarrow-\frac{1}{2}$ for type A,
and $(B\cdot B-\frac{N^{2}-1}{4})\rightarrow-\frac{N^{2}-1}{4}$ for
type B in the large $N$ limit. In particular, only the type A solutions
will have a finite contribution \begin{equation}
S_{\mathrm{fluxon}}=\frac{V}{g^{2}\cN}\left(\frac{n}{4R^{4}}+\frac{2n}{R^{4}}\frac{N^{2}-3}{4}\right)\;\;\rightarrow\frac{8\pi^{2}}{g^{2}}\; n\label{defect-action}\end{equation}
 to the action%
\footnote{A finite action can also be obtained for the type $B$ solution using
a slightly modified action (\ref{action-2}), as discussed below.%
}, which for $N\rightarrow\infty$ is only due to the field strength.
We will see below that these type A solutions can be interpreted as
a localized flux or vortex, and we will call them fluxons since they
will lead in the scaling limit to solutions on $\R_{\theta}^{4}$
which we denoted as such \cite{Polychronakos:2000zm,Gross:2000ph,Harvey:2000jb}.

One can now combine these fluxon solutions with the monopole solutions
(\ref{monopole-ansatz}) in the form \begin{eqnarray}
B_{i}^{L} & = & \left(\begin{array}{cc}
\alpha^{L}\;\lambda_{i}^{N-m_{L}}\tens\one_{N-m_{R}} & 0\\
0 & \mathrm{diag}(d_{i,1}^{L},...,d_{i,n}^{L})\end{array}\right),\label{eq: R instantons on S2 S2}\\
B_{i}^{R} & = & \left(\begin{array}{cc}
\alpha^{R}\;\one_{N-m_{L}}\otimes\lambda_{i}^{N-m_{R}} & 0\\
0 & \mathrm{diag}(d_{i,1}^{R},...,d_{i,n}^{R})\end{array}\right).\nonumber \end{eqnarray}
 These are now matrices of size $\cN=(N-m_{L})(N-m_{R})+n$, which
must agree with $\cN=N^{2}$. This is clearly possible for \begin{equation}
m_{L}=-m_{R}=m,\quad n=m^{2},\label{m-constraints}\end{equation}
 while for $m_{L}\neq-m_{R}$ the contribution from the fluxons would
be infinite since $n$ would be of order $N$. To understand these
solutions, we can compute the gauge field from (\ref{B-A-relation}),
\begin{equation}
A_{i}^{L}=\frac{1}{R}\left(B_{i}^{L}-\la_{i}^{N}\tens\one_{N}\right)=A_{i}^{L}(x^{L},x^{R}).\end{equation}
 To evaluate this, we first have to choose a gauge, i.e. a unitary
transformation $U$ for (\ref{eq: R instantons on S2 S2}) which allows
to express e.g. $\la_{i}^{N-m_{L}}\tens\one_{N-m_{R}}$ in terms of
$x_{i}^{L}\propto\la_{i}^{N}\tens\one_{N}$ and $x_{i}^{R}\propto\one_{N}\tens\la_{i}^{N}$.
For example, in the case $m_{L}=-m_{R}=m$ this can be done using
a unitary map \begin{equation}
U:\;\;\C^{N-m}\tens\C^{N+m}\oplus\C^{m^{2}}\rightarrow\C^{N}\tens\C^{N},\end{equation}
 mapping a $(N-m)\times(N+m)$ matrix into a $N\times N$ matrix by
trivially matching the upper-left corner in the obvious way, and fitting
$\C^{m^{2}}$ into the remaining lower-right corner. With this being
understood, one can write \begin{eqnarray}
RA_{i}^{L}(x^{L},x^{R}) & = & (\a^{L}\la_{i}^{N-m}-\la_{i}^{N})\tens\one_{N+m}\\
 &  & +\la_{i}^{N}\tens(\one_{N+m}-\one_{N})+\mbox{(d-terms)}\nonumber \\
 & = & A_{i}^{(m_{L})}(x^{L})\;\;+\;\mathrm{sing}(x_{3}^{L}=-R,x_{3}^{R}=-R)\nonumber \end{eqnarray}
 where $A_{i}^{(m_{L})}(x^{L})$ is indeed the gauge field of a monopole
with charge $m$ on $S_{L}^{2}$ in the large $N$ limit, as was checked
explicitly in \cite{Steinacker:2003sd}. Here $\mathrm{sing}(x_{3}^{L}=-R,x_{3}^{R}=-R)$
indicates a field which is singular for large $N$ and localized at
the south pole of $S_{L}^{2}$ and $S_{R}^{2}$. It originates both
from cutting and pasting the bottom and right border of the above
matrices using $U$ (leading to singular gauge fields but regular
field strength at the south poles), as well as the $d$-block (leading
to a singular field strength). To see this recall that in general
for the standard representation (\ref{standard-rep}) of fuzzy spheres,
entries in the lower-right block of the matrices correspond to functions
localized at $x_{3}=-R$, cp. (\ref{fuzzydelta}). The gauge field
near this singularity will be studied in more detail in chapter \ref{sec:fluxon-limit}.
The field strength is \begin{equation}
F_{iLjL}=-\frac{m^{L}}{2R^{3}}\epsilon_{ijk}x_{k}^{L}+\epsilon_{ijk}\frac{1}{R^{2}}\;\sum_{i=1}^{n}\; d_{k,i}^{L}P_{i}\label{eq: F instantons on S2 S2}\end{equation}
 in the commutative limit, where $P_{i}$ are projectors in the algebra
of functions on $S_{N}^{2}\times S_{N}^{2}$ of rank $1$; recalling
(\ref{fuzzydelta}), they should be interpreted as delta-functions
$P_{i}=\frac{V}{N^{2}}\;\d{^{(4)}}(x_{3}=-R)$. Similar formulae hold
for $A_{i}^{R}(x^{L},x^{R})$ and $F_{iRjR}$, while $F_{LR}=0$.

We assumed above that these delta-functions are localized at the south
poles $x_{3}^{L}=x_{3}^{R}=-R$. However, the location of these delta-functions
can be chosen freely using gauge transformations. This can be seen
by applying suitable successive gauge transformations using $N-k$-dimensional
irreps of $SU(2)$ for $k=0,1,...,m-1$, which from the classical
point of view all correspond to global rotations, successively moving
the individual delta-peaks. Therefore the solution (\ref{eq: R instantons on S2 S2})
should in general be interpreted as a monopole on $S^{2}\times S^{2}$
with monopole number $m_{L}=-m_{R}=m$, combined with a localized
singular field strength characterized by its position and a vector
$d_{k,i}^{L}$. We will see in chapter \ref{sec:scaling} that it
becomes the fluxon solution in the planar limit $\R_{\theta}^{4}$.

The total action of these solutions (\ref{eq: R instantons on S2 S2})
is the sum of the contributions from the monopole field plus the contribution
from the fluxons (\ref{defect-action}), which both give the same
contribution \begin{equation}
S_{(m)}=\frac{4\pi^{2}}{g^{2}}\left(2m^{2}+2m^{2}\right)\label{action-fluxon-total}\end{equation}
 in the large $N$ limit, using (\ref{m-constraints}). The first
term is due to the global monopole field (\ref{monopole-field}),
and the second term is the contribution of the fluxons through the
localized field strength.

The interpretation of these solutions depends on the scaling limit
$N\rightarrow\infty$ which we want to consider. We have seen that
in the commutative limit keeping $R=\mathrm{const}$, these solutions
become commutative monopoles on $S^{2}\times S^{2}$ with magnetic
charges $m_{L}=-m_{R}$, plus additional localized fluxon degrees
of freedom. For large $R$, the field strength of the monopoles vanishes,
leaving only the localized fluxons. In particular, we will see in
the following chapter that in the scaling limit $S_{N}^{2}\times S_{N}^{2}\rightarrow\R_{\theta}^{4}$
only the fluxons survive and become well-known solutions for gauge
theory on $\R_{\theta}^{4}$. 

A final remark is in order: if we fix the size $\cN$ of the matrices,
only certain fluxon and monopole numbers are allowed, given by (\ref{m-constraints}).
Otherwise the number $n$ of fluxons and hence the action would diverge
with $N$. This can be seen as an interesting feature of our model:
viewed as a regularization of gauge theory on $\R_{\theta}^{4}$,
this points to possible subtleties of defining the admissible field
configurations in infinite-dimensional Hilbert spaces and relations
with topological terms in the action. On the other hand, we could
accommodate the most general solutions including also type B solutions
(\ref{singular-solutionsAB}) by modifying the action similar as in
\cite{Steinacker:2003sd}. For example, \begin{equation}
S=\frac{1}{g^{2}}\int\Big(\frac{4B_{i}^{L}B_{i}^{L}}{N^{2}R^{4}}(B_{i}^{L}B_{i}^{L}-\frac{N_{L}^{2}-1}{4})^{2}+\frac{4B_{i}^{R}B_{i}^{R}}{R^{4}}(B_{i}^{R}B_{i}^{R}-\frac{N_{R}^{2}-1}{4})^{2}+\frac{1}{2}F_{ia,jb}F_{ia,jb}\Big)\label{action-2}\end{equation}
 leads to the same commutative action, but with a vanishing action
for the Dirac string in the type B solutions.

\subsection{Spherical branes}

Consider the following solutions \begin{eqnarray}
B_{i}^{L} & = & \left(\begin{array}{cc}
\alpha^{L}\;\lambda_{i}^{N-m} & 0\\
0 & \mathrm{diag}(d_{i,1},...,d_{i,m})\end{array}\right)\tens\one_{N},\label{S2-branes}\\
B_{i}^{R} & = & \one_{N}\tens\lambda_{i}^{N}\nonumber \end{eqnarray}
 which are matrices of size $\cN=N^{2}$. The corresponding field
strength is \begin{eqnarray}
\quad F_{iLjL} & = & -\frac{m}{2R^{3}}\epsilon_{ijk}x_{k}^{L}+\epsilon_{ijk}\frac{1}{R^{2}}\;\sum_{i=1}^{m}\; d_{k,i}P_{i}\\
F_{RR} & = & F_{LR}=0\nonumber \end{eqnarray}
 where $P_{i}$ are projectors in the algebra of functions on $S_{L}^{2}$
of rank $1$ which should be interpreted as delta-functions $P_{i}=\frac{4\pi R^{2}}{N}\;\d{^{(2)}}(x_{3}=-R)$.
In particular the gauge field $\cA$ vanishes on $S_{R}^{2}$, while
on $S_{L}^{2}$ there is a monopole field together with a singularity
at a point. This is similar to the fluxons of the previous chapter,
but now only on $S_{L}^{2}$. This leads to the interpretation as
a 2-dimensional brane located at a point on $S_{L}^{2}$. The action
for these solutions is infinite. In the limit $S_{N}^{2}\times S_{N}^{2}\rightarrow\R_{\theta}^{4}$,
the flux will be located at a 2-dimensional hyperplane. Such solutions
for gauge theory on $\R_{\theta}^{4}$ were found in \cite{Aganagic:2000mh,Gross:2000ss},
which would be recovered in the scaling limit $S_{N}^{2}\times S_{N}^{2}\rightarrow\R_{\theta}^{4}$
.

\chapter{Gauge theory on $\mathbb{R}_{\theta}^{4}$ from $S_{N}^{2}\times S_{N}^{2}$}

\label{sec:scaling}

We saw in chapter (\ref{planelimit}) that $\mathbb{R}_{\theta}^{4}$
can be obtained as a scaling limit of fuzzy $S_{N_{L}}^{2}\times S_{N_{R}}^{2}$.
Here we will extend this scaling also to the covariant coordinates
$B_{\mu}$, thereby relating the gauge theory on $S_{N_{L}}^{2}\times S_{N_{R}}^{2}$
to that on $\mathbb{R}_{\theta}^{4}$ and hence providing a regularization
for the latter. We will in particular relate the instanton solutions
on these two spaces.

On noncommutative $\mathbb{R}_{\theta}^{2}$, all $U(1)$-instantons
were constructed and classified in \cite{Gross:2000ss}. One can indeed
recover these instantons from corresponding solutions on $S_{N}^{2}$,
as we will show below. However, since we are mainly interested in
the 4-dimensional case here, we will only present the corresponding
constructions on $S_{N_{L}}^{2}\times S_{N_{R}}^{2}$ resp. $\mathbb{R}_{\theta}^{4}$
here, without discussing the 2-dimensional case separately. It can
be recovered in an obvious way from the considerations below.

The situation on $\R_{\theta}^{4}$ is more complicated, and there
are different types of non-trivial $U(1)$ instanton solutions on
$\mathbb{R}_{\theta}^{4}$. The instantons found by solving the noncommutative
version of the ADHM equations \cite{Nekrasov:1998ss,Furuuchi:1999kv,Chu:2001cx,Hamanaka:2001dr,Ivanova:2005fh}
are hard to find in the fuzzy case, as this construction relies heavily
on selfduality, a notion which isn't naturally available in our formulation
of $S^{2}\times S^{2}$ embedded in $\mathbb{R}^{6}$. But the four-dimensional
fluxon solutions discussed in detail in chapter \ref{se: instantons}
can be recovered as scaling limits of the solutions (\ref{eq: R instantons on S2 S2})
on $S_{N_{L}}^{2}\times S_{N_{R}}^{2}$. In particular, the moduli
of the fluxon solutions on $\mathbb{R}_{\theta}^{4}$ will be related
to the free parameters $d_{i}^{L,R}$ in (\ref{eq: R instantons on S2 S2}).
This supports our suggestion to use gauge theory on $S_{N_{L}}^{2}\times S_{N_{R}}^{2}$
as a regularization for gauge theory on $\mathbb{R}_{\theta}^{4}$.

\section{The action }

\label{sec:action-scaling}

We saw in chapter \ref{planelimit} that the fuzzy space $S_{N_{L}}^{2}\times S_{N_{R}}^{2}$has
a scaling limit to $\mathbb{R}_{\theta}^{4}$, with $\theta$ cast
in the following form: \begin{equation}
\theta_{\mu\nu}=\left(\begin{array}{cccc}
0 & \theta_{12} & 0 & 0\\
-\theta_{12} & 0 & 0 & 0\\
0 & 0 & 0 & \theta_{34}\\
0 & 0 & -\theta_{34} & 0\end{array}\right)\:.\end{equation}
This scaling can also be applied to the covariant coordinates $B_{\mu}$,
connecting the gauge theory on $S_{N_{L}}^{2}\times S_{N_{R}}^{2}$to
that on $\mathbb{R}_{\theta}^{4}$ and therefore providing it with
a regularisation. For the selfdual case (i.e. $\theta_{12}>0$ and
$\theta_{34}>0$) we can define \begin{eqnarray}
X_{1,2} & := & \sqrt{\frac{2\theta_{12}}{N_{L}}}\; B_{1,2}^{L}\;,\label{eq: scaling limit1}\\
X_{3,4} & := & \sqrt{\frac{2\theta_{34}}{N_{R}}}\; B_{1,2}^{R}\;,\label{eq: scaling limit2}\\
\phi^{L,R} & := & B_{3}^{L,R}-\frac{N_{L,R}}{2}+\frac{1}{N_{L,R}}((B_{1}^{L,R})^{2}+(B_{2}^{L,R})^{2})\;,\label{eq: scaling limit3}\end{eqnarray}
The antiselfdual case ($\theta_{34}<0$) can easily be reached by
setting e.g.\begin{equation}
X_{4/3}:=\sqrt{\frac{2\theta_{34}}{N_{R}}}B_{1/2}^{R},\end{equation}
but for simplicity we will limit us to the selfdual case in the following.
The $X$ will become the covariant coordinates on $\mathbb{R}_{\theta}^{4}$
in the limit $N_{L/R}\rightarrow\infty$, and the $\phi$ an auxiliary
field. To see this we now blow up the spheres by setting\begin{equation}
R^{2}=\frac{1}{2}N_{L}\theta_{34}=\frac{1}{2}N_{R}\theta_{12}.\label{R-thetaLR-2}\end{equation}
 With this double scaling limit $R,\, N\rightarrow\infty$ keeping
$\theta$ fixed we calculate for the field strength \begin{eqnarray}
\frac{1}{R^{2}}([B_{1}^{L},B_{1}^{R}]) & = & \frac{1}{\theta_{12}\theta_{34}}[X_{1},X_{3}],\quad\mbox{etc.,}\\
\frac{1}{R^{2}}(B_{1}^{L}+i[B_{2}^{L},B_{3}^{L}]) & = & \sqrt{\frac{1}{\theta_{12}\theta_{34}R^{2}}}\;\Big(X_{1}+i[X_{2},\phi^{L}]-\frac{i}{2\theta_{12}}[X_{2},(X_{1})^{2}]\Big)\nonumber \\
\frac{1}{R^{2}}(B_{2}^{L}+i[B_{3}^{L},B_{1}^{L}]) & = & \sqrt{\frac{1}{\theta_{12}\theta_{34}R^{2}}}\;\Big(X_{2}+i[X_{1},\phi^{L}]-\frac{i}{2\theta_{12}}[X_{1},(X_{2})^{2}]\Big)\nonumber \\
\frac{1}{R^{2}}(B_{3}^{L}+i[B_{1}^{L},B_{2}^{L}]) & = & \frac{1}{\theta_{12}\theta_{34}}\Big(\theta_{12}+i[X_{1},X_{2}]\nonumber \\
 &  & \,\,\,\,\;\;\;\;\;\;\;\;\;\;\;+\frac{\theta_{12}\theta_{34}}{R^{2}}\phi_{L}-\frac{\theta_{12}\theta_{34}^{2}}{2R^{4}}((X_{1})^{2}+(X_{2})^{2})\Big)\,.\nonumber \end{eqnarray}
 Analogous expressions hold for $B_{i}^{R}$. For the potential term
we get \begin{eqnarray}
\frac{1}{R^{2}}(B_{i}^{L}B_{i}^{L}-\frac{N_{L}^{2}-1}{4}) & = & \frac{1}{\theta_{34}}\phi^{L}+\frac{2}{R^{2}}((\phi^{L})^{2}+\frac{1}{4})\\
 &  & -\frac{1}{\theta_{12}R^{2}}\{\phi^{L},(X_{1})^{2}+(X_{2})^{2}\}\nonumber \\
 &  & +\frac{1}{\theta_{12}^{2}R^{2}}((X_{1})^{2}+(X_{2})^{2})^{2}.\nonumber \end{eqnarray}
 We immediately see that the only terms from action (\ref{action})
involving $\phi^{L,R}$ are \begin{equation}
\frac{1}{\theta_{34}^{2}}(\phi^{L})^{2}+\frac{1}{\theta_{12}^{2}}(\phi^{R})^{2}+O(\frac{1}{R}),\end{equation}
 and therefore we can integrate them out in the limit $R\rightarrow\infty$.
In the leading order in $R$ the remaining terms give the standard
action \begin{equation}
S=-\frac{1}{2g^{2}\theta_{12}^{2}\theta_{34}^{2}}\int([X_{\mu},X_{\nu}]-i\theta_{\mu\nu})^{2}\end{equation}
 for a gauge theory on $\mathbb{R}_{\theta}^{4}$ for general $\theta_{\mu\nu}$.
The $X_{\mu}$ are interpreted as covariant coordinates, which can
be written as%
\footnote{we do not distinguish between upper and lower indices%
} \begin{equation}
X_{\mu}=x_{\mu}+A_{\mu}.\end{equation}
 Hence the gauge fields $A_{\mu}$ describe the fluctuations around
the vacuum. In particular, note that our regularization procedure
clearly fixes the rank of the gauge group, unlike in the naive definition
on $\R_{\theta}^{d}$ as discussed in chapter \ref{sec: matrix canonical case}.
The generalization to the $U(n)$ case is obvious.

\section{Instantons on $\mathbb{R}_{\theta}^{4}$ from $S_{N}^{2}\times S_{N}^{2}$}

\label{sec:fluxon-limit}

With the scaling limit of chapter \ref{sec:action-scaling}, the gauge
theory on $S_{N}^{2}\times S_{N}^{2}$ provides us with a regularization
for the gauge theory on $\mathbb{R}_{\theta}^{4}$. Of course, such
a regularization might affect the topological features of the theory,
an effect we want to investigate in this chapter. For this, we will
map the topologically nontrivial solutions found in chapter \ref{se: monopoles}
on $S_{N}^{2}\times S_{N}^{2}$ to $\mathbb{R}_{\theta}^{4}$.

Consider again the solutions (\ref{eq: R instantons on S2 S2}) that
combine the fluxon solutions with the monopoles, with the fluxons
at the north pole instead of the south pole because we want to study
their structure. Their scaling limit as in (\ref{eq: scaling limit1})
gives \begin{eqnarray}
X_{i} & = & \sqrt{\frac{2\theta}{N}}\left(\begin{array}{cc}
\mathrm{diag}(d_{i,1}^{L},...,d_{i,n}^{L}) & 0\\
0 & \alpha^{L}\;\lambda_{i}^{N-m}\otimes\one\end{array}\right),\label{scaled fluxons L}\\
X_{i+2} & = & \sqrt{\frac{2\theta}{N}}\left(\begin{array}{cc}
\mathrm{diag}(d_{i,1}^{R},...,d_{i,n}^{R}) & 0\\
0 & \alpha^{R}\;\one\otimes\lambda_{i}^{N+m}\end{array}\right)\label{scaled fluxons R}\end{eqnarray}
 for $i=1,2$. Recalling that the rescaled $\lambda_{1,2}$ on $S_{N_{L}}^{2}\times S_{N_{R}}^{2}$
become the $x_{\pm}$'s on $\mathbb{R}_{\theta}^{4}$ in the scaling
limit \[
\sqrt{\frac{2\theta}{N}}(\lambda_{1}^{L,R}\pm i\lambda_{2}^{L,R})\rightarrow x_{\pm L,R},\]
 we see that (\ref{scaled fluxons L}) and (\ref{scaled fluxons R})
become the instantons (\ref{eq: sol for instantons on R4 L}, \ref{eq: sol for instantons on R4 R})
on $\mathbb{R}_{\theta}^{4}$, \begin{eqnarray}
X_{1}+iX_{2} & \rightarrow & X_{+L}^{(n)}=Sx_{+L}S^{\dagger}+\sum_{k=1}^{n}\gamma_{k}^{L}|i_{k},j_{k}\rangle\langle i_{k},j_{k}|,\label{limit to instantons 1}\\
X_{3}+iX_{4} & \rightarrow & X_{+R}^{(n)}=Sx_{+R}S^{\dagger}+\sum_{k=1}^{n}\gamma_{k}^{R}|i_{k},j_{k}\rangle\langle i_{k},j_{k}|.\label{limit to instantons 2}\end{eqnarray}
 Here the $(d_{i})$-block acting on a basis $|i_{k},j_{k}\rangle$
of $V_{n}\subset\mathcal{H}\cong\C^{\cN}$ becomes the projector part
of (\ref{limit to instantons 1}, \ref{limit to instantons 2}) with
\begin{eqnarray}
\sqrt{\frac{2\theta}{N}}\; d_{1,k}^{L,R} & \rightarrow & \mathrm{Re}\gamma_{k}^{L,R},\label{d-gamma-relation}\\
\sqrt{\frac{2\theta}{N}}\; d_{2,k}^{L,R} & \rightarrow & \mathrm{Im}\gamma_{k}^{L,R},\nonumber \end{eqnarray}
 and the monopole block becomes $Sx_{+}S^{\dagger}$ where $S$ is
a partial isometry from $\mathcal{H}$ to $\mathcal{H}$\textbackslash{}$V_{n}$.
Note that we can recover any value for the $\gamma$'s in this scaling,
solving the constraint $d_{i}d_{i}=\frac{N^{2}-3}{4}$ by $d_{3}\approx\frac{N}{2}$.
Therefore the full moduli space of the fluxon solutions (\ref{eq: sol for instantons on R4 L},
\ref{eq: sol for instantons on R4 R}) on $\mathbb{R}_{\theta}^{4}$
can be recovered in this way. Furthermore, the meaning of the parameters
$\g^{L,R}$ is easy to understand in our approach: Note first that
using a rotation (which acts also on the indices) followed by a gauge
transformation, the $d_{i}$ can be fixed to be radial at the north
pole, $d_{i}^{L,R}\sim(0,0,N/2)$. This is a fluxon localized at the
north pole. Now apply a translation at the north pole, which corresponds
to a suitable rotation on the sphere. As the $\gamma_{k}^{1,2}$,
according to (\ref{d-gamma-relation}), are the projections of the
vectors $d_{i}^{L,R}$ onto the surface of the spheres, rotating the
vector $d_{i}^{L,R}$ in the scaling limit amounts to a translation
of the $\gamma_{k}^{1,2}$, which therefore parametrize the position
of the fluxons.

It has been noted \cite{Douglas:2001ba} that the $Sx_{+}S^{\dagger}$
correspond to a pure (but topologically nontrivial) gauge, which can
qualitatively be seen already in two dimensions. There, the partial
isometry $S:|k\rangle\rightarrow|k+n\rangle$ is basically $(\frac{x_{-}}{\sqrt{x_{-}x_{+}}})^{n}\sim(\frac{x-iy}{r})^{n}\sim e^{in\varphi}$
and therefore the gauge field $\mc A_{i}=S\dd_{i}S^{\dagger}$ has
a winding number $n$. The topological nature of the $Sx_{+}S^{\dagger}$
is even more evident in our setting, as they are the limit of the
monopole solutions (\ref{monopoles L},\ref{monopole-ansatz}) on
$S_{N}^{2}\times S_{N}^{2}$. Moreover, note that their contribution
to the action (\ref{action-fluxon-total}) survives the scaling: even
though the field strength vanishes as $R\rightarrow\infty$, the integral
gives a finite contribution equal to the contribution of the fluxon
part. This topological {}``surface term'' is usually omitted in
the literature on $\mathbb{R}_{\theta}^{4}$, but becomes apparent
in the regularized theory.

So it seems that we recovered all the instantons of chapter \ref{se: instantons},
but in fact there is an important detail that we haven't discussed
jet. It is the embedding of the $n$-dimensional fluxons and the $(N-m)(N+m)$-dimensional
monopole solutions into the $N^{2}$-dimensional matrices of the ground
state. Such an embedding is clearly only possible for $n=m^{2}$.
This means that the regularized theory has a superselection rule for
the dimension of the allowed instantons, a rule that did not exist
in the unregularized theory%
\footnote{Note that this is different in two dimensions. There, a rank $n$
fluxon can be combined with a $(N-n)$-dimensional monopole block
and all the instantons on $\mathbb{R}_{\theta}^{2}$ can be recovered.
Furthermore, the actions for the fluxons and the monopoles scale differently
with $N$. Therefore, in two dimensions, the action for the monopoles
vanishes in the scaling limit that produces a gauge theory on $\mathbb{R}_{\theta}^{2}$
with rescaled coupling constant.%
}.

One way to allow arbitrary instanton numbers is to allow the size
$\cN$ of the matrices to vary. However, this is less satisfactory
as it destroys the unification of topological sectors, which is a
beautiful feature of noncommutative gauge theory. On the other hand,
the type B solutions (\ref{singular-solutionsAB}) together with the
changed action (\ref{action-2}) might allow the construction of the
missing instantons. The idea is to fill up the unnecessary $m^{2}-n$
places with $d_{i}=0$. The changed action would not suppress such
solutions any more, and in fact they would not even contribute to
the action. This amounts to adding a discrete sector to the theory
which accommodates these type B solutions, but decouples from the
rest of the model. Whether or not one wants to do this appears to
be a matter of choice. This emphasizes again the importance of a careful
regularization of the theory. It would be very interesting to see
what happens in other regularizations e.g. using gauge theory on noncommutative
tori or fuzzy $\C P^{2}$.

\cleardoublepage

\fancyhead[LE,RO]{\thepage} 

\fancyhead[LO,RE]{}

\addcontentsline{toc}{part}{Appendix}

\begin{center}

\Huge \bf $\mbox{}$ \vspace{180pt}

Appendix

\end{center}

\newpage

\thispagestyle{empty}

\begin{appendix}

\fancyhead[LE,RO]{\thepage} 

\fancyhead[LO,RE]{\leftmark}

\chapter{Brackets, forms and frames}

\section{Definitions of the brackets}

\subsection{\label{Schouten-Nijenhuis bracket}The Schouten-Nijenhuis bracket}

The Schouten-Nijenhuis bracket for multivectorfields $\pi_{s}=\pi_{s}^{i_{1}\ldots i_{k_{s}}}\partial_{i_{1}}\wedge\ldots\wedge\partial_{i_{k_{s}}}$
can be written as (\cite{Arnal:2000hy},IV.2.1):

\begin{equation}
{}[\pi_{1},\pi_{2}]_{S}=(-1)^{k_{1}-1}\pi_{1}\bullet\pi_{2}-(-1)^{k_{1}(k_{2}-1)}\pi_{2}\bullet\pi_{1},\end{equation}

\begin{equation}
\pi_{1}\bullet\pi_{2}=\sum_{l=1}^{k_{1}}(-1)^{l-1}\pi_{1}^{i_{1}\ldots i_{k_{1}}}\partial_{l}\pi_{2}^{j_{1}\ldots j_{k_{2}}}\partial_{i_{1}}\wedge\ldots\wedge\widehat{{\partial_{i_{l}}}}\wedge\ldots\wedge\partial_{i_{k_{1}}}\wedge\partial_{j_{1}}\wedge\ldots\wedge\partial_{j_{k_{2}}},\end{equation}
 where the hat marks an omitted derivative.

For a function $g$, vectorfields $X=X^{k}\partial_{k}$ and $Y=Y^{k}\partial_{k}$
and a bivectorfield $\pi=\frac{1}{2}\pi^{kl}\partial_{k}\wedge\partial_{l}$
we get:

\begin{eqnarray}
{}[X,g]_{S} & = & X^{k}\partial_{k}g,\\
{}[\pi,g]_{S} & = & -\pi^{kl}\partial_{k}g\partial_{l},\nonumber \\
{}[X,\pi]_{S} & = & \frac{1}{2}(X^{k}\partial_{k}\pi^{ij}-\pi^{ik}\partial_{k}X^{j}+\pi^{jk}\partial_{k}X^{i})\partial_{i}\wedge\partial_{j},\nonumber \\
{}[\pi,\pi]_{S} & = & \frac{1}{3}(\pi^{kl}\partial_{l}\pi^{ij}+\pi^{il}\partial_{l}\pi^{jk}+\pi^{jl}\partial_{l}\pi^{ki})\partial_{k}\wedge\partial_{i}\wedge\partial_{j}.\nonumber \end{eqnarray}

\subsection{\label{Gerstenhaber bracket}The Gerstenhaber bracket}

The Gerstenhaber bracket for polydifferential operators $A_{s}$ can
be written as (\cite{Arnal:2000hy},IV.3):

\begin{equation}
{}[A_{1},A_{2}]_{G}=A_{1}\circ A_{2}-(-1)^{(|A_{1}|-1)(|A_{2}|-1)}A_{2}\circ A_{1},\end{equation}
\begin{eqnarray}
\lefteqn{(A_{1}\circ A_{2})(f_{1},\ldots\, f_{m_{1}+m_{2}-1})}\\
 & = & \sum_{j=1}^{m_{1}}(-1)^{(m_{2}-1)(j-1)}A_{1}(f_{1},\ldots\, f_{j-1},A_{2}(f_{j},\ldots,f_{j+m_{2}-1}),f_{j+m_{2}},\ldots,f_{m_{1}+m_{2}-1}),\nonumber \end{eqnarray}
where $|A_{s}|$ is the degree of the polydifferential operator $A_{s}$,
i.e. the number of functions it is acting on.

For functions $g$ and $f$, differential operators $D_{1}$and $D_{2}$
of degree one and $P$ of degree two we get

\begin{eqnarray}
{}[D,g]_{G} & = & D(g),\label{gerst 2 1}\\
{}[P,g]_{G}(f) & = & P(g,f)-P(f,g),\nonumber \\
{}[D_{1},D_{2}]_{G}(g) & = & D_{1}(D_{2}(g))-D_{2}(D_{1}(g)),\nonumber \\
{}[P,D]_{G}(f,g) & = & P(D(f),g)+P(f,D(g))-D(P(f,g)).\nonumber \end{eqnarray}

\section{\label{sec:Noncommutative-forms}Noncommutative forms}

We are now able to introduce noncommutative forms as well. If we have
a map $\delta$ from the Poisson vector fields to the derivations
of the $\star$-product algebra, we have seen that there is a natural
Lie-algebra structure\begin{equation}
[\delta_{X},\delta_{Y}]=\delta_{[X,Y]_{\star}},\end{equation}
 over the space of these derivations. On this we can easily construct
the Chevalley cohomology. Further, again with the map $\delta$, we
can lift derivations of the Poisson structure to derivations of the
$\star$-product. Therefore it should be possible to pull back the
Chevalley cohomology from the space of derivations to the Poisson
vector fields. This will be done in the following.

A deformed $k$-form is defined to map $k$ Poisson vector fields
to a function and has to be skew-symmetric and linear over $\mathbb{C}$.
This is a generalization of the undeformed case, where a form has
to be linear over the algebra of functions. Functions are defined
to be $0$-forms. The space of forms $\Omega_{\star}M$ is now a $\star$-bimodule
via \begin{equation}
(f\star\omega\star g)(X_{1},\dots,X_{k})=f\star\omega(X_{1},\dots,X_{k})\star g.\label{bimodule_structure_forms}\end{equation}
 As expected, the exterior differential is defined with the help of
the map $\delta$.\begin{eqnarray}
\lefteqn{\delta\omega(X_{0},\dots,X_{k})=}\label{nc_differential}\\
 &  & \,\;\;\;\;\;\;\sum_{i=0}^{k}(-1)^{i}\,\delta_{X_{i}}\omega(X_{0},\dots,\hat{X}_{i},\dots,X_{k})\nonumber \\
 &  & +\sum_{0\leq i<j\leq k}(-1)^{i+j}\omega([X_{i},X_{j}]_{\star},X_{0},\dots,\hat{X}_{i},\dots,\hat{X}_{j},\dots,X_{k}).\nonumber \end{eqnarray}
 With the properties of $\delta$ and $[\cdot,\cdot]_{\star}$ it
follows that \begin{equation}
\delta^{2}\omega=0.\end{equation}
 To be more explicit we give formulas for a function $f$, a one form
$A$ and a two form $F$\begin{eqnarray}
\delta f(X) & = & \delta_{X}f,\\
\delta A(X,Y) & = & \delta_{X}A_{Y}-\delta_{Y}A_{X}-A_{[X,Y]_{\star}},\nonumber \\
\delta F(X,Y,Z) & = & \delta_{X}F_{Y,Z}-\delta_{Y}F_{X,Z}+\delta_{Z}F_{X,Y},\nonumber \\
 &  & -F_{[X,Y]_{\star},Z}+F_{[X,Z]_{\star},Y}-F_{[Y,Z]_{\star},X}.\nonumber \end{eqnarray}
 A wedge product may be defined \begin{eqnarray}
\lefteqn{\omega_{1}\wedge\omega_{2}(X_{1},\dots,X_{p+q})=}\\
 &  & \frac{1}{p!q!}\sum_{I,J}\varepsilon(I,J)\,\omega_{1}(X_{i_{1}},\dots,X_{i_{p}})\star\omega_{2}(X_{j_{1}},\dots,X_{j_{q}})\nonumber \end{eqnarray}
 where $(I,J)$ is a partition of $(1,\dots,p+q)$ and $\varepsilon(I,J)$
is the sign of the corresponding permutation. The wedge product is
linear and associative and generalizes the bimodule structure (\ref{bimodule_structure_forms}).
We note that it is no more graded commutative. We again give some
formulas.\begin{eqnarray}
(f\wedge A)_{X} & = & f\star A_{X},\\
(A\wedge f)_{X} & = & A_{X}\star f,\nonumber \\
(A\wedge B)_{X,Y} & = & A_{X}\star B_{Y}-A_{Y}\star B_{X}.\nonumber \end{eqnarray}
 The differential (\ref{nc_differential}) fulfills the graded Leibniz
rule\begin{equation}
\delta(\omega_{1}\wedge\omega_{2})=\delta\omega_{1}\wedge\omega_{2}+(-1)^{k_{2}}\,\omega_{1}\wedge\delta\omega_{2}.\end{equation}

\section{\label{sec: Frames}Frames}

We will now propose a method how to find frames and Poisson structures
of quantum groups that are compatible. On several quantum spaces deformed
derivations have been constructed \cite{Wess:1991vh,Lorek:1997eh,Cerchiai:1998ef}.
In most cases the deformed Leibniz rule may be written in the following
form\begin{equation}
\hat{\partial}_{\mu}(\hat{f}\hat{g})=\hat{\partial}_{\mu}\hat{f}\hat{g}+\hat{T}_{\mu}{}^{\nu}(\hat{f})\hat{\partial}_{\nu}\hat{g},\end{equation}
 where $\hat{T}$ is an algebra morphism from the quantum space to
its matrix ring\begin{equation}
\hat{T}_{\mu}{}^{\nu}(\hat{f}\hat{g})=\hat{T}_{\mu}{}^{\alpha}(\hat{f})\hat{T}_{\alpha}{}^{\nu}(\hat{g}).\end{equation}
 Again in some cases it is possible to implement this morphism with
some kind of inner morphism\begin{equation}
\hat{T}_{\mu}{}^{\nu}(\hat{f})=\hat{e}_{\mu}{}^{a}\hat{f}\hat{e}_{a}{}^{\nu},\end{equation}
 where $\hat{e}_{a}{}^{\mu}$ is an invertible matrix with entries
from the quantum space. If we define\begin{equation}
\hat{e}_{a}=\hat{e}_{a}{}^{\mu}\hat{\partial}_{\mu},\end{equation}
 the $\hat{e}_{a}$ are derivations\begin{equation}
\hat{e}_{a}(\hat{f}\hat{g})=\hat{e}_{a}(\hat{f})\hat{g}+\hat{f}\hat{e}_{a}(\hat{g}).\end{equation}
 The dual formulation of this with covariant differential calculi
on quantum spaces is the formalism with commuting frames investigated
for example in \cite{Dimakis:1996,Madore:1999bi,Cerchiai:2000qi,Madore:2000aq}.
There one can additionally find how our formalism fits into the language
of Connes' spectral triples.

We can now represent the quantum space with the help of a $\star$-product.
For example, we can use the Weyl ordered $\star$-product constructed
in chapter \ref{wely_star_construction}. Further we can calculate
the action of the operators $\hat{e}_{a}$ on functions. Since the
$\hat{e}_{a}$ are now derivations of a $\star$-product, there necessarily
exist Poisson vector fields $e_{a}$ with \begin{equation}
\delta_{e_{a}}=\hat{e}_{a}.\end{equation}

\chapter{Calculation of the SW-map to all orders}

\section{\label{Commutator theta theta and theta X}Calculation of $[\theta_{t},\theta_{t}]$
and $[\theta_{t},X_{t}]$}

We want to show that $\theta_{t}$ is still a Poisson tensor and that
$X_{t}$ still commutes with $\theta_{t}$. For this we first define
$\theta(n)_{l}^{k}=(\theta f)^{n}=\theta^{ki}f_{ij}\ldots\theta^{rs}f_{sl}=f_{li}\theta^{ij}\ldots f_{rs}\theta^{sk}=(f\theta)^{n}$
and $\theta(n)^{kl}=\theta(f\theta)^{n}=\theta^{ki}f_{ij}\ldots f_{rs}\theta^{sl}$.
In the calculations to follow we will sometimes drop the derivatives
of the polyvectorfields and associate $\pi^{k_{1}\ldots k_{n}}$ with
$\pi^{k_{1}\ldots k_{n\:\:}}\frac{1}{n}\partial_{k_{1}}\wedge\ldots\wedge\partial_{k_{n}}$for
simplicity. All the calculations are done locally.

We evaluate

\begin{eqnarray}
{[}\theta_{t},\theta_{t}]_{S} & = & \theta_{t}^{kl}\partial_{l}\theta_{t}^{ij}+\mbox{c.p.}\;\mbox{in}\;(kij)\\
 & = & \sum_{n,m=0}^{\infty}\sum_{o=0}^{m}(-t)^{n+m}\theta(n)_{r}^{k}\theta(o)_{s}^{i}\theta(m-o)_{p}^{j}\theta^{rl}\partial_{l}\theta^{sp}+\mbox{c.p.}\;\mbox{in}\;(kij)\nonumber \\
 &  & +\sum_{n,m=0}^{\infty}\sum_{o=0}^{m}(-t)^{n+m+1}\theta(n)^{kl}\theta(o)^{is}\theta(m-o)^{pj}\partial_{l}f_{sp}+\mbox{c.p.}\;\mbox{in}\;(kij)\nonumber \\
 & = & \sum_{n,m,o=0}^{\infty}(-t)^{n+m+o}\theta(n)_{r}^{k}\theta(o)_{s}^{i}\theta(m)_{p}^{j}\theta^{rl}\partial_{l}\theta^{sp}+\mbox{c.p.}\;\mbox{in}\;(kij)\nonumber \\
 &  & -\sum_{n,m,o=0}^{\infty}(-t)^{n+m+o+1}\theta(n)^{kl}\theta(o)^{is}\theta(m)^{jp}\partial_{l}f_{sp}+\mbox{c.p.}\;\mbox{in}\;(kij).\nonumber \end{eqnarray}
The first part vanishes because $\theta_{t}$ is a Poisson tensor,
i.e.\begin{equation}
{[}\theta,\theta]_{S}=\theta^{kl}\partial_{l}\theta^{ij}+\mbox{c.p.}\;\mbox{in}\;(kij)=0,\label{poisson}\end{equation}
the second part because of\begin{equation}
\partial_{k}f_{ij}+\mbox{c.p.}\;\mbox{in}\;(kij)=0.\label{cyclic f}\end{equation}
To prove that $X_{t}$ still commutes with $\theta_{t}$, we first
note that\begin{equation}
X_{t}=X\sum_{n=0}^{\infty}(-tf\theta)=X(1-tf\theta_{t}).\end{equation}
With this we can write

\begin{eqnarray}
{[}X_{t},\theta_{t}] & = & [X,\theta_{t}]-t[Xf\theta_{t},\theta_{t}]\label{comm X theta}\\
 & = & X^{n}\partial_{n}\theta_{t}^{kl}-\theta_{t}^{kn}\partial_{n}X^{l}+\theta_{t}^{ln}\partial_{n}X^{k}\nonumber \\
 &  & -tX^{m}f_{mi}\theta_{t}^{in}\partial_{n}\theta_{t}^{kl}+t\theta_{t}^{kn}\partial_{n}(X^{m}f_{mi}\theta_{t}^{il})-t\theta_{t}^{ln}\partial_{n}(X^{m}f_{mi}\theta_{t}^{ik})\nonumber \\
 & = & X^{n}\partial_{n}\theta_{t}^{kl}-\theta_{t}^{kn}\partial_{n}X^{l}+\theta_{t}^{ln}\partial_{n}X^{k}\nonumber \\
 &  & +t\theta_{t}^{kn}\partial_{n}X^{m}f_{mi}\theta_{t}^{il}-t\theta_{t}^{ln}\partial_{n}X^{m}f_{mi}\theta_{t}^{ik}\nonumber \\
 &  & +t\theta_{t}^{kn}X^{m}\partial_{n}f_{mi}\theta_{t}^{il}-t\theta_{t}^{ln}X^{m}\partial_{n}f_{mi}\theta_{t}^{ik}.\nonumber \end{eqnarray}
In the last step we used (\ref{poisson}). To go on we note that

\begin{equation}
t\theta_{t}^{kn}X^{m}\partial_{n}f_{mi}\theta_{t}^{il}-t\theta_{t}^{ln}X^{m}\partial_{n}f_{mi}\theta_{t}^{ik}=tX^{n}\theta_{t}^{km}\partial_{n}f_{mi}\theta_{t}^{il},\end{equation}
where we used (\ref{cyclic f}). Making use of the power series expansion
and the fact that $X$ commutes with $\theta,$ i.e. \begin{equation}
{[}X,\theta]=X^{n}\partial_{n}\theta^{kl}-\theta^{kn}\partial_{n}X^{l}+\theta^{ln}\partial_{n}X^{k}=0,\end{equation}
 we further get

\begin{eqnarray}
X^{n}\partial_{n}\theta_{t}^{kl}+tX^{n}\theta_{t}^{km}\partial_{n}f_{mi}\theta_{t}^{il} & = & \sum_{r,s=0}^{\infty}(-t)^{r+s}\theta(r)_{i}^{k}X^{n}\partial_{n}\theta^{ij}\theta(s)_{j}^{l}\\
 & = & \sum_{r,s=0}^{\infty}(-t)^{r+s}\theta(r)_{i}^{k}\theta^{in}\partial_{n}X^{j}\theta(s)_{j}^{l}\nonumber \\
 &  & -\sum_{r,s=0}^{\infty}(-t)^{r+s}\theta(r)_{i}^{k}\theta^{jn}\partial_{n}X^{i}\theta(s)_{j}^{l}.\nonumber \end{eqnarray}
Therefore (\ref{comm X theta}) reads

\begin{eqnarray}
{[}X_{t},\theta_{t}] & = & \sum_{r,s=0}^{\infty}(-t)^{r+s}\theta(r)_{i}^{k}\theta(s)_{j}^{l}\theta^{in}\partial_{n}X^{j}\\
 &  & -\sum_{r,s=0}^{\infty}(-t)^{r+s}\theta(r)_{i}^{k}\theta(s)_{j}^{l}\theta^{jn}\partial_{n}X^{i}\nonumber \\
 &  & -\theta_{t}^{kn}\partial_{n}X^{l}+\theta_{t}^{ln}\partial_{n}X^{k}+t\theta_{t}^{kn}\partial_{n}X^{m}f_{mi}\theta_{t}^{il}-t\theta_{t}^{ln}\partial_{n}X^{m}f_{mi}\theta_{t}^{ik}\nonumber \\
 & = & 0.\nonumber \end{eqnarray}

\section{Calculation of the commutators}

\subsection{\label{Semi-classical commutators}Semi-classical construction}

We calculate the commutator (\ref{comm rel 1}) (see also \cite{Jurco:2001my}),
dropping the t-subscripts on $\theta_{t}$ for simplicity and using
local expressions.

\begin{eqnarray}
{[}a_{\theta},d_{\theta}(g)] & = & -\theta^{ij}a_{j}\partial_{i}\theta^{kl}\partial_{k}g\partial_{l}-\theta^{ij}a_{j}\theta^{kl}\partial_{i}\partial_{k}g\partial_{l}\\
 &  & +\theta^{kl}\partial_{k}g\partial_{l}\theta^{ij}a_{j}\partial_{i}+\theta^{kl}\partial_{k}g\theta^{ij}\partial_{l}a_{j}\partial_{i}\nonumber \\
 & = & -\theta^{kl}\partial_{k}\theta^{ij}a_{j}\partial_{i}g\partial_{l}-\theta^{kl}\theta^{ij}a_{j}\partial_{k}\partial_{i}g\partial_{l}-\theta^{kl}\theta^{ij}\partial_{j}a_{k}\partial_{i}g\partial_{l}\nonumber \\
 & = & +\theta^{ij}f_{jk}\theta^{kl}\partial_{i}g\partial_{l}-\theta^{kl}\partial_{k}(\theta^{ij}a_{j}\partial_{i}g)\partial_{l}\nonumber \\
 & = & -d_{\theta f\theta}g+d_{\theta}(a_{\theta}(g))\nonumber \\
 & = & -\partial_{t}(d_{\theta})g+d_{\theta}(a_{\theta}(g)).\nonumber \end{eqnarray}
For (\ref{comm rel}) we get

\begin{eqnarray}
{[}a_{\theta},X_{t}] & = & \theta^{ij}a_{j}\partial_{i}X^{k}\partial_{k}-X^{k}\partial_{k}\theta^{ij}a_{j}\partial_{i}-X^{k}\theta^{ij}\partial_{k}a_{j}\partial_{i}\\
 & = & -\theta^{ij}X^{k}\partial_{k}a_{j}\partial_{i}-\theta^{ik}\partial_{k}X^{j}a_{j}\partial_{i}\nonumber \\
 & = & X^{k}f_{ki}\theta^{ij}\partial_{j}+\theta^{ij}\partial_{i}(X^{k}a_{k})\partial_{j}\nonumber \\
 & = & -\partial_{t}X-d_{\theta}(X^{k}a_{k}).\nonumber \end{eqnarray}

\subsection{\label{Quantum commutators}Quantum construction}

In \cite{Manchon:2000hy}, (\ref{comm phi f star},\ref{comm delta X star},\ref{comm delta X Phi g})
have already been calculated, unluckily (and implicitly) using a different
sign convention for the brackets of polyvectorfields. In \cite{Jurco:2001my},
again a different sign convention is used, coinciding with the one
in \cite{Manchon:2000hy} in the relevant cases. In order to keep
our formulas consistent with the ones used in \cite{Manchon:2000hy,Jurco:2001my},
we define our bracket on polyvectorfields $\pi_{1}$ and $\pi_{2}$
as in \cite{Manchon:2000hy} to be\begin{equation}
{}[\pi_{1},\pi_{2}]=-[\pi_{2},\pi_{1}]_{S},\end{equation}
giving an extra minus sign for $\pi_{1}$ and $\pi_{2}$ both even.
The bracket on polydifferential operators is always the Gerstenhaber
bracket.

With these conventions and\begin{equation}
d_{\star}=-[\cdot,\star]\end{equation}

we rewrite the formulas (\ref{comm delta X Phi g},\ref{comm delta X delta Y},\ref{comm phi f star},\ref{comm delta X star})
so we can use them in the following

\begin{eqnarray}
{}[\Phi(X),\Phi(g)]_{G} & = & \Phi([X,g])+\Psi([\theta,g],X)-\Psi([\theta,X],g),\label{Phi X Phi g}\\
{}[\Phi(X),\Phi(Y)]_{G} & = & d_{\star}\Psi(X,Y)\label{Phi X, Phi Y}\\
 &  & +\Phi([X,Y])+\Psi([\theta,Y],X)-\Psi([\theta,X],Y),\nonumber \\
d_{\star}\Phi(g) & = & \Phi(d_{\theta}(g)),\\
d_{\star}\Phi(X) & = & \Phi(d_{\theta}(X)).\label{d star x}\end{eqnarray}
For the calculation of the commutators of the quantum objects we first
define\begin{equation}
a_{\star}=\Phi(a_{\theta_{t}})\end{equation}
and \begin{equation}
f_{\star}=\Phi(f_{\theta_{t}}).\end{equation}
With (\ref{d star x}) we get the quantum version of (\ref{f theta})\begin{equation}
f_{\star}=d_{\star}a_{\star}.\end{equation}
For functions $f$ and $g$ we get\begin{eqnarray}
\partial_{t}(f\star g) & = & \sum_{n=0}^{\infty}\frac{1}{n!}\partial_{t}U_{n}(\theta_{t},\,\ldots,\theta_{t})(f,g)\\
 & = & \sum_{n=1}^{\infty}\frac{1}{(n-1)!}U_{n}(f_{\theta},\,\ldots,\theta_{t})(f,g)\nonumber \\
 & = & f_{\star}(f,g)\nonumber \end{eqnarray}

With these two formulas we can now calculate the quantum version of
(\ref{comm rel 1}) as in \cite{Jurco:2001my}. On two functions $f$
and $g$ we have

\begin{eqnarray}
\partial_{t}(f\star g) & = & f_{\star}(f,g)\\
 & = & d_{\star}a_{\star}(f,g)\nonumber \\
 & = & -[a_{\star},\star](f,g)\nonumber \\
 & = & -a_{\star}(f\star g)+a_{\star}(f)\star g+f\star a_{\star}(g),\nonumber \end{eqnarray}
where we used (\ref{gerst 2 1}) in the last step. Therefore\begin{eqnarray}
{}[a_{\star},d_{\star}(g)](f) & = & a_{\star}(d_{\star}(g)(f))-d_{\star}(g)(a_{\star}(f))\\
 & = & a_{\star}([f\stackrel{\star}{,}g])-[a_{\star}(f)\stackrel{\star}{,}g]\nonumber \\
 & = & -\partial_{t}[f\stackrel{\star}{,}g]-[a_{\star}(g)\stackrel{\star}{,}f]\nonumber \\
 & = & -\partial_{t}d_{\star}(g)(f)+d_{\star}(a_{\star}(g))(f).\nonumber \end{eqnarray}
For a function $g$ which might also depend on $t$ the quantum version
of (\ref{comm rel 1}) now reads\begin{equation}
{}[a_{\star}+\partial_{t},d_{\star}(g)]=d_{\star}(a_{\star}(g)).\end{equation}
We go on to calculate the quantum version of (\ref{comm rel}). We
first note that\begin{equation}
\partial_{t}\Phi(X_{t})=\sum_{n=1}^{\infty}\frac{1}{(n-1)!}\partial_{t}U_{n}(X_{t},\theta_{t},\,\ldots,\theta_{t})=\Phi(\partial_{t}X_{t})+\Psi(f_{\theta},X_{t}).\end{equation}
With this we get\begin{eqnarray}
{}[\Phi(a_{\theta}),\Phi(X_{t})] & = & d_{\star}\Psi(a_{\theta},X_{t})+\Phi([a_{\theta},X_{t}])\\
 &  & -\Psi([\theta_{t}\, a_{\theta}])+\Psi([\theta_{t},X_{t}],a_{\theta})\nonumber \\
 & = & d_{\star}\Psi(a_{\theta},X_{t})+\Phi(-d_{\theta}(X_{t}^{k}a_{k}))+\Phi(-\partial_{t}X_{t})-\Psi(f_{\theta},X_{t})\nonumber \\
 & = & -d_{\star}(\Phi(X_{t}^{k}a_{k})-\Psi(a_{\theta},X_{t}))-\partial_{t}\Phi(X_{t}),\nonumber \end{eqnarray}
where we have used (\ref{Phi X, Phi Y}).

\section{\label{Transformation properties of K}The transformation properties
of $K_{t}$}

To calculate the transformation properties of $K_{t}(X_{t}^{k}a_{k})$,
we first evaluate

\begin{eqnarray}
\delta_{\lambda}((a_{\theta}+\partial_{t})^{n})X^{k}a_{k} & = & \sum_{i=0}^{n-1}(a_{\theta}+\partial_{t})^{i}d_{\theta}(\lambda)(a_{\theta}+\partial_{t})^{n-1-i}X^{k}a_{k}\\
 & = & \sum_{i=0}^{n-1}\sum_{l=0}^{i}{{i \choose l}}d_{\theta}((a_{\theta}+\partial_{t})^{l}(\lambda))(a_{\theta}+\partial_{t})^{n-1-l}X^{k}a_{k}\nonumber \end{eqnarray}
and

\begin{eqnarray}
\lefteqn{(a_{\theta}+\partial_{t})^{n}\delta_{\lambda}(X^{k}a_{k})}\\
 & = & (a_{\theta}+\partial_{t})^{n}X^{k}\partial_{k}\lambda\nonumber \\
 & = & X^{k}\partial_{k}(a_{\theta}+\partial_{t})^{n}-\sum_{i=0}^{n-1}(a_{\theta}+\partial_{t})^{i}d_{\theta}(X^{k}a_{k})(a_{\theta}+\partial_{t})^{n-1-i}\lambda\nonumber \\
 & = & X^{k}\partial_{k}(a_{\theta}+\partial_{t})^{n}-\sum_{i=0}^{n-1}\sum_{j=0}^{n-1-i}{{n-1-i \choose j}}(-1)^{n-1-i-j}(a_{\theta}+\partial_{t})^{i+j}\times\nonumber \\
\nonumber \\ &  & \;\;\;\;\;\;\;\;\;\;\;\;\;\;\;\;\;\;\;\;\;\;\;\;\;\;\;\;\;\;\;\;\;\;\;\;\;\;\;\;\;\;\; d_{\theta}((a_{\theta}+\partial_{t})^{n-1-i-j}(X^{k}a_{k}))(\lambda)\nonumber \\
\nonumber \\ & = & X^{k}\partial_{k}(a_{\theta}+\partial_{t})^{n}+\sum_{i=0}^{n-1}\sum_{j=0}^{n-1-i}{{n-1-i \choose j}}(-1)^{n-1-i-j}(a_{\theta}+\partial_{t})^{i+j}\times\nonumber \\
\nonumber \\ &  & \;\;\;\;\;\;\;\;\;\;\;\;\;\;\;\;\;\;\;\;\;\;\;\;\;\;\;\;\;\;\;\;\;\;\;\;\;\;\;\;\;\;\; d_{\theta}(\lambda)((a_{\theta}+\partial_{t})^{n-1-i-j}(X^{k}a_{k}))\nonumber \\
\nonumber \\ & = & X^{k}\partial_{k}(a_{\theta}+\partial_{t})^{n}+\sum_{i=0}^{n-1}\sum_{j=0}^{n-1-i}\sum_{l=0}^{i+j}{{n-1-i \choose j}}{{i+j \choose l}}(-1)^{n-1-i-j}\times\nonumber \\
\nonumber \\ &  & \;\;\;\;\;\;\;\;\;\;\;\;\;\;\;\;\;\;\;\;\;\;\;\;\;\;\;\;\;\;\;\;\;\;\;\;\;\;\;\;\;\;\;\;\;\;\;\; d_{\theta}((a_{\theta}+\partial_{t})^{l}(\lambda))((a_{\theta}+\partial_{t})^{n-1-l}(X^{k}a_{k})).\nonumber \\
\nonumber \end{eqnarray}
We go on by simplifying these expressions. Using\begin{equation}
{{i \choose l}}={{i-1 \choose l}}+{{i-1 \choose l-1}}\;\;\;\;\mbox{for}\;\;\;\; i>l\label{binomial}\end{equation}
we get\begin{equation}
\sum_{m=l}^{n-1}\sum_{i=0}^{m}{{n-1-i \choose m-i}}{{m \choose l}}(-1)^{n-1-m}=\sum_{m=l}^{n-1}{{n \choose m}}{{m \choose l}}(-1)^{n-1-m}.\end{equation}
 Using (\ref{binomial}) again two times and then using induction
we go on to\begin{equation}
\sum_{m=l}^{n-1}{{n \choose m}}{{m \choose l}}(-1)^{n-1-m}=\sum_{i=0}^{l}{{n-1-i \choose n-1-l}},\end{equation}
 giving, after using (\ref{binomial}) again\begin{equation}
\sum_{i=0}^{l}{{n-1-i \choose n-1-l}}={{n \choose l}}.\end{equation}
Together with\begin{equation}
\sum_{i=l}^{n-1}{{i \choose l}}={{n \choose l+1}}\end{equation}
these formulas add up to give\begin{equation}
\sum_{m=l}^{n-1}\sum_{i=0}^{m}{{n-1-i \choose m-i}}{{m \choose l}}(-1)^{n-1-m}+\sum_{i=l}^{n-1}{{i \choose l}}={{n+1 \choose l+1}}\end{equation}
and therefore

\begin{equation}
\delta_{\lambda}(K_{t}(X^{k}a_{k}))=X^{k}\partial_{k}(K_{t}(\lambda))+d_{\theta}(K_{t}(\lambda))K_{t}(X^{k}a_{k}).\end{equation}

\chapter{Representations}

\section{The standard representation of the fuzzy sphere}

\label{sec:useful}

The irreducible $N$-dimensional representation of the $su(2)$ algebra
$\lambda_{i}$ (\ref{lambda-algebra}) is given by \begin{eqnarray}
(\lambda_{3})_{kl} & = & \d{_{kl}}\;\frac{N+1-2k}{2},\label{reps}\\
\nonumber \\(\lambda_{+})_{kl} & = & \d{_{k+1,l}}\sqrt{(N-k)k},\label{standard-rep}\\
\nonumber \end{eqnarray}
 where $k,l=1,...,N$ and $\lambda_{\pm}=\lambda_{1}\pm i\lambda_{2}$.

\section{Representation of the $SO(6)$- intertwiners and Clifford algebra}

\label{app:gamma-mat}

Latin indices $i,\: j$ will run from $1$ to $3$, whereas Greek
indices $\mu,\:\nu,\:$ ... denote all the six dimensions, i.e. both
the three left and the three right indices. We will use the Pauli
matrices \begin{equation}
\sigma^{1}=\left(\begin{array}{cc}
0 & 1\\
1 & 0\end{array}\right),\hspace{0,8cm}\sigma^{2}=\left(\begin{array}{cc}
0 & -i\\
i & 0\end{array}\right),\hspace{0,8cm}\sigma^{3}=\left(\begin{array}{cc}
1 & 0\\
0 & -1\end{array}\right),\end{equation}
 which satisfy \begin{equation}
\sigma^{i}\sigma^{j}=\d{^{ij}}+i\vare^{ijk}\sigma^{k}.\end{equation}
 With these we define the $4$-dimensional antisymmetric matrices
\begin{equation}
\begin{array}{llll}
 & \gamma_{L}^{1}=\sigma^{1}\otimes\sigma^{2}, & \gamma_{L}^{2}=\sigma^{2}\otimes1, & \gamma_{L}^{3}=\sigma^{3}\otimes\sigma^{2},\\
 & \gamma_{R}^{1}=i\;\sigma^{2}\otimes\sigma^{1}, & \gamma_{R}^{2}=i\;1\otimes\sigma^{2}, & \gamma_{R}^{3}=i\;\sigma^{2}\otimes\sigma^{3}.\end{array}\label{gamma-wolfg}\end{equation}
 They are the intertwiners between $SU(4)\otimes SU(4)$ and $SO(6)$
and fulfill the following relations: \begin{eqnarray}
(\gamma_{L}^{i})^{\dagger} & = & \gamma_{L}^{i},\\
(\gamma_{R}^{i})^{\dagger} & = & -\gamma_{R}^{i}\nonumber \end{eqnarray}
 and \begin{eqnarray}
\gamma_{L}^{i}\gamma_{L}^{j} & = & \delta^{ij}+i\epsilon_{k}^{ij}\gamma_{L}^{k},\\
\gamma_{R}^{i}\gamma_{R}^{j} & = & -\delta^{ij}-\epsilon_{k}^{ij}\gamma_{R}^{k},\nonumber \\
{}[\gamma_{L}^{i},\gamma_{R}^{j}] & = & 0.\nonumber \end{eqnarray}
 We can now define the $8$-dimensional representation of the $SO(6)$-Clifford
algebra as \begin{equation}
\Gamma^{\mu}=\left(\begin{array}{cc}
0 & \gamma^{\mu}\\
\gamma^{\mu\dagger} & 0\end{array}\right),\end{equation}
 with the desired anticommutation relations \begin{equation}
\{\Gamma^{\mu},\Gamma^{\nu}\}=\left(\begin{array}{cc}
\gamma^{\mu}\gamma^{\nu\dagger}+\gamma^{\nu}\gamma^{\mu\dagger} & 0\\
0 & \gamma^{\mu\dagger}\gamma^{\nu}+\gamma^{\nu\dagger}\gamma^{\mu}\end{array}\right)=2\delta^{\mu\nu}.\end{equation}
 The chirality operator in this basis is \begin{equation}
\Gamma=i\Gamma_{L}^{1}\Gamma_{L}^{2}\Gamma_{L}^{3}\Gamma_{R}^{1}\Gamma_{R}^{2}\Gamma_{R}^{3}=\left(\begin{array}{cc}
-1 & 0\\
0 & 1\end{array}\right).\end{equation}
 The $8$-dimensional $SO(6)$-rotations are generated by\begin{equation}
\Sigma_{8}^{\mu\nu}=-\frac{i}{4}[\Gamma^{\mu},\Gamma^{\nu}]=-\frac{i}{4}\left(\begin{array}{cc}
\gamma^{\mu}\gamma^{\nu\dagger}-\gamma^{\nu}\gamma^{\mu\dagger} & 0\\
0 & \gamma^{\mu\dagger}\gamma^{\nu}-\gamma^{\nu\dagger}\gamma^{\mu}\end{array}\right).\end{equation}
 If we define\begin{equation}
\Sigma^{\mu\nu}=-\frac{i}{4}(\gamma^{\mu}\gamma^{\nu\dagger}-\gamma^{\nu}\gamma^{\mu\dagger})\;\;\;\mbox{and}\;\;\;\overline{\Sigma}^{\mu\nu}=-\frac{i}{4}(\gamma^{\mu\dagger}\gamma^{\nu}-\gamma^{\nu\dagger}\gamma^{\mu}),\end{equation}
 the Clifford algebra transforms as\begin{equation}
{}[\Sigma_{8}^{\mu\nu},\Gamma^{\sigma}]=\left(\begin{array}{cc}
0 & \Sigma^{\mu\nu}\gamma^{\sigma}-\gamma^{\sigma}\overline{\Sigma}^{\mu\nu}\\
\overline{\Sigma}^{\mu\nu}\gamma^{\sigma\dagger}-\gamma^{\sigma\dagger}\Sigma^{\mu\nu} & 0\end{array}\right).\end{equation}
 Explicitly we have\begin{eqnarray}
\Sigma^{iL\: jL}= & -\frac{i}{4}[\gamma_{L}^{i},\gamma_{L}^{j}] & =\overline{\Sigma}^{iL\: jL},\label{sigma L L}\\
\Sigma^{iR\: jR}= & \frac{i}{4}[\gamma_{R}^{i},\gamma_{R}^{j}] & =\overline{\Sigma}^{iR\: jR},\label{sigma R R}\\
\Sigma^{iR\: jL}= & -\frac{i}{4}\{\gamma_{R}^{i},\gamma_{L}^{j}\} & =-\overline{\Sigma}^{iR\: jL}\label{sigma R L}\end{eqnarray}
 and therefore\begin{eqnarray}
{[}\Sigma_{8}^{iL\: jL},\Gamma^{\sigma}{]} & = & \left(\begin{array}{cc}
0 & {[}\Sigma^{iL\: jL},\gamma^{\sigma}]\\
{[}\Sigma^{iL\: jL},\gamma^{\sigma\dagger}] & 0\end{array}\right),\label{trans gamma LL}\\
{[}\Sigma_{8}^{iR\: jR},\Gamma^{\sigma}] & = & \left(\begin{array}{cc}
0 & {[}\Sigma^{iR\: jR},\gamma^{\sigma}]\\
{[}\Sigma^{iR\: jR},\gamma^{\sigma\dagger}] & 0\end{array}\right),\label{trans gamma RR}\\
{[}\Sigma_{8}^{iR\: jL},\Gamma^{\sigma}] & = & \left(\begin{array}{cc}
0 & \{\Sigma^{iR\: jL},\gamma^{\sigma}\}\\
-\{\Sigma^{iR\: jL},\gamma^{\sigma\dagger}\} & 0\end{array}\right).\label{trans gamma rl}\end{eqnarray}

\chapter{Calculations for the matrix model approach}

\section{Alternative formulation using $4\cN\times4\cN$ matrices}

\label{sec:4Nmatrices}

Let us rewrite the action (\ref{eq: action S_6 - S_break}) in terms
of the $4\cN\times4\cN$ matrices $B_{L},B_{R}$ (\ref{B-L-R}). Noting
that \begin{equation}
C_{L}C_{R}+C_{R}C_{L}=\left(\begin{array}{cc}
-[B_{L},B_{R}] & 0\\
0 & [B_{L},B_{R}]\end{array}\right)\end{equation}
 we can rewrite $S_{6}$ (\ref{action-so6}) as \begin{equation}
S_{6}=2\mathrm{Tr}\left(B_{L}^{2}-B_{R}^{2}-\frac{N^{2}}{2}\right)^{2}+2\mathrm{Tr}\left([B_{L},B_{R}]^{2}\right),\end{equation}
 where the trace is now over $4\cN\times4\cN$ matrices. Similarly
\begin{eqnarray}
S_{\mathrm{break}} & = & -4\mathrm{Tr}\left(B_{L}^{2}-\frac{N^{2}}{4}\right)\left(-B_{R}^{2}-\frac{N^{2}}{4}\right)\end{eqnarray}
 and combined we recover (\ref{action}) as \begin{equation}
S=S_{6}-S_{\mathrm{break}}=2\mathrm{Tr}\left((B_{L}^{2}-\frac{N^{2}}{4})^{2}+(-B_{R}^{2}-\frac{N^{2}}{4})^{2}+[B_{L},B_{R}]^{2}\right).\label{action-BLR}\end{equation}
 This looks like a 2-matrix model, however the degrees of freedom
$B_{L},B_{R}$ are still very much constrained and span only a small
subspace of the $4\cN\times4\cN$ matrices. We would like to find
an intrinsic characterization without using the $\g_{\mu}$ explicitly.
One possibility is to choose the $\g_{\mu}$ to be completely anti-symmetric
matrices, see Appendix \ref{app:gamma-mat}. However this does not
extend to $B$, since the $B_{\mu}$ should be Hermitian and not necessarily
symmetric, and moreover the $\g_{\mu}$ are not Hermitian (the conjugate
being the intertwiner $(6)\subset(\obar4)\tens(\obar4)$). Another
possibility is provided by the following representation of the $\g$-matrices:
\begin{equation}
\g_{L}^{i}=\sigma^{i}\tens\one_{2\times2},\qquad\g_{R}^{i}=\one_{2\times2}\tens i\sigma^{i}.\label{gamma-rep-1}\end{equation}
 They satisfy the relations (\ref{g-conj}) -- (\ref{g-L-R}), but
are not antisymmetric. Now note that \begin{equation}
\g_{R}^{i}=iP\g_{L}^{i}P\end{equation}
 where \begin{equation}
P=\left(\begin{array}{cccc}
1 & 0 & 0 & 0\\
0 & 0 & 1 & 0\\
0 & 1 & 0 & 0\\
0 & 0 & 0 & 1\end{array}\right)=\frac{1}{2}(1+\sigma^{i}\tens\sigma^{i})\label{P-explicit}\end{equation}
 permutes the two tensor factors and satisfies \begin{equation}
P^{2}=1.\end{equation}
 Therefore we can characterize the degrees of freedom in terms of
2 Hermitian $2\cN\times2\cN$ matrices \begin{equation}
X_{L}=B_{L}^{i}\sigma_{i}+\frac{1}{2},\qquad X_{R}=B_{R}^{i}\sigma_{i}+\frac{1}{2}\end{equation}
 which are arbitrary up to the constraint that $X_{L,R}^{0}=\frac{1}{2}$.
Then \begin{equation}
B_{L}=X_{L}\tens\one_{2\times2},\quad B_{R}=iP(X_{R}\tens\one_{2\times2})P;\end{equation}
 they could be extracted from a single complex matrix $\tilde{B}=(X_{L}+iX_{R})\tens\one_{2\times2}$.
Furthermore, matrices of the form $X\tens\one_{2\times2}$ are characterized
through their spectrum, which is doubly degenerate; indeed any such
Hermitian matrix can be cast into the above form using suitable unitary
$SU(4\cN)$ transformations. Similarly, $P$ can also be characterized
intrinsically: any matrix $P$ written as \begin{equation}
P=P_{0}\tens\one_{2\times2}+P_{i}\tens\sigma^{i}\end{equation}
 which satisfies the constraints \begin{equation}
P_{0}=\frac{1}{2},\quad P^{2}=\one\end{equation}
 is given by (\ref{P-explicit}) up to an irrelevant unitary transformation
$U\tens\one$. We could therefore write down the action (\ref{action-BLR})
in terms of three matrices $B_{L},-iPB_{R}P$ and $P$, all of which
are characterized by their spectrum and constraints of the form $(..)_{0}=\frac{1}{2}$.
The hope is that such a reformulation may allow to apply some of the
powerful methods from random matrix theory, in the spirit of \cite{Steinacker:2003sd}.

\section{Stability analysis of the $SO(6)$ - invariant action (\ref{action-so6})}

\label{app:stability}

Consider the action (\ref{action-so6}). We will split off the radial
degrees of freedom for large $N$ by setting%
\footnote{the fact that this leads to non-hermitian fields for finite $N$ is
not essential here%
} \begin{equation}
B_{iL}=\lambda_{iL}+A_{iL}=\lambda_{iL}+\mc A_{iL}+x_{iL}\Phi_{L}\end{equation}
 requiring that $\lambda_{iL}\mc A_{iL}=0$, and similarly for $B_{iR}$,
The stability of our geometry will depend on the behavior of $\Phi^{L}$
and $\Phi^{R}$. We calculate that

\begin{equation}
B_{\mu}B_{\mu}-\frac{N^{2}-1}{2}=N(\Phi_{L}+\Phi_{R})+\Phi_{L}\Phi_{L}+\Phi_{R}\Phi_{R}+\mc A_{\mu}\mc A_{\mu}-[\lambda_{\mu},\mc A_{\mu}]+\mc O(\frac{1}{N}),\end{equation}
 where we used that $\lambda_{ia}\mc A_{ia}=0$ and therefore both
$\mc A_{ia}x_{ia}=\mc O(\frac{1}{N})$ and $\mc A_{ia}[\lambda_{ia},\:\cdot\:]=\mc O(\frac{1}{N})$
for $a=L,R$. Setting \begin{eqnarray}
\Phi_{L}+\Phi_{R} & = & \Phi_{1},\\
\Phi_{L}-\Phi_{R} & = & \Phi_{2}\nonumber \end{eqnarray}
 we get \begin{equation}
B_{\mu}B_{\mu}-\frac{N^{2}-1}{2}=N\Phi_{1}+\Phi_{1}\Phi_{1}+\Phi_{2}\Phi_{2}+\mc A_{\mu}\mc A_{\mu}-[\lambda_{\mu},\mc A_{\mu}]+\mc O(\frac{1}{N}).\label{B_i B_i}\end{equation}
 In the limit $N\rightarrow\infty$ we can integrate out $\Phi_{1}$,
as it acquires an infinite mass. Alternatively we can rescale $\Phi_{1}$
by setting $\phi_{1}=\frac{1}{N}\Phi_{1}$. Then, all the terms involving
$\phi_{1}$ but the first one in (\ref{B_i B_i}) will be of order
$\frac{1}{N}$ and we can equally integrate out $\phi_{1}$.

The terms from \begin{equation}
F_{iL}F_{iL}+F_{iR}F_{iR}-[B_{iL},B_{iR}]^{2}\end{equation}
 with $F_{iL}=\frac{1}{2}\epsilon_{ijk}F_{jL\, kL}$ etc. involving
the remaining $\Phi_{2}$ will be (in the limit $N\rightarrow\infty$)
\begin{equation}
\frac{1}{2}\Phi_{2}\Phi_{2}-J_{\mu}(\Phi_{2})J_{\mu}(\Phi_{2})-F_{iL}x_{iL}\Phi_{2}+F_{iR}x_{iR}\Phi_{2}\end{equation}
 with the tangential derivatives $J_{ia}=-i\epsilon_{ijk}x_{ja}\dd_{ka}$.
Calculating that \begin{equation}
J_{\mu}\Phi_{2}J_{\mu}\Phi_{2}=-\dd_{\mu}\Phi_{2}\dd_{\mu}\Phi_{2}-x_{iL}\dd_{iL}\Phi_{2}x_{jL}\dd_{jL}\Phi_{2}-x_{iR}\dd_{iR}\Phi_{2}x_{jR}\dd_{jR}\Phi_{2}\end{equation}
 and using partial integration under the integral this gives \begin{equation}
\frac{1}{2}\Phi_{2}\Phi_{2}-\Phi_{2}\dd_{\mu}\dd_{\mu}\Phi_{2}-x_{iL}\dd_{iL}\Phi_{2}x_{jL}\dd_{jL}\Phi_{2}-x_{iR}\dd_{iR}\Phi_{2}x_{jR}\dd_{jR}\Phi_{2}-F_{iL}x_{iL}\Phi_{2}+F_{iR}x_{iR}\Phi_{2}\end{equation}
 Expanding both $\Phi_{2}$ and $F$ in left and right spherical harmonics
as \begin{equation}
\Phi_{2}=\sum_{klmn}c_{klmn}Y_{km}^{L}Y_{ln}^{R}\;\;\;\;\mbox{and}\;\;\;\; F_{ia}x_{ia}=\sum_{klmn}f_{klmn}^{a}Y_{km}^{L}Y_{ln}^{R}\end{equation}
 we get for fixed $klmn$, setting $c=c_{klmn}$, $f^{a}=f_{klmn}^{a}$
and $p=\frac{1}{2}+l(l+1)+k(k+1)$ the following expression \begin{equation}
pc^{2}-cf^{L}+cf^{R}=p(c-\frac{1}{2p}f^{L}+\frac{1}{2p}f^{R})^{2}-\frac{1}{4p}(f^{L}-f^{R})^{2}.\end{equation}
 Integrating out the $c$'s and putting everything back this leaves
us with the additional term \begin{equation}
-(F_{iL}x_{iL}-F_{iR}x_{iR})\frac{1}{4(\frac{1}{2}-\dd_{\mu}\dd_{\mu})}(F_{iL}x_{iL}-F_{iR}x_{iR})\end{equation}
 in the action (\ref{action-so6}).

\section{The\label{appendix b} Dirac operator in spherical coordinates}

For a general Riemannian manifold with metric\begin{equation}
g=g_{\mu\nu}dx^{\mu}dx^{\nu}\end{equation}
 the Christoffel symbols are given by\begin{equation}
\Gamma_{\mu\nu}^{\sigma}=\frac{1}{2}g^{\sigma\lambda}(\pp\mu g_{\lambda\nu}+\pp\nu g_{\lambda\mu}-\pp\lambda g_{\mu\nu}).\label{Christoffel symbol}\end{equation}
 We can change to a non-coordinate basis (labeled by Latin indices
in contrast to the Greek indices for the coordinates) by introducing
the vielbeins $e_{a}^{\mu}$ with \begin{eqnarray}
e_{\mu}^{a}e_{b}^{\mu} & = & \delta_{b}^{a},\\
g_{\mu\nu} & = & e_{\mu}^{a}e_{\nu}^{b}\delta_{ab},\qquad g^{\mu\nu}=e_{a}^{\mu}e_{b}^{\nu}\delta^{ab}.\nonumber \end{eqnarray}
 With these, the Dirac operator is given by\begin{equation}
D=-i\gamma^{a}e_{a}^{\mu}(\pp\mu+\frac{1}{4}\omega_{\mu ab}[\gamma^{a},\gamma^{b}]),\end{equation}
 where the $\gamma^{a}$ form a flat Clifford algebra, i. e.\begin{equation}
\{\gamma^{a},\gamma^{b}\}=2\delta^{ab}\;\;\;\;,\;\;\;\;\gamma^{a\dagger}=\gamma^{a}\end{equation}
 and the spin connection $\omega$ fulfills \begin{equation}
\pp\mu e_{\nu}^{a}-\Gamma_{\mu\nu}^{\lambda}e_{\lambda}^{a}+\omega_{\mu\:\: b}^{\:\: a}\: e_{\nu}^{b}=0.\label{spinor connection}\end{equation}

\subsection{The Dirac operator on $\mathbb{R}^{6}$ in spherical coordinates}

We will now write down the flat $SO(6)$ Dirac operator $D_{6}$ by
splitting $\mathbb{R}^{6}$ into $\mathbb{R}_{L}^{3}\times\mathbb{R}_{R}^{3}$
and introducing spherical coordinates on both the left and right hand
side. The flat metric becomes\begin{eqnarray}
g_{6} & = & r_{L}^{2}\; d\theta_{L}\otimes d\theta_{L}+r_{L}^{2}\sin^{2}\theta_{L}\; d\phi_{L}\otimes d\phi_{L}+dr_{L}\otimes dr_{L}\label{metric 6D}\\
 &  & +r_{R}^{2}\; d\theta_{R}\otimes d\theta_{R}+r_{R}^{2}\sin^{2}\theta_{R}\; d\phi_{R}\otimes d\phi_{R}+dr_{R}\otimes dr_{R}.\nonumber \end{eqnarray}
 Looking at the formula for the Christoffel symbols (\ref{Christoffel symbol}),
we see that all the symbols with both right and left indices vanish.
For the symbols with only right or only left indices we get\begin{eqnarray}
\Gamma_{\phi\phi}^{\theta} & = & -\sin\theta\cos\theta,\label{christoffel 6D}\\
\Gamma_{\theta\phi}^{\phi} & = & \frac{\cos\theta}{\sin\theta}=\Gamma_{\phi\theta,}^{\phi}\nonumber \\
\Gamma_{\theta\theta}^{r} & = & -r,\nonumber \\
\Gamma_{\phi\phi}^{r} & = & -r\sin^{2}\theta,\nonumber \\
\Gamma_{r\theta}^{\theta} & = & \frac{1}{r}=\Gamma_{\theta r}^{\theta},\nonumber \\
\Gamma_{r\phi}^{\phi} & = & \frac{1}{r}=\Gamma_{\phi r}^{\phi},\nonumber \end{eqnarray}
 where we have dropped the left or right subscript for simplicity.
All other symbols vanish. We want to go to a non-coordinate basis
by introducing the vielbeins\begin{eqnarray}
e_{\theta_{L}}^{1_{L}}=r_{L}; & e_{\phi_{L}}^{2_{L}}=r_{L}\sin\theta_{L}; & e_{r_{L}}^{3_{L}}=1;\label{vielbein 6D}\\
e_{\theta_{R}}^{1_{R}}=r_{L}; & e_{\phi_{R}}^{2_{R}}=r_{R}\sin\theta_{R}; & e_{r_{R}}^{3_{R}}=1.\label{vielbein 6D 2}\end{eqnarray}
 Calculating the spinor connection by (\ref{spinor connection}),
we again see that all the terms with both left and right indices vanish.
The terms with only left or only right indices are \begin{eqnarray}
\omega_{\phi\:\:2}^{\:\:1}= & -\cos\theta & =-\omega_{\phi\:\:1}^{\:\:2},\label{spin conn 6D}\\
\omega_{\phi\:\:3}^{\:\:2}= & \sin\theta & =-\omega_{\phi\:\:2}^{\:\:3},\nonumber \\
\omega_{\theta\:\:3}^{\:\:1}= & 1 & =-\omega_{\theta\:\:1}^{\:\:3},\nonumber \end{eqnarray}
 where we again dropped the left or right subscripts. Putting all
this together we see that $D_{6}$ splits up into a left part $D_{3L}$
and a right part $D_{3R}$ as \begin{equation}
D_{6}=D_{3L}+D_{3R}\label{D6-3plus3}\end{equation}
 with \begin{eqnarray}
D_{3L} & = & -i\overline{\Gamma}_{L}^{1}\frac{1}{r_{L}}(\dd_{\theta_{L}}+\frac{\cos\theta_{L}}{\sin\theta_{L}})-i\overline{\Gamma}_{L}^{2}\frac{1}{r_{L}\sin\theta_{L}}\dd_{\phi_{L}}-i\overline{\Gamma}_{L}^{3}(\dd_{r_{L}}+\frac{1}{r_{L}}),\label{D3L spherical}\\
D_{3R} & = & -i\overline{\Gamma}_{R}^{1}\frac{1}{r_{R}}(\dd_{\theta_{R}}+\frac{\cos\theta_{R}}{\sin\theta_{R}})-i\overline{\Gamma}_{R}^{2}\frac{1}{r_{R}\sin\theta_{R}}\dd_{\phi_{R}}-i\overline{\Gamma}_{R}^{3}(\dd_{r_{R}}+\frac{1}{r_{R}}).\label{D3R spherical}\end{eqnarray}
 where the $\overline{\Gamma}$ have to form a $SO(6)$ Clifford algebra.

\subsection{The Dirac operator on $S^{2}\times S^{2}$ }

We now want to calculate the curved Dirac operator $D_{4}$ on $S^{2}\times S^{2}$
in the spherical coordinates of the spheres (they are the same spherical
coordinates we used before, now restricted to the spheres). The metric
on $S^{2}\times S^{2}$ with radii $r_{L}$ and $r_{R}$ is\begin{eqnarray}
g_{4} & = & r_{L}^{2}\; d\theta_{L}\otimes d\theta_{L}+r_{L}^{2}\sin^{2}\theta_{L}\; d\phi_{L}\otimes d\phi_{L}\\
 &  & +r_{R}^{2}\; d\theta_{R}\otimes d\theta_{R}+r_{R}^{2}\sin^{2}\theta_{R}\; d\phi_{R}\otimes d\phi_{R}.\nonumber \end{eqnarray}
 The metric is the same as (\ref{metric 6D}) restricted to the spheres,
so the Christoffel symbols are the same as (\ref{christoffel 6D}).
Again introducing the vielbeins\begin{eqnarray}
e_{\theta_{L}}^{1_{L}}=r_{L}; & e_{\phi_{L}}^{2_{L}}=r_{L}\sin\theta_{L};\label{vielbein 4D}\\
e_{\theta_{R}}^{1_{R}}=r_{L}; & e_{\phi_{R}}^{2_{R}}=r_{R}\sin\theta_{R},\end{eqnarray}
 we see that also the spin connection is the same as (\ref{spin conn 6D}),
and therefore we can again split $D_{4}$ into a right part $D_{2R}$
and a left part $D_{2L}$ as $D_{4}=D_{2L}+D_{2R}$ with

\begin{eqnarray}
D_{2L} & = & -i\widetilde{\Gamma}_{L}^{1}\frac{1}{r_{L}}(\dd_{\theta_{L}}+\frac{\cos\theta_{L}}{\sin\theta_{L}})-i\widetilde{\Gamma}_{L}^{2}\frac{1}{r_{L}\sin\theta_{L}}\dd_{\phi_{L}},\label{D2-L}\\
D_{2R} & = & -i\widetilde{\Gamma}_{R}^{1}\frac{1}{r_{R}}(\dd_{\theta_{R}}+\frac{\cos\theta_{R}}{\sin\theta_{R}})-i\widetilde{\Gamma}_{R}^{2}\frac{1}{r_{R}\sin\theta_{R}}\dd_{\phi_{R},}\label{D2-R}\end{eqnarray}
 where the $\widetilde{\Gamma}$ form a flat $SO(4)$ Clifford algebra.

\subsection{$SO(3)\times SO(3)$-covariant form of the Dirac operator on $S^{2}\times S^{2}$ }

\label{sec:so6-dirac-4}

The flat $SO(6)$ Dirac operator $D_{6}$ can be split into a left
part $D_{3L}$ and a right part $D_{3R}$ using spherical coordinates
in \ref{D6-3plus3}. Of course, $D_{6}$ can also be written in the
usual Euclidian coordinates as \begin{equation}
D_{6}=-i\Gamma^{\mu}\pp\mu,\end{equation}
 where again we can split it into a left and a right part as \begin{equation}
D_{6}=D_{3L}+D_{3R}\end{equation}
 with\begin{eqnarray}
 &  & D_{3L}=-i\Gamma_{L}^{i}\pp i,\qquad D_{3R}=-i\Gamma_{R}^{i}\pp i,\label{D3 euklid}\\
 &  & \qquad\qquad\{ D_{3L},D_{3R}\}=0.\nonumber \end{eqnarray}
 We have left open which representation of the $SO(6)$ Clifford algebra
we want to use for the $\overline{\Gamma}$ in (\ref{D3L spherical},\ref{D3R spherical}),
but $\Gamma$ in (\ref{D3 euklid}) is really the representation given
by (\ref{Gamma}). We will now relate the two expressions for the
Clifford algebra and the Dirac operator by first defining \begin{equation}
J_{iL}=-i\epsilon_{ijk}x_{jL}\dd_{kL}\;\;\;\;\mbox{and}\;\;\;\; J_{iR}=-i\epsilon_{ijk}x_{jR}\dd_{kR}\end{equation}
 and noting that \begin{equation}
\left(\frac{\Gamma_{L}^{i}x_{iL}}{r_{L}}\right)^{2}=\left(\frac{\Gamma_{R}^{i}x_{iR}}{r_{R}}\right)^{2}=1.\label{eins calc}\end{equation}
 We calculate that \begin{eqnarray}
\left(\frac{\Gamma_{L}^{j}x_{jL}}{r_{L}}\right)^{2}\Gamma_{L}^{i}\dd_{iL} & = & \left(\frac{\Gamma_{L}^{j}x_{jL}}{r_{L}}\right)\left(\frac{x_{iL}\dd_{iL}}{r_{L}}-\frac{1}{r_{L}}\left(\begin{array}{cc}
\gamma_{L}^{i} & 0\\
0 & \gamma_{L}^{i}\end{array}\right)J_{iL}\right)\!\!,\label{calc gamma L}\\
\left(\frac{\Gamma_{R}^{j}x_{jR}}{r_{R}}\right)^{2}\Gamma_{R}^{i}\dd_{iR} & = & \left(\frac{\Gamma_{R}^{j}x_{jR}}{r_{R}}\right)\left(\frac{x_{iR}\dd_{iR}}{r_{R}}+\frac{i}{r_{R}}\left(\begin{array}{cc}
\gamma_{R}^{i} & 0\\
0 & \gamma_{R}^{i}\end{array}\right)J_{iR}\right)\!\!,\label{calc gamma R}\end{eqnarray}
 and therefore \begin{eqnarray}
D_{3L} & = & -i\left(\frac{\Gamma_{L}^{j}x_{jL}}{r_{L}}\right)\left(\dd_{r_{L}}-\frac{1}{r_{L}}\left(\begin{array}{cc}
\gamma_{L}^{i} & 0\\
0 & \gamma_{L}^{i}\end{array}\right)J_{iL}\right)\!\!,\label{calc D3L}\\
D_{3R} & = & -i\left(\frac{\Gamma_{R}^{j}x_{jR}}{r_{R}}\right)\left(\dd_{r_{R}}+\frac{i}{r_{R}}\left(\begin{array}{cc}
\gamma_{R}^{i} & 0\\
0 & \gamma_{R}^{i}\end{array}\right)J_{iR}\right)\!\!.\label{calc D3R}\end{eqnarray}
 Comparing this with (\ref{D3L spherical},\ref{D3R spherical}) we
see that \begin{equation}
\overline{\Gamma}_{L}^{3}=\left(\frac{\Gamma_{L}^{i}x_{iL}}{r_{L}}\right)\;\;\;\;\mbox{and}\;\;\;\;\overline{\Gamma}_{R}^{3}=\left(\frac{\Gamma_{R}^{i}x_{iR}}{r_{R}}\right)\!\!,\label{Gamma 3 calc}\end{equation}
 as the $J_{L}$ and $J_{R}$ have no radial components. From (\ref{calc D3L},\ref{calc D3R})
we can also deduce that \begin{equation}
[\overline{\Gamma}_{L}^{i},\left(\begin{array}{cc}
0 & 1\\
1 & 0\end{array}\right)]=0=[\overline{\Gamma}_{R}^{i},\left(\begin{array}{cc}
0 & 1\\
-1 & 0\end{array}\right)]\label{comm}\end{equation}
 and\begin{equation}
\{\overline{\Gamma}_{R}^{i},\left(\begin{array}{cc}
0 & 1\\
1 & 0\end{array}\right)\}=0=\{\overline{\Gamma}_{L}^{i},\left(\begin{array}{cc}
0 & 1\\
-1 & 0\end{array}\right)\}.\label{anti}\end{equation}
 The curved Dirac operator $D_{4}$ on $S^{2}\times S^{2}$ expressed
in the spherical coordinates of the spheres also splits up as $D_{4}=D_{2L}+D_{2R}$
with right part $D_{2R}$ and left part $D_{2L}$ given in (\ref{D2-L},\ref{D2-R}).
Comparing this with (\ref{D3L spherical},\ref{D3R spherical}), we
see that the dependence on the tangential coordinates is the same
in both expressions. With (\ref{comm},\ref{anti}) we see that the
matrices $-i\left(\begin{array}{cc}
0 & 1\\
1 & 0\end{array}\right)\overline{\Gamma}_{L}^{3}\overline{\Gamma}_{L}^{i}$ and $\left(\begin{array}{cc}
0 & 1\\
-1 & 0\end{array}\right)\overline{\Gamma}_{R}^{3}\overline{\Gamma}_{R}^{j}$ for $i,j=1,2$ form a $SO(4)$ Clifford algebra and can therefore
be used as the $\widetilde{\Gamma}$. Note that this representation
is still reducible, a problem we deal with in chapter \ref{chapter on projections}.
Now we can get a simple relation between the $D_{3}$ restricted on
the spheres and the $D_{2}$\begin{eqnarray}
-\left(\begin{array}{cc}
0 & 1\\
1 & 0\end{array}\right)(i\overline{\Gamma}_{L}^{3}D_{3L}|_{res.}-\frac{1}{r_{L}}) & = & D_{2L},\\
-i\left(\begin{array}{cc}
0 & 1\\
-1 & 0\end{array}\right)(i\overline{\Gamma}_{R}^{3}D_{3R}|_{res.}-\frac{1}{r_{R}}) & = & D_{2R}.\end{eqnarray}
 Inserting (\ref{calc D3L},\ref{calc D3R}) and using (\ref{Gamma 3 calc})
together with (\ref{eins calc}) we find that \begin{eqnarray}
D_{2L} & = & \frac{1}{r_{L}}(\Gamma_{L}^{i}J_{iL}+\left(\begin{array}{cc}
0 & 1\\
1 & 0\end{array}\right)),\label{D 2L final}\\
D_{2R} & = & \frac{1}{r_{R}}(\Gamma_{R}^{i}J_{iR}+i\left(\begin{array}{cc}
0 & 1\\
-1 & 0\end{array}\right)).\label{D 2R final}\end{eqnarray}
 Setting $r_{L}=r_{R}=1$ for simplicity, the Dirac operator $D_{4}$
on $S^{2}\times S^{2}$ takes the form (\ref{Dirac operator}).

\end{appendix}

\newpage

\fancyhead[LE,RO]{\thepage} 

\fancyhead[LO,RE]{\nouppercase{\leftmark}}

\bibliographystyle{neuuu}
\bibliography{mainbib}

\begin{thebibliography}{100}

\bibitem{Aganagic:2000mh}
M.~Aganagic, R.~Gopakumar, S.~Minwalla and A.~Strominger,
\newblock {\em Unstable solitons in noncommutative gauge theory},
\newblock JHEP {\bf 04}, 001 (2001),
\newblock [hep-th/0009142].

\bibitem{Alexanian:2001qj}
G.~Alexanian, A.~P. Balachandran, G.~Immirzi and B.~Ydri,
\newblock {\em Fuzzy {$CP(2)$}},
\newblock J. Geom. Phys. {\bf 42}, 28 (2002),
\newblock [hep-th/0103023].

\bibitem{Ambjorn:1999ts}
J.~Ambjorn, Y.~M. Makeenko, J.~Nishimura and R.~J. Szabo,
\newblock {\em Finite N matrix models of noncommutative gauge theory},
\newblock JHEP {\bf 11}, 029 (1999),
\newblock [hep-th/9911041].

\bibitem{Ambjorn:2000cs}
J.~Ambjorn, Y.~M. Makeenko, J.~Nishimura and R.~J. Szabo,
\newblock {\em Lattice gauge fields and discrete noncommutative Yang-Mills
  theory},
\newblock JHEP {\bf 05}, 023 (2000),
\newblock [hep-th/0004147].

\bibitem{Ambjorn:2000nb}
J.~Ambjorn, Y.~M. Makeenko, J.~Nishimura and R.~J. Szabo,
\newblock {\em Nonperturbative dynamics of noncommutative gauge theory},
\newblock Phys. Lett. {\bf B480}, 399 (2000),
\newblock [hep-th/0002158].

\bibitem{Aoki:2002fq}
H.~Aoki, S.~Iso and K.~Nagao,
\newblock {\em Ginsparg-Wilson relation, topological invariants and finite
  noncommutative geometry},
\newblock Phys. Rev. {\bf D67}, 085005 (2003),
\newblock [hep-th/0209223].

\bibitem{Arnal:2000hy}
D.~Arnal, D.~Manchon and M.~Masmoudi,
\newblock {\em Choix des signes pour la formalite de M. Kontsevich},
\newblock [math.qa/0003003].

\bibitem{Aschieri:2005yw}
P.~Aschieri {\em et~al.},
\newblock {\em A gravity theory on noncommutative spaces},
\newblock [hep-th/0504183].

\bibitem{Balachandran:2000du}
A.~P. Balachandran, T.~R. Govindarajan and B.~Ydri,
\newblock {\em The fermion doubling problem and noncommutative geometry. II},
\newblock [hep-th/0006216].

\bibitem{Balachandran:2003ay}
A.~P. Balachandran and G.~Immirzi,
\newblock {\em The fuzzy {Ginsparg-Wilson} algebra: A solution of the fermion
  doubling problem},
\newblock Phys. Rev. {\bf D68}, 065023 (2003),
\newblock [hep-th/0301242].

\bibitem{Bayen:1977ha}
F.~Bayen, M.~Flato, C.~Fronsdal, A.~Lichnerowicz and D.~Sternheimer,
\newblock {\em Deformation Theory and Quantization. 1. Deformations of
  Symplectic Structures},
\newblock Ann. Phys. {\bf 111}, 61 (1978).

\bibitem{Bayen:1977hb}
F.~Bayen, M.~Flato, C.~Fronsdal, A.~Lichnerowicz and D.~Sternheimer,
\newblock {\em Deformation Theory and Quantization. 2. Physical Applications},
\newblock Ann. Phys. {\bf 111}, 111 (1978).

\bibitem{Behr:2002wx}
W.~Behr {\em et~al.},
\newblock {\em The Z --> gamma gamma, g g decays in the noncommutative standard
  model},
\newblock Eur. Phys. J. {\bf C29}, 441 (2003),
\newblock [hep-ph/0202121].

\bibitem{Behr:2005wp}
W.~Behr, F.~Meyer and H.~Steinacker,
\newblock {\em Gauge theory on fuzzy $S^2 \times S^2$ and regularization on
  noncommutative $R^4$},
\newblock accepted by JHEP  (2005),
\newblock [hep-th/0503041].

\bibitem{Behr:2003qc}
W.~Behr and A.~Sykora,
\newblock {\em Construction of gauge theories on curved noncommutative
  spacetime},
\newblock Nucl. Phys. {\bf B698}, 473 (2004),
\newblock [hep-th/0309145].

\bibitem{Behr:2003hg}
W.~Behr and A.~Sykora,
\newblock {\em NC Wilson lines and the inverse Seiberg-Witten map for
  nondegenerate star products},
\newblock Eur. Phys. J. {\bf C35}, 145 (2004),
\newblock [hep-th/0312138].

\bibitem{Brace:2001fj}
D.~Brace, B.~L. Cerchiai, A.~F. Pasqua, U.~Varadarajan and B.~Zumino,
\newblock {\em A cohomological approach to the non-Abelian Seiberg-Witten map},
\newblock JHEP {\bf 06}, 047 (2001),
\newblock [hep-th/0105192].

\bibitem{Calmet:2001na}
X.~Calmet, B.~Jurco, P.~Schupp, J.~Wess and M.~Wohlgenannt,
\newblock {\em The standard model on non-commutative space-time},
\newblock Eur. Phys. J. {\bf C23}, 363 (2002),
\newblock [hep-ph/0111115].

\bibitem{Calmet:2003jv}
X.~Calmet and M.~Wohlgenannt,
\newblock {\em Effective field theories on non-commutative space-time},
\newblock Phys. Rev. {\bf D68}, 025016 (2003),
\newblock [hep-ph/0305027].

\bibitem{Carow-Watamura:2004ct}
U.~Carow-Watamura, H.~Steinacker and S.~Watamura,
\newblock {\em Monopole bundles over fuzzy complex projective spaces},
\newblock [hep-th/0404130].

\bibitem{Carow-Watamura:1998jn}
U.~Carow-Watamura and S.~Watamura,
\newblock {\em Noncommutative geometry and gauge theory on fuzzy sphere},
\newblock Commun. Math. Phys. {\bf 212}, 395 (2000),
\newblock [hep-th/9801195].

\bibitem{Castro-Villarreal:2004vh}
P.~Castro-Villarreal, R.~Delgadillo-Blando and B.~Ydri,
\newblock {\em A gauge-invariant {UV-IR} mixing and the corresponding phase
  transition for {$U(1)$} fields on the fuzzy sphere},
\newblock Nucl. Phys. {\bf B704}, 111 (2005),
\newblock [hep-th/0405201].

\bibitem{Cerchiai:2000qi}
B.~L. Cerchiai, G.~Fiore and J.~Madore,
\newblock {\em Frame formalism for the N-dimensional quantum Euclidean spaces},
\newblock [math.qa/0007044].

\bibitem{Cerchiai:1998ef}
B.~L. Cerchiai, R.~Hinterding, J.~Madore and J.~Wess,
\newblock {\em The Geometry of a $q$-Deformed Phase Space},
\newblock Eur. J. Phys. {\bf C8}, 533 (1999),
\newblock [math.qa/9807123].

\bibitem{Cerchiai:2002ss}
B.~L. Cerchiai, A.~F. Pasqua and B.~Zumino,
\newblock {\em The Seiberg-Witten map for noncommutative gauge theories},
\newblock [hep-th/0206231].

\bibitem{Chu:1998qz}
C.-S. Chu and P.-M. Ho,
\newblock {\em Noncommutative open string and D-brane},
\newblock Nucl. Phys. {\bf B550}, 151 (1999),
\newblock [hep-th/9812219].

\bibitem{Chu:2001cx}
C.-S. Chu, V.~V. Khoze and G.~Travaglini,
\newblock {\em Notes on noncommutative instantons},
\newblock Nucl. Phys. {\bf B621}, 101 (2002),
\newblock [hep-th/0108007].

\bibitem{Chu:2001xi}
C.-S. Chu, J.~Madore and H.~Steinacker,
\newblock {\em Scaling limits of the fuzzy sphere at one loop},
\newblock JHEP {\bf 08}, 038 (2001),
\newblock [hep-th/0106205].

\bibitem{Connes:1996gi}
A.~Connes,
\newblock {\em Gravity coupled with matter and the foundation of non-
  commutative geometry},
\newblock Commun. Math. Phys. {\bf 182}, 155 (1996),
\newblock [hep-th/9603053].

\bibitem{Correa:2004cm}
D.~H. Correa, C.~D. Fosco, F.~A. Schaposnik and G.~Torroba,
\newblock {\em On coordinate transformations in planar noncommutative
  theories},
\newblock JHEP {\bf 09}, 064 (2004),
\newblock [hep-th/0407220].

\bibitem{Das:2000md}
S.~R. Das and S.-J. Rey,
\newblock {\em Open Wilson lines in noncommutative gauge theory and tomography
  of holographic dual supergravity},
\newblock Nucl. Phys. {\bf B590}, 453 (2000),
\newblock [hep-th/0008042].

\bibitem{Dimakis:1996}
A.~Dimakis and J.~Madore,
\newblock {\em Differential calculi an linear connections},
\newblock J. Math. Phys. {\bf 37}, 4647 (1996).

\bibitem{Dimitrijevic:2003wv}
M.~Dimitrijevic {\em et~al.},
\newblock {\em Deformed field theory on kappa-spacetime},
\newblock Eur. Phys. J. {\bf C31}, 129 (2003),
\newblock [hep-th/0307149].

\bibitem{Dimitrijevic:2003pn}
M.~Dimitrijevic, F.~Meyer, L.~Moller and J.~Wess,
\newblock {\em Gauge theories on the kappa-Minkowski spacetime},
\newblock Eur. Phys. J. {\bf C36}, 117 (2004),
\newblock [hep-th/0310116].

\bibitem{Dimitrijevic:2004vv}
M.~Dimitrijevic, L.~Moller and E.~Tsouchnika,
\newblock {\em Derivatives, forms and vector fields on the kappa-deformed
  Euclidean space},
\newblock J. Phys. {\bf A37}, 9749 (2004),
\newblock [hep-th/0404224].

\bibitem{Doplicher:1994tu}
S.~Doplicher, K.~Fredenhagen and J.~E. Roberts,
\newblock {\em The Quantum structure of space-time at the Planck scale and
  quantum fields},
\newblock Commun. Math. Phys. {\bf 172}, 187 (1995),
\newblock [hep-th/0303037].

\bibitem{Douglas:2001ba}
M.~R. Douglas and N.~A. Nekrasov,
\newblock {\em Noncommutative field theory},
\newblock Rev. Mod. Phys. {\bf 73}, 977 (2001),
\newblock [hep-th/0106048].

\bibitem{Felder:2000hy}
G.~Felder and B.~Shoikhet,
\newblock {\em Deformation quantization with traces},
\newblock [math.qa/0002057].

\bibitem{Filk:1996dm}
T.~Filk,
\newblock {\em Divergencies in a field theory on quantum space},
\newblock Phys. Lett. {\bf B376}, 53 (1996).

\bibitem{Fosco:2004yz}
C.~D. Fosco and G.~Torroba,
\newblock {\em Noncommutative theories and general coordinate transformations},
\newblock Phys. Rev. {\bf D71}, 065012 (2005),
\newblock [hep-th/0409240].

\bibitem{Furuuchi:1999kv}
K.~Furuuchi,
\newblock {\em Instantons on noncommutative {$R^4$} and projection operators},
\newblock Prog. Theor. Phys. {\bf 103}, 1043 (2000),
\newblock [hep-th/9912047].

\bibitem{Furuuchi:2000vc}
K.~Furuuchi,
\newblock {\em Topological charge of {$U(1)$} instantons on noncommutative
  {$R^4$}},
\newblock Prog. Theor. Phys. Suppl. {\bf 144}, 79 (2001),
\newblock [hep-th/0010006].

\bibitem{Ginsparg:1981bj}
P.~H. Ginsparg and K.~G. Wilson,
\newblock {\em A REMNANT OF CHIRAL SYMMETRY ON THE LATTICE},
\newblock Phys. Rev. {\bf D25}, 2649 (1982).

\bibitem{Griguolo:2001ce}
L.~Griguolo, D.~Seminara and P.~Valtancoli,
\newblock {\em Towards the solution of noncommutative {YM(2)}: Morita
  equivalence and large {N}-limit},
\newblock JHEP {\bf 12}, 024 (2001),
\newblock [hep-th/0110293].

\bibitem{Griguolo:2003kq}
L.~Griguolo and D.~Seminara,
\newblock {\em Classical solutions of the {TEK} model and noncommutative
  instantons in two dimensions},
\newblock JHEP {\bf 03}, 068 (2004),
\newblock [hep-th/0311041].

\bibitem{Griguolo:2004jp}
L.~Griguolo, D.~Seminara and R.~J. Szabo,
\newblock {\em Instantons, fluxons and open gauge string theory},
\newblock [hep-th/0411277].

\bibitem{Groenewold:1946kp}
H.~J. Groenewold,
\newblock {\em On the Principles of elementary quantum mechanics},
\newblock Physica {\bf 12}, 405 (1946).

\bibitem{Gross:2000ba}
D.~J. Gross, A.~Hashimoto and N.~Itzhaki,
\newblock {\em Observables of non-commutative gauge theories},
\newblock Adv. Theor. Math. Phys. {\bf 4}, 893 (2000),
\newblock [hep-th/0008075].

\bibitem{Gross:2000ph}
D.~J. Gross and N.~A. Nekrasov,
\newblock {\em Dynamics of strings in noncommutative gauge theory},
\newblock JHEP {\bf 10}, 021 (2000),
\newblock [hep-th/0007204].

\bibitem{Gross:2000ss}
D.~J. Gross and N.~A. Nekrasov,
\newblock {\em Solitons in noncommutative gauge theory},
\newblock JHEP {\bf 03}, 044 (2001),
\newblock [hep-th/0010090].

\bibitem{Grosse:1995pr}
H.~Grosse, C.~Klimcik and P.~Presnajder,
\newblock {\em Field theory on a supersymmetric lattice},
\newblock Commun. Math. Phys. {\bf 185}, 155 (1997),
\newblock [hep-th/9507074].

\bibitem{Grosse:1999ci}
H.~Grosse and A.~Strohmaier,
\newblock {\em Noncommutative geometry and the regularization problem of 4D
  quantum field theory},
\newblock Lett. Math. Phys. {\bf 48}, 163 (1999),
\newblock [hep-th/9902138].

\bibitem{Grosse:2004wm}
H.~Grosse and H.~Steinacker,
\newblock {\em Finite gauge theory on fuzzy {$\mathbb{C} P^2$}},
\newblock Nucl. Phys. {\bf B707}, 145 (2005),
\newblock [hep-th/0407089].

\bibitem{Grosse:2004ik}
H.~Grosse and R.~Wulkenhaar,
\newblock {\em Renormalisation of phi**4 theory on noncommutative R**4 to all
  orders},
\newblock [hep-th/0403232].

\bibitem{Hamanaka:2001dr}
M.~Hamanaka,
\newblock {\em {ADHM/Nahm} construction of localized solitons in noncommutative
  gauge theories},
\newblock Phys. Rev. {\bf D65}, 085022 (2002),
\newblock [hep-th/0109070].

\bibitem{Harvey:2000jb}
J.~A. Harvey, P.~Kraus and F.~Larsen,
\newblock {\em Exact noncommutative solitons},
\newblock JHEP {\bf 12}, 024 (2000),
\newblock [hep-th/0010060].

\bibitem{Imai:2003vr}
T.~Imai, Y.~Kitazawa, Y.~Takayama and D.~Tomino,
\newblock {\em Quantum corrections on fuzzy sphere},
\newblock Nucl. Phys. {\bf B665}, 520 (2003),
\newblock [hep-th/0303120].

\bibitem{Ishibashi:1996xs}
N.~Ishibashi, H.~Kawai, Y.~Kitazawa and A.~Tsuchiya,
\newblock {\em A large-{N} reduced model as superstring},
\newblock Nucl. Phys. {\bf B498}, 467 (1997),
\newblock [hep-th/9612115].

\bibitem{Ishibashi:1999hs}
N.~Ishibashi, S.~, H.~Kawai and Y.~Kitazawa,
\newblock {\em Wilson loops in noncommutative Yang-Mills},
\newblock Nucl. Phys. {\bf B573}, 573 (2000),
\newblock [hep-th/9910004].

\bibitem{Iso:2001mg}
S.~Iso, Y.~Kimura, K.~Tanaka and K.~Wakatsuki,
\newblock {\em Noncommutative gauge theory on fuzzy sphere from matrix model},
\newblock Nucl. Phys. {\bf B604}, 121 (2001),
\newblock [hep-th/0101102].

\bibitem{Ivanova:2005fh}
T.~A. Ivanova and O.~Lechtenfeld,
\newblock {\em Noncommutative instantons in 4k dimensions},
\newblock [hep-th/0502117].

\bibitem{Jambor:2004kc}
C.~Jambor and A.~Sykora,
\newblock {\em Realization of algebras with the help of *-products},
\newblock [hep-th/0405268].

\bibitem{Jurco:2001rq}
B.~Jurco, L.~Moller, S.~Schraml, P.~Schupp and J.~Wess,
\newblock {\em Construction of non-Abelian gauge theories on noncommutative
  spaces},
\newblock Eur. Phys. J. {\bf C21}, 383 (2001),
\newblock [hep-th/0104153].

\bibitem{Jurco:2000ja}
B.~Jurco, S.~Schraml, P.~Schupp and J.~Wess,
\newblock {\em Enveloping algebra valued gauge transformations for non- Abelian
  gauge groups on non-commutative spaces},
\newblock Eur. Phys. J. {\bf C17}, 521 (2000),
\newblock [hep-th/0006246].

\bibitem{Jurco:2000fs}
B.~Jurco, P.~Schupp and J.~Wess,
\newblock {\em Noncommutative gauge theory for Poisson manifolds},
\newblock Nucl. Phys. {\bf B584}, 784 (2000),
\newblock [hep-th/0005005].

\bibitem{Jurco:2001my}
B.~Jurco, P.~Schupp and J.~Wess,
\newblock {\em Nonabelian noncommutative gauge theory via noncommutative extra
  dimensions},
\newblock Nucl. Phys. {\bf B604}, 148 (2001),
\newblock [hep-th/0102129].

\bibitem{Jurco:2001kp}
B.~Jurco, P.~Schupp and J.~Wess,
\newblock {\em Noncommutative line bundle and Morita equivalence},
\newblock Lett. Math. Phys. {\bf 61}, 171 (2002),
\newblock [hep-th/0106110].

\bibitem{Karabali:2001te}
D.~Karabali, V.~P. Nair and A.~P. Polychronakos,
\newblock {\em Spectrum of {Schroedinger} field in a noncommutative magnetic
  monopole},
\newblock Nucl. Phys. {\bf B627}, 565 (2002),
\newblock [hep-th/0111249].

\bibitem{Kimura:2002nq}
Y.~Kimura,
\newblock {\em Noncommutative gauge theory on fuzzy four-sphere and matrix
  model},
\newblock Nucl. Phys. {\bf B637}, 177 (2002),
\newblock [hep-th/0204256].

\bibitem{Kitazawa:2002xj}
Y.~Kitazawa,
\newblock {\em Matrix models in homogeneous spaces},
\newblock Nucl. Phys. {\bf B642}, 210 (2002),
\newblock [hep-th/0207115].

\bibitem{Kontsevich:1997vb}
M.~Kontsevich,
\newblock {\em Deformation quantization of Poisson manifolds, I},
\newblock Lett. Math. Phys. {\bf 66}, 157 (2003),
\newblock [q-alg/9709040].

\bibitem{Lorek:1997eh}
A.~Lorek, W.~Weich and J.~Wess,
\newblock {\em Non-commutative Euclidean and Minkowski structures},
\newblock Z. Phys. {\bf C76}, 375 (1997).

\bibitem{Madore:1991bw}
J.~Madore,
\newblock {\em The fuzzy sphere},
\newblock Class. Quant. Grav. {\bf 9}, 69 (1992).

\bibitem{Madore:1992}
J.~Madore,
\newblock {\em Gravity on fuzzy space-time},
\newblock Class. Quant. Grav. {\bf 9}, 69 (1992).

\bibitem{Madore:1997ta}
J.~Madore,
\newblock {\em The Fuzzy Sphere},
\newblock [gr-qc/9709002].

\bibitem{Madore:1999bi}
J.~Madore,
\newblock {\em Noncommutative geometry for pedestrians},
\newblock [gr-qc/9906059].

\bibitem{Madore:2000aq}
J.~Madore,
\newblock {\em An introduction to noncommutative differential geometry and its
  physical applications},
\newblock Lond. Math. Soc. Lect. Note Ser. {\bf 257}, 1 (2000),
\newblock Cambridge Univ. Pr.

\bibitem{Madore:2000en}
J.~Madore, S.~Schraml, P.~Schupp and J.~Wess,
\newblock {\em Gauge theory on noncommutative spaces},
\newblock Eur. Phys. J. {\bf C16}, 161 (2000),
\newblock [hep-th/0001203].

\bibitem{Manchon:2000hy}
D.~Manchon,
\newblock {\em Poisson bracket, deformed bracket and gauge group actions in
  Kontsevich deformation quantization},
\newblock [math.qa/0003004].

\bibitem{Melic:2005hb}
B.~Melic, K.~Passek-Kumericki and J.~Trampetic,
\newblock {\em Quarkonia decays into two photons induced by the space-time
  non-commutativity},
\newblock [hep-ph/0503133].

\bibitem{Melic:2005fm}
B.~Melic, K.~Passek-Kumericki, J.~Trampetic, P.~Schupp and M.~Wohlgenannt,
\newblock {\em The standard model on non-commutative space-time: Electroweak
  currents and Higgs sector},
\newblock [hep-ph/0502249].

\bibitem{Melic:2005am}
B.~Melic, K.~Passek-Kumericki, J.~Trampetic, P.~Schupp and M.~Wohlgenannt,
\newblock {\em The standard model on non-commutative space-time: Strong
  interactions included},
\newblock [hep-ph/0503064].

\bibitem{Meyer:2003wj}
F.~Meyer and H.~Steinacker,
\newblock {\em Gauge field theory on the E(q)(2)-covariant plane},
\newblock Int. J. Mod. Phys. {\bf A19}, 3349 (2004),
\newblock [hep-th/0309053].

\bibitem{Mikulovic:2003sq}
D.~Mikulovic,
\newblock {\em Seiberg-Witten map for superfields on canonically deformed N =
  1, d = 4 superspace},
\newblock JHEP {\bf 01}, 063 (2004),
\newblock [hep-th/0310065].

\bibitem{Mikulovic:2004qj}
D.~Mikulovic,
\newblock {\em Seiberg-Witten map for superfields on N = (1/2,0) and N =
  (1/2,1/2) deformed superspace},
\newblock JHEP {\bf 05}, 077 (2004),
\newblock [hep-th/0403290].

\bibitem{Minwalla:1999px}
S.~Minwalla, M.~Van~Raamsdonk and N.~Seiberg,
\newblock {\em Noncommutative perturbative dynamics},
\newblock JHEP {\bf 02}, 020 (2000),
\newblock [hep-th/9912072].

\bibitem{Moyal:1949sk}
J.~E. Moyal,
\newblock {\em Quantum mechanics as a statistical theory},
\newblock Proc. Cambridge Phil. Soc. {\bf 45}, 99 (1949).

\bibitem{Nekrasov:1998ss}
N.~Nekrasov and A.~Schwarz,
\newblock {\em Instantons on noncommutative {$R^4$} and (2,0) superconformal
  six dimensional theory},
\newblock Commun. Math. Phys. {\bf 198}, 689 (1998),
\newblock [hep-th/9802068].

\bibitem{Ohl:2004tn}
T.~Ohl and J.~Reuter,
\newblock {\em Testing the noncommutative standard model at a future photon
  collider},
\newblock Phys. Rev. {\bf D70}, 076007 (2004),
\newblock [hep-ph/0406098].

\bibitem{Okawa:2001mv}
Y.~Okawa and H.~Ooguri,
\newblock {\em An exact solution to Seiberg-Witten equation of noncommutative
  gauge theory},
\newblock Phys. Rev. {\bf D64}, 046009 (2001),
\newblock [hep-th/0104036].

\bibitem{Paniak:2002fi}
L.~D. Paniak and R.~J. Szabo,
\newblock {\em Instanton expansion of noncommutative gauge theory in two
  dimensions},
\newblock Commun. Math. Phys. {\bf 243}, 343 (2003),
\newblock [hep-th/0203166].

\bibitem{Paniak:2003xm}
L.~D. Paniak and R.~J. Szabo,
\newblock {\em Lectures on two-dimensional noncommutative gauge theory. {II}:
  Quantization},
\newblock [hep-th/0304268].

\bibitem{Pinzul:2005nh}
A.~Pinzul and A.~Stern,
\newblock {\em A perturbative approach to fuzzifying field theories},
\newblock [hep-th/0502018].

\bibitem{Polychronakos:2000zm}
A.~P. Polychronakos,
\newblock {\em Flux tube solutions in noncommutative gauge theories},
\newblock Phys. Lett. {\bf B495}, 407 (2000),
\newblock [hep-th/0007043].

\bibitem{Seiberg:1999vs}
N.~Seiberg and E.~Witten,
\newblock {\em String theory and noncommutative geometry},
\newblock JHEP {\bf 09}, 032 (1999),
\newblock [hep-th/9908142].

\bibitem{Steinacker:2003sd}
H.~Steinacker,
\newblock {\em Quantized gauge theory on the fuzzy sphere as random matrix
  model},
\newblock Nucl. Phys. {\bf B679}, 66 (2004),
\newblock [hep-th/0307075].

\bibitem{Sykora:2004pt}
A.~Sykora,
\newblock {\em The application of *-products to noncommutative geometry and
  gauge theory},
\newblock [hep-th/0412012].

\bibitem{Szabo:2001kg}
R.~J. Szabo,
\newblock {\em Quantum field theory on noncommutative spaces},
\newblock Phys. Rept. {\bf 378}, 207 (2003),
\newblock [hep-th/0109162].

\bibitem{Wess:1991vh}
J.~Wess and B.~Zumino,
\newblock {\em Covariant Differential Calculus on the Quantum Hyperplane},
\newblock Nucl. Phys. Proc. Suppl. {\bf 18B}, 302 (1991).

\bibitem{Weyl:1927vd}
H.~Weyl,
\newblock {\em Quantum mechanics and group theory},
\newblock Z. Phys. {\bf 46}, 1 (1927).

\bibitem{Wimmer:2005bz}
R.~Wimmer,
\newblock {\em D0-D4 brane tachyon condensation to a BPS state and its
  excitation spectrum in noncommutative super Yang-Mills theory},
\newblock JHEP {\bf 05}, 022 (2005),
\newblock [hep-th/0502158].

\bibitem{Ydri:2002nt}
B.~Ydri,
\newblock {\em Noncommutative chiral anomaly and the {Dirac-Ginsparg-Wilson}
  operator},
\newblock JHEP {\bf 08}, 046 (2003),
\newblock [hep-th/0211209].

\bibitem{Ydri:2004im}
B.~Ydri,
\newblock {\em Exact solution of noncommutative {$U(1)$} gauge theory in 4-
  dimensions},
\newblock Nucl. Phys. {\bf B690}, 230 (2004),
\newblock [hep-th/0403233].

\bibitem{Ydri:2004vq}
B.~Ydri,
\newblock {\em Noncommutative {$U(1)$} gauge theory as a non-linear sigma
  model},
\newblock Mod. Phys. Lett. {\bf 19}, 2205 (2004),
\newblock [hep-th/0405208].

\end{thebibliography}

\end{document}